\documentclass[11pt]{amsart}
\usepackage{epsfig}

\usepackage{graphicx}

\newtheorem{theorem}{Theorem}[section]
\newtheorem{lemma}[theorem]{Lemma}
\newtheorem{proposition}[theorem]{Proposition}
\newcommand{\op}{\mbox{Op}}
\newcommand{\jacob}{{\mathcal J}({\mathbb T}^2)}
\newcommand{\lieg}{{\mathfrak{g}}}

\newcounter{figs}

\newcommand{\heis}{{\bf H}}
\newcommand{\grpalg}{{\mathcal A}_N}

\newcommand{\ctor}{{\mathcal L}_t({\mathbb T}^2\times [0,1])}
\newcommand{\rctor}{\widetilde{{\mathcal L}}_t({\mathbb T}^2\times [0,1])}
\newcommand{\cstor}{{\mathcal L}_t(S^1\times {\mathbb D}^2)}
\newcommand{\rcstor}{\widetilde{{\mathcal L}}_t({S^1\times \mathbb D}^2)}
\newtheorem{cor}[theorem]{Corollary}
\theoremstyle{definition}

\theoremstyle{remark}
\newtheorem{remark}[theorem]{Remark}

\numberwithin{equation}{section}

\title[Quantum mechanics and theta functions]{Quantum mechanics  
  and non-abelian theta functions for the gauge group $SU(2)$}
\author{R{\u{a}}zvan Gelca}
\address{Department of Mathematics and Statistics, 
Texas Tech University, Lubbock, TX 79409 }
\email{rgelca@gmail.com}
\thanks{Research of the first author supported by 
the NSF, award No. DMS 0604694}
\author{Alejandro Uribe}
\address{Department of Mathematics, 
University of Michigan, Ann Arbor, MI 48109}
\thanks{Research of the second author supported by the NSF, award No.
DMS 0805878}
\email{uribe@math.lsa.umich.edu}

\date{}


\begin{document}

\begin{abstract}
This paper outlines an approach to the non-abelian theta
functions of the $SU(2)$-Chern-Simons theory 
with the methods used by A. Weil for studying classical
theta functions. First we translate in  knot  theoretic language 
classical theta functions, the action of the finite Heisenberg 
group, and the discrete Fourier transform. Then we explain how
the non-abelian counterparts of these 
arise in the framework of the  quantum group quantization of the 
moduli space of flat $SU(2)$-connections on a surface, in the guise
of the non-abelian theta functions, the action 
of a skein algebra, and the Reshetikhin-Turaev
representation of the mapping class group. 
We  prove a Stone-von Neumann theorem on the moduli space of flat 
$SU(2)$-connections on the torus, and using it we deduce the existence
and the formula for the Reshetikhin-Turaev representation on the
torus from quantum mechanical considerations. We show how
one can derive in a quantum mechanical setting
the skein that allows handle slides,
which is the main ingredient in the construction of quantum
$3$-manifold invariants.  
\end{abstract}

\maketitle

\tableofcontents

\section{Introduction}\label{sec:1}
This paper outlines a study of the non-abelian theta functions that arise
in Chern-Simons theory by adapting the method used by Andr\'{e} Weil  for
studying classical theta functions \cite{weil}. We discuss the case of
the gauge group $SU(2)$, which is important because it corresponds to 
the Witten-Reshetikhin-Turaev topological quantum field theory, and
hence is  related to the Jones polynomial of knots \cite{witten},
\cite{reshetikhinturaev}. The methods can
be applied to more general gauge groups, which will be done in subsequent
work.

In Weil's approach, classical theta functions come with an action of the
finite Heisenberg group and a projective representation of the mapping class
 group. By analogy, our point of view  is 
 that the theory of non-abelian theta functions
consists of:

$\bullet$ the Hilbert space of non-abelian theta functions, namely the holomorphic
sections of the Chern-Simons line bundle;

$\bullet$  an irreducible representation on the space of theta functions of the
algebra generated by quantized Wilson lines (i.e. of the quantizations of traces of 
holonomies of simple closed curves);

$\bullet$ a projective representation of the mapping class group of
the surface on the space of non-abelian theta functions.

\noindent
The representation of the mapping class group intertwines the quantized
Wilson lines; in this sense the two representations satisfy the exact Egorov
identity. 

Non-abelian theta functions, quantized Wilson lines, and the projective
representations of the mapping class groups of surfaces
have each been studied  
separately;  we suggest that they should be studied together.
One should recall one instance when these  were considered together: 
the proof of the asymptotic faithfulness
of the Reshetikhin-Turaev representation 
\cite{freedmanwang}, \cite{andersen1}.

Our prototype  is the quantization of a one-dimensional particle.
The  paradigm is that {\em the quantum group quantization
of the moduli space of flat $SU(2)$-connections on a surface and
the Reshetikhin-Turaev representation of the mapping class group are the
analogues of the Schr\"{o}dinger representation of the Heisenberg group
and of the metaplectic representation}. The Schr\"{o}dinger representation 
arises from the quantization of the position and the momentum of a 
one-dimensional
free particle, and is a consequence of a fundamental postulate in quantum
mechanics. It is a unitary irreducible representation of the 
Heisenberg group, and the  Stone-von Neumann theorem shows that it is unique.
This uniqueness implies that linear
changes of coordinates (which act as outer 
automorphisms of the Heisenberg group)
are also quantizable,
and their quantization yields an infinite dimensional 
representation of the metaplectic group. 
 
  Weil \cite{weil} observed that a finite Heisenberg group 
 acts on classical theta functions, and that the
well known Hermite-Jacobi action of the modular group $SL(2,{\mathbb Z})$
is induced via a Stone-von Neumann theorem. 
Then it was noticed that classical theta functions, the action of
the Heisenberg group, and of the modular group  arise from the Weyl 
quantization of Jacobian varieties.  
As such,  classical theta functions are the holomorphic
 sections of a line bundle over the moduli space of flat $u(1)$-connections
 on a surface, and by analogy, the holomorphic sections 
of the similar line bundle
 over the moduli space of flat ${\lieg}$-connections over a surface (where
 ${\lieg}$ is the Lie algebra of a compact simple
 Lie group)  were called non-abelian theta functions.  Witten \cite{witten}  
placed non-abelian theta functions in the context of Chern-Simons theory, 
related them to the Jones polynomial \cite{jones} and conformal field theory,
and gave new methods for studying them.
 This had a great impact in the guise of
 the Verlinde formula which computes the dimension of the space of non-abelian
 theta functions. We explain how within Witten's theory one can
 find the non-abelian analogues of Weil's constructs. This is done  for
 the group $SU(2)$.

 Because the work joins methods from the theory
of theta functions, quantum mechanics,
representation theory, and low dimensional topology, we made it 
as self-contained as possible. At the heart of the paper lies a 
comparison between classical theta functions on the Jacobian variety
of a complex torus and the non-abelian
theta functions of the gauge group $SU(2)$.

We first review the prototype: the Schr\"{o}dinger
and metaplectic representations. Then, we recall the necessary
facts about classical theta functions  from \cite{gelcauribe2}, just for
the case of the torus. We point out two  properties of the finite
Heisenberg group: the Stone-von Neumann theorem, and the fact that
its group algebra is symmetric with respect to the action of the mapping
class group of the torus. Each is responsible
for the existence of the Hermite-Jacobi action by discrete Fourier
transforms on  theta functions. These properties have non-abelian
 counterparts, which we reveal later in the paper. 
We then rephrase  theta functions,  the Schr\"{o}dinger representation,
and the  discrete Fourier transform
in topological language following \cite{gelcauribe2}, using 
 skein modules \cite{turaev2}, \cite{przytycki}. 

The fact that skein modules can be used to describe classical theta functions
is a corollary of Witten's Feynman integral approach to the Chern-Simons
theory for the group $U(1)$. In fact, Witten has explained in \cite{witten}
that the topological quantum field theory of any Lie group gives rise
to skein relations, hence it should have a skein theoretic version. For
example, the skein theoretic version for the gauge group
$SU(2)$  was constructed in \cite{BHMV}. We  showed in 
\cite{gelcauribe2} that the skein modules of the abelian theory
arise naturally from quantum mechanics,
without relying on quantum field theory.

We then recall the construction of 
 non-abelian theta functions from the 
quantization of the moduli space of flat connections
on a surface, and present in detail the case of the torus, 
where the moduli space is the pillow case.
The paper continues with a description of the quantum group quantization of 
the moduli space of flat $SU(2)$-connections on a surface, 
rephrased  into the language of the skein modules. We deduce that
the quantum group  quantization is defined by the left action 
of a skein algebra on a quotient of itself. Because we are interested
in the multiplicative structure, we use the skein relations of the
Reshetikhin-Turaev invariant instead of those of the Kauffman bracket,
since the latter introduce sign discrepancies. We recall
our previous result \cite{gelcauribe} that on the pillow case, 
the quantum group  and Weyl quantizations coincide. Then we prove 
a Stone-von Neumann theorem on the pillow case.

An application  is to deduce the existence
of the Reshetikhin-Turaev representation on the torus as a consequence
of this Stone-von Neumann theorem. We show how the
explicit formula for this representation can be computed from quantum
mechanical considerations. As such we arrive at the
element $\Omega$ by which one colors knots as to allow Kirby
moves along them. This element is the fundamental building
block in the construction of quantum $3$-manifold invariants (see
\cite{reshetikhinturaev}, \cite{BHMV}, \cite{lickorish}), and
we give a natural way to derive its existence.

We conclude  by explaining how the Reshetikhin-Turaev
representation of the mapping class group of a surface is a non-abelian
analogue of the Hermite-Jacobi action given by discrete Fourier transforms.

\section{The prototype}\label{sec:2}

\subsection{The Schr\"{o}dinger representation}\label{sec:2.1}
In this section we review briefly the Schr\"{o}\-din\-ger and the
metaplectic representations.
For a detailed discussion we suggest  \cite{lionvergne} and \cite{polishchuk}.

In the canonical formalism, a classical mechanical system is described
by a symplectic manifold $(M^{2n},\omega)$, which is the phase space of the
system. The classical observables are $C^{\infty}$ functions on $M$. To each
$f\in C^{\infty}(M^{2n},{\mathbb R})$ one associates a Hamiltonian vector
field $X_f$ on $M^{2n}$ by
$df(\cdot)=\omega (X_f,\cdot).$
This vector field defines a Hamiltonian flow on the manifold which preserves
the form $\omega$.
The symplectic  form  defines a Poisson bracket by $\{f,g\}=
\omega (X_f,X_g)$. There is a special observable $H$,  called  the Hamiltonian
(total energy) of the system. The time evolution of an observable is described
by the equation $$\frac{df}{dt}=\{f,H\}.$$

Quantization replaces the symplectic manifold by a Hilbert space, real-valued
observables $f$
by self-adjoint operators $\op(f)$ called quantum observables,  Hamiltonian flows by 
one-parameter groups of unitary operators, and the  Poisson bracket of $\{f,g\}$ 
by $\frac{\hbar}{i}[\op(f),\op(g)]$, where $\hbar$ is Planck's constant and
$[\cdot,\cdot]$ is the commutator of operators. 
Dirac's conditions should hold: $\op(1)=Id$ and  $$\op(\{f,g\})
=\frac{i}{\hbar}[\op(f),\op(g)]+O(\hbar).$$ The second condition is phrased by
that the quantization is performed in the direction of the Poisson
 bracket. The time evolution of a quantum observable is described 
by Schr\"{o}dinger's equation
\begin{eqnarray*}
i\hbar\frac{d\op(f)}{dt}=[\op(f),\op(H)].
\end{eqnarray*}

A fundamental example is that of a 
  particle in a $1$-dimensional space, which we  discuss in the
case where Planck's constant is  equal to $1$.
The phase space is ${\mathbb
  R}^2$, with coordinates the position $x$ and the momentum $y$, and symplectic
form $\omega=dx\wedge dy$. The associated 
Poisson bracket is
\begin{eqnarray*}
\{f,g\}=\frac{\partial f}{\partial x}\frac{\partial g}{\partial y}-
\frac{\partial f}{\partial y}\frac{\partial g}{\partial x}.
\end{eqnarray*}
The symplectic form $\omega$ induces a nondegenerate antisymmetric bilinear
form on ${\mathbb R}^2$, also denoted by $\omega$, given by $\omega((x,y),(x',y'))=xy'-x'y$.

The Lie algebra of observables has a subalgebra 
generated by $Q(x,y)=x$,
$P(x,y)=y$, and $E(x,y)=1$, called the Heisenberg Lie algebra. 
Abstractly, this algebra is defined by
$[Q,P]=E, [P,E]=[Q,E]=0$.

It is a  postulate of quantum mechanics that the quantization of the
position,  the momentum, and the constant functions 
is the representation of the Heisenberg Lie algebra on
$L^2({\mathbb R}, dx)$  defined by
\begin{eqnarray*}
Q\rightarrow  M_x,\quad 
P\rightarrow \frac{1}{i}\frac{d }{d x},\quad
E\rightarrow i Id.
\end{eqnarray*}
Here $M_x$ denotes the operator of multiplication by the variable:
$\phi(x)\rightarrow x\phi(x)$. 
This is  the Schr\"{o}dinger representation of the Heisenberg Lie algebra.

The Lie group of the Heisenberg Lie algebra is the Heisenberg group. 
It is defined as ${\mathbb R}^3$ with the multiplication rule
\begin{eqnarray*}
(x,y,t)(x',y',t')=\left(x+x',y+y',t+t'+\frac{1}{2}\omega((x,y),(x',y'))\right).
\end{eqnarray*}
It is standard to denote $\exp(xQ+yP+tE)=(x,y,t)$.
By exponentiating the Schr\"{o}dinger representation of the Lie algebra one
 obtains the Schr\"{o}dinger representation of the Heisenberg group:
\begin{eqnarray*}
\exp(Q)\rightarrow e^{2\pi i M_x}, \quad 
\exp(P)\rightarrow e^{-\frac{d}{d x}},
\quad
\exp(E)\rightarrow e^{2\pi i  Id}.
\end{eqnarray*}
Specifically, for $\phi\in L^2({\mathbb R},dx)$, 
\begin{eqnarray*}
&\exp(y_0P)\phi(x)=\phi(x-y_0),\quad
\exp(x_0Q)\phi(x)=e^{2\pi i xx_0}\phi(x),\\
&\exp(tE)\phi(x)=e^{2\pi i t}\phi(x),
\end{eqnarray*}
meaning that $\exp(y_0P)$ acts as a translation, $\exp(x_0Q)$ acts as
the multiplication by a character, and $\exp(tE)$ acts as the multiplication
by a scalar. 
This is the rule for quantizing  exponential
functions. Specifically, $\exp (x_0Q+y_0P+tE)$ is the 
quantization of the function $f(x,y)=\exp(2\pi i(x_0x+y_0y+t))$.

Extending by linearity one obtains the quantization of  the group
ring 
of the Heisenberg group. This was further
generalized by Hermann Weyl, who gave
a method for quantizing all functions $f\in C^{\infty}({\mathbb R}^2)$
by using the Fourier transform
\begin{eqnarray*}
\hat{f}(\xi,\eta)=\iint f(x,y)\exp(-2\pi ix\xi-2\pi iy\eta)dxdy
\end{eqnarray*}
and then defining
\begin{eqnarray*}
\op(f)=\iint \hat{f}(\xi,\eta)\exp2\pi i(\xi Q+\eta P)d\xi d\eta,
\end{eqnarray*}
where for $\exp(\xi Q+\eta P)$ he used the Schr\"{o}dinger representation.

\medskip

\noindent 
{\bf Theorem} (Stone-von Neumann) The Schr\"{o}dinger representation of the
Heisenberg group is the unique irreducible unitary  representation
 of this group such that $\exp(tE)$ acts as $e^{2\pi it}Id$ 
for all $t\in {\mathbb R}$. 

\medskip

There are two other important realizations of the irreducible representation that this
theorem characterizes.  One comes from the quantization of the plane in
a holomorphic polarization. 
The  Hilbert space is the Bargmann space,
\begin{eqnarray*}
\mbox{Bargmann}({\mathbb C})=\left\{f: {\mathbb C}\to {\mathbb C} \mbox{ holomorphic }, \int_{\mathbb C}
|f(z)|^2e^{-2\pi|\mbox{Im }z|^2}dz\wedge d\bar{z}<\infty\right\} ,
\end{eqnarray*}
where the Heisenberg group acts by 
\begin{eqnarray*}
&\exp(x_0P)f(z)=f(z-x_0),\quad
\exp(y_0Q)f(z)=e^{\pi  (y_0^2-2iy_0z)}f(z+iy_0),\\
&\exp(tE)f(z)=e^{2\pi i  t}f(z).
\end{eqnarray*}

For the other  one has to choose a Lagrangian subspace
${\bf L}$ of ${\mathbb R}P+{\mathbb R}Q$ (which in this case is just
a one-dimensional subspace). Then $\exp ({\bf L}+{\mathbb R}E)$  is
a maximal abelian subgroup of the Heisenberg group. Consider
the character of this subgroup defined by 
$\chi_{\bf L}(\exp(l+tE))=e^{2\pi i t},  l\in {\bf L}.$
The Hilbert space of the quantization,  
${\mathcal H}({\bf L})$, is defined as the
 space of functions $\phi(u)$ on $\heis ({\mathbb R}) $ 
satisfying 
\begin{eqnarray*}
\phi (uu')=\chi_L(u')^{-1}\phi(u) \mbox{ for all }u'\in \exp({\bf L}+{\mathbb R}E) 
\end{eqnarray*}
and
 such that $u\rightarrow |\phi(u)|$ is a
square integrable function on the left equivalence classes modulo
$\exp({\bf L}+{\mathbb R}E)$.
The representation of the Heisenberg group is given by 
\begin{eqnarray*}
u_0\phi(u)=\phi(u_0^{-1}u).
\end{eqnarray*}

If we choose an algebraic complement ${\bf L}'$ of ${\bf L}$, meaning that
we write ${\mathbb R}P+{\mathbb R}Q={\bf L}+{\bf L}'={\mathbb R}+{\mathbb R}$,
 then ${\mathcal H}({\bf L})$ can be realized as
 $L^2({\bf L}')\cong L^2({\mathbb R})$. 
Under a natural isomorphism, 
\begin{eqnarray*}
&&\exp (x_0)\phi(x)=\phi(x-x_0) , \quad x,x_0\in {\bf L}'\\
&& \exp (y_0)\phi(x)=e^{2\pi i \omega (x,y_0)}\phi(x), \quad x\in {\bf L}',
y_0\in {\bf L}\\
&&\exp (tE)\phi(x)=e^{2\pi i t}\phi(x), \quad x\in {\bf L'}
\end{eqnarray*}
 where $\omega $ is the standard symplectic form on 
${\mathbb R}P+{\mathbb R}Q$.
For ${\bf L}={\mathbb R}P$ and ${\bf L}'={\mathbb R}Q$,
 one obtains the standard
Schr\"{o}dinger representation in the position representation.
 For $L={\mathbb R}Q$ and $L'={\mathbb R}P$, one obtains
the Schr\"{o}dinger representation in 
the momentum representation: $\exp(y_0P)\phi(x)=e^{-2\pi i xy_0}\phi(x)$,
$\exp(x_0Q)\phi(x)=\phi(x-x_0)$, $\exp(tE)\phi(x)=e^{2\pi i t}\phi(x)$.


 \subsection{ The metaplectic representation}\label{sec:2.2}

The Stone-von Neumann theorem implies that if we change coordinates by
a linear symplectomorphism and then quantize, we obtain a unitary equivalent
representation of the Heisenberg group.  
Hence linear symplectomorphisms can be quantized, giving rise to unitary
operators, although they do not arise from Hamiltonian flows.
 Irreducibility implies, by Schur's lemma, that these
operators are unique up to a multiplication by a constant.
Hence we have a projective representation $\rho$  of the linear 
symplectic group $SL(2,{\mathbb R})$ on $L^2({\mathbb R})$.
This  can be made into a true representation by passing
to the double cover
 of $SL(2,{\mathbb R})$, namely to  the
metaplectic group $Mp(2,{\mathbb R})$. 
The representation of the metaplectic group
is known as the {\em metaplectic representation} or the
 Segal-Shale-Weil representation.

The fundamental symmetry that Weyl quantization 
has is that, if $h\in Mp(2,{\mathbb R})$, then
\begin{eqnarray*}
\op(f\circ h^{-1})=\rho(h)\op(f)\rho(h)^{-1},
\end{eqnarray*}
for every observable $f\in C^{\infty}({\mathbb R}^2)$, where $\op(f)$
is the operator associated to $f$ through Weyl quantization. 
For other quantization models this relation holds only
 mod $O(\hbar)$, ({\em Egorov's theorem}). 
When it is satisfied with equality, as it is  in our case,
it is called the {\em exact} Egorov identity. 

An elegant way to define the metaplectic representation is
to use the third version of the Schr\"{o}dinger representation
discussed in the previous section,
which identifies the metaplectic representation as a  Fourier 
transform (see \cite{lionvergne}). Let $h$ be a linear symplectomorphism of
the plane, then let  ${\bf L}_1$ be a Lagrangian
subspace of ${\mathbb R}P+{\mathbb R}Q$ and  ${\bf L}_2=h(L_1)$. 
Define the quantization of $h$ as 
$\rho(h):{\mathcal H}({\bf L}_1)\rightarrow {\mathcal H}({\bf L}_2)$,
\begin{eqnarray*}
(\rho(h)\phi)(u)=\int_{\exp {\bf L}_2/\exp ({\bf L}_1\cap {\bf L}_2)}
\phi(uu_2)\chi_{{\bf L}_2}(u_2)d\mu(u_2),
\end{eqnarray*}
where $d\mu$ is the measure induced on the space of equivalence
classes by the Haar measure on $\heis({\mathbb R})$.

To write explicit formulas for $\rho(h)$ one needs to choose 
the algebraic complements ${\bf L}_1'$ and ${\bf L}_2'$ of ${\bf L}_1$ and
${\bf L}_2$
and unfold the isomorphism $L^2({\bf L}')\cong L^2({\mathbb R})$.
For example, for
\begin{eqnarray*}
S=\left( 
\begin{array}{rr}
0& 1\\
-1& 0
\end{array}
\right),
\end{eqnarray*}
if we set ${\bf L}_1={\mathbb R}P$ with variable $y$ and $L_2=S({\bf L}_1)
={\mathbb R}Q$
with variable $x$ and ${\bf L}_1'={\bf L}_2$ and ${\bf L}_2'=S({\bf L}_1')
={\bf L}_1$, then 
\begin{eqnarray*}
\rho(S)f(x)=\int_{\mathbb R} f(y)e^{-2\pi i xy}dy,
\end{eqnarray*}
 the usual Fourier transform, which establishes the unitary
equivalence between the position and the momentum representations. 
Similarly, for  \begin{eqnarray*}
T_a=\left(
\begin{array}{rr}
1& a\\
0 &1
\end{array}
\right),
\end{eqnarray*}
if we set ${\bf L}_1={\bf L}_2={\mathbb R}P=$, 
${\bf L}_1'={\mathbb R}Q$, and ${\bf L}_2'={\mathbb R}(P+Q)$, then 
\begin{eqnarray*}
\rho(T_a)f(x)=  e^{2\pi i x^2a}f(x)
\end{eqnarray*}
which is the multiplication by the exponential of a quadratic function.
 These are the well known formulas that define the action
of  the metaplectic group on $L^2({\mathbb R},dx)$.


The cocycle of the projective representation of the symplectic group
is 
\begin{eqnarray*}
c_L(h',h)=e^{-\frac{i\pi}{ 4}{\boldsymbol \tau}({\bf L},h({\bf L}),h'\circ h({\bf L}))}
\end{eqnarray*}
where ${\boldsymbol \tau}$ is the Maslov index. This means that
\begin{eqnarray*}
\rho(h'h)=c_{\bf L}(h',h)\rho(h')\rho(h)
\end{eqnarray*}
 for $h,h'\in SL(2,{\mathbb R})$.

\section{Classical theta functions}\label{sec:3}

\subsection{Classical theta functions from the quantization of the torus}\label{sec:3.1}

For an extensive treatment of  theta functions the reader can consult
\cite{mumford}, \cite{lionvergne}, \cite{polishchuk}.
We  consider the simplest situation, that of theta functions
on the Jacobian variety of a 2-dimensional complex torus ${\mathbb T}^2$.
Our discussion is sketchy, the details can be found, for the
 case of all closed Riemann surfaces, in \cite{gelcauribe2}.

\begin{figure}[h]
\centering
\scalebox{.35}{\includegraphics{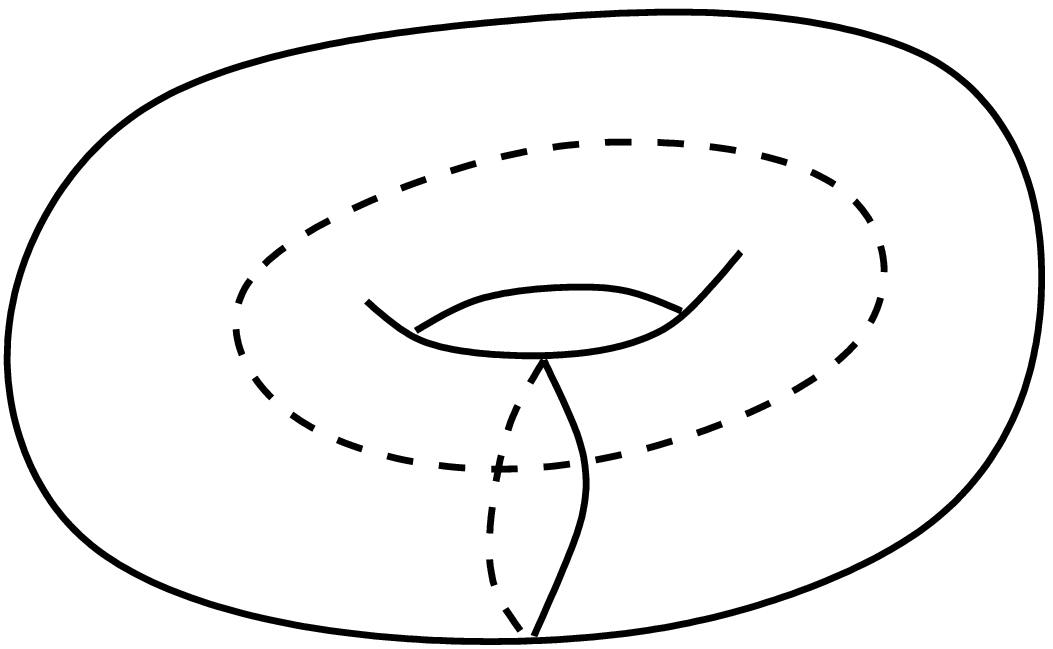}}
\caption{}
\label{rigtorus}
\end{figure}

Given the complex torus and two simple closed curves $a$ and $b$ (see 
Figure~\ref{rigtorus})
which define a canonical basis 
of $H_1({\mathbb T}^2,{\mathbb R})$ (or equivalently of $\pi_1({\mathbb
  T}^2)$),
 consider a holomorphic 1-form $\zeta$ such that $\int_a\zeta=1$.
Then the complex number $\tau=\int_b\zeta$, which 
 depends on the complex structure,
has  positive imaginary part. The {\em Jacobian variety} associated
to ${\mathbb T}^2$, denoted by $\jacob$, is a 2-dimensional
torus with complex structure obtained by viewing $\tau$ as an element
in its Teichm\"{u}ller space. Equivalently, 
\begin{eqnarray*}
\jacob={\mathbb C}/{\mathbb Z}+{\mathbb Z}\tau .
\end{eqnarray*}  
We introduce real coordinates $(x,y)$ on $\jacob$ by setting
$z=x+\tau y$. In these coordinates, $\jacob$ is the quotient of the plane
by the integer lattice.  
The Jacobian variety is endowed with the canonical symplectic form
$\omega =dx\wedge dy$,
which is a generator of 
$H^2({\mathbb T}^2,{\mathbb Z})$. $\jacob$
 with its complex structure
and this symplectic form is a K\"{a}hler manifold.

In short, classical theta functions and the action of the Heisenberg group
on them can be obtained by applying Weyl quantization to $\jacob$ in the 
holomorphic polarization.   
To obtain theta functions, we apply 
the procedure of geometric quantization. We start by 
setting $\hbar = \frac{1}{N}$, with $N$ a positive even integer. 

The Hilbert space of the quantization consists of the {\em classical 
theta functions}, which are the holomorphic 
sections of a line bundle over the Jacobian variety.
This line bundle is the tensor product of a line bundle of curvature $-2\pi i
N\omega$ and a half-density. By pulling back the line bundle to ${\mathbb C}$,
we can view  these sections as entire functions satisfying certain
periodicity conditions. The line bundle with curvature $2\pi i N\omega$
is unique up to tensoring with a flat bundle. Choosing the latter
appropriately, we can ensure that the periodicity conditions are
\begin{eqnarray*}
f(z+m+n\tau)=e^{-2\pi iN(\tau n^2+2nz)}f(z).
\end{eqnarray*}
 An orthonormal basis of the space of classical theta functions  is given by
the {\em theta series}
\begin{equation}\label{thetabasis}
\begin{array}{ll}
\theta_j^\tau (z)=
  \sum_{n\in {\mathbb Z}}e^{2\pi i N\left[\frac{\tau}{2}\left(\frac{j}{N}+n\right)^2 +
z\left(\frac{j}{N}+n\right)\right]},\quad j=0,1,2,\ldots, N-1.
\end{array}
\end{equation}
It will be convenient to extend this  definition to all
indices $j$ by the periodicity condition
$\theta_{j+N}^\tau(z)=\theta_j^\tau(z)$,
namely to take indices modulo $N$.

Given a complex torus  one doesn't get automatically
the theta series,  one needs a pair of
generators of the fundamental group. 
Here, the generators $a$ and $b $ of the fundamental group of the
original torus give rise to the curves on the Jacobian variety 
that are the images of the curves in ${\mathbb C}$
from $0$ to $1$ respectively from $0$ to $\tau$.
The complex structure and these two generators of 
$\pi_1(\jacob)$ define a point in the 
Teichm\"{u}ller space of $\jacob$, which is  parametrized
 by the complex number $\tau$.

Let us  turn to the operators.
The only exponentials on the plane that are double periodic, and therefore
give rise to functions on the torus, are
\begin{eqnarray*}
f(x,y)=\exp 2\pi i (mx+ny), \quad m,n \in {\mathbb Z}.
\end{eqnarray*}
Since the torus is a quotient of the plane by a discrete group, we 
can apply the Weyl quantization procedure. In the complex
polarization Weyl quantization is defined as follows (see \cite{folland}):
A fundamental domain of the torus is the unit square $[0,1]\times [0,1]$
(this is done in the $(x,y)$ coordinates, in the complex plane it
is actually a parallelogram).
The value of a theta function is completely determined by its values
on this unit square. The Hilbert space of classical theta functions
can be isometrically embedded into $L^2([0,1]\times [0,1])$ with
the inner product
\begin{eqnarray*}
\left<f,g\right>=\left(-i N(\tau-\bar{\tau})\right)^{1/2}
\int_0^1\int_0^1f(x,y)\overline{g(x,y)}
e^{iN(\tau-\bar{\tau})\pi y^2}dxdy.
\end{eqnarray*}

The operator associated by Weyl quantization to a function $f$ on the torus
is the Toeplitz operator with symbol
$e^{-\frac{\hbar\Delta _\tau}{4}}f$, where $\Delta_\tau$ is the Laplacian,
which in the $(x,y)$ coordinates is given by the formula
\begin{equation}\label{delta}
\begin{array}{ll}
\Delta_\tau & = \frac{i}{\pi(\tau-\bar{\tau})}\left[\tau\bar{\tau}\frac{\partial ^2}{
\partial x^2}-(\tau+\bar{\tau})\frac{\partial ^2}{\partial x\partial y}+
\frac{\partial ^2}{\partial y^2}\right].
\end{array}
\end{equation}
This means that the  Weyl quantization of $f$ maps a classical theta function 
$g$ to the orthogonal
projection onto the Hilbert space of classical theta functions
of $(e^{-\frac{\Delta _\tau}{4N}}f)g$. The following result is standard; see
\cite{gelcauribe2} for a proof.

\begin{proposition}  The Weyl quantization of the exponentials is given by
\begin{eqnarray*}
\mbox{Op}\left(e^{2\pi i(px+qy)}\right)\theta_j^\tau(z)=
e^{-\frac{\pi i}{N}pq-\frac{2\pi i}{N}jq}\theta_{j+p}^\tau(z).
\end{eqnarray*}
\end{proposition}

The Weyl quantization of the exponentials gives rise to the Schr\"{o}dinger
 representation
of the Heisenberg group with integer entries $\heis ({\mathbb Z})$
onto the space of theta functions. This Heisenberg group is 
\begin{eqnarray*}
\heis ({\mathbb Z})=\{(p,q,k), \; p,q,k\in{\mathbb Z}\}
\end{eqnarray*}
with multiplication 
\begin{eqnarray*}
(p,q,k)(p',q',k')=(p+p',q+q',k+k'+(pq'-qp')).
\end{eqnarray*}
The proposition implies that
\[
(p,q,k)\mapsto \ \mbox{the Weyl quantization of }e^{\frac{\pi i}{N}k}
\exp2\pi i(px+qy)
\]
is a group morphism.  This is the Schr\"odinger representation.

The Schr\"odinger representation of 
 $\heis ({\mathbb Z})$ is far from faithful.
Because of this we factor it out by its kernel. The kernel
is the subgroup consisting
 of the elements of the form $(p,q,k)^N$, with $k$ even \cite{gelcauribe2}.
Let $\heis ({\mathbb Z}_N)$ be the finite Heisenberg group
obtained by factoring  $\heis({\mathbb Z})$ by
this  subgroup, and let $\exp (pP+qQ+kE)$ be the image of $(p,q,k)$
in it. Then
\begin{eqnarray*}
\exp(pP)\theta_j^\tau=\theta_{j+p}^\tau,\quad
\exp(qQ)\theta_j^\tau=e^{-\frac{2\pi i}{N}qj}\theta_j^\tau,\quad
  \exp(kE)\theta_j^\tau=
e^{\frac{\pi i}{N}k}\theta_j^\tau 
\end{eqnarray*}
for all $p,q,k,j\in {\mathbb Z}$.

The following is a well known result (see for example \cite{gelcauribe2} for a
proof). 

\begin{theorem}
(Stone-von Neumann)  The Schr\"{o}dinger representation
of $\heis ({\mathbb Z}_N)$ is the {\em unique} irreducible
unitary representation of this group with the property that 
$\exp(kE)$ acts as $e^{\frac{\pi i}{N}k}Id$ for all $k\in {\mathbb Z}$.
\end{theorem}

We must mention another important representation of 
the finite Heisenberg group,
which by the Stone-von Neumann theorem is unitary equivalent to this
one. It comes from the quantization of the torus in
a real polarization. An orthonormal basis of the Hilbert space
is given by  Bohr-Sommerfeld leaves of the polarization, 
and so the  Hilbert space can be identified with $L^2({\mathbb Z}_{N})$. 
If the polarization is given by ``$Q$ curves", the
finite Heisenberg group  acts by 
\begin{eqnarray*}
\exp(pP)f(j)=f(j-p), \exp(qQ)f(j)=e^{-\frac{2\pi i}{N}qj}f(j),
\exp(kE)f(j)=e^{\frac{\pi i}{N}k}f(j).
\end{eqnarray*}
Thus $\exp(pP)$ acts as translation and $\exp(qQ)$ acts as a multiplication
by a character of ${\mathbb Z}_{N}$. 
The  characteristic functions of the singletons: $\delta _i(j)=\delta_{ij}$,
 $i=0, 1,2,\ldots, N-1$ correspond to the theta series 
 $\theta_0^\tau,\theta_1^\tau,\ldots, \theta_{N-1}^\tau$ through the
unitary isomorphisms that identifies the two representations. Note that a 
left shift in the index corresponds to a right shift in the variable.
\begin{remark}
There is a sign discrepancy between these formulas
and the ones from \S~\ref{sec:2.1}, 
which shows up in the exponent of the formula
for $\exp (qQ)$. This is due to a 
disagreement between the standard notations in quantum mechanics and 
topology, in which the roles of the letters $p$ and $q$ are exchanged. Because
Chern-Simons theory has been studied extensively by topologists,
we use the convention for $p$ and $q$  from topology. We
point out that the above formulas describe the action of the Heisenberg group
in the the momentum representation.    
\end{remark}

The Schr\"{o}dinger representation of the finite Heisenberg group can be
extended by linearity to a representation of the group algebra with
coefficients in ${\mathbb C}$ of the finite Heisenberg group, 
${\mathbb C}[\heis ({\mathbb Z}_N)]$. Since the elements of 
$\exp({\mathbb Z}E)$ act as multiplications by constants, 
this is in fact a representation of the algebra $\grpalg$ obtained by
factoring ${\mathbb C}[\heis ({\mathbb Z}_N)]$ by the relations
$\exp(kE)-e^{\frac{\pi i }Nk}$ for all $k\in {\mathbb Z}$. By abuse
of language, we will call this representation the Schr\"{o}dinger
representation as well. The Schr\"{o}dinger representation of $\grpalg$  
defines the quantizations of trigonometric polynomials on the torus. 

\begin{proposition}\label{allspace}
a) The algebra of Weyl quantizations of trigonometric polynomials is
the algebra of all linear operators on the space of theta functions.\\
b) The Schr\"{o}dinger  representation of the 
algebra ${\grpalg}$ on theta functions is faithful. 
\end{proposition}

\begin{proof} 
For a proof of part a) see \cite{gelcauribe2}. Part b) follows from the fact
that
 $\exp(pP+qQ)$, $p,q=0,1,\ldots, N-1$, form a basis of  
${\grpalg}$ as a vector space. 
\end{proof}

Because of this result, we may identify $\grpalg$ with the algebra
of Weyl quantizations of trigonometric polynomials.
$\grpalg$ can also  be described in terms of the noncommutative torus.
(The relevance of the noncommutative torus for Chern-Simons
theory was first revealed in \cite{frohmangelca} for  the
gauge group $SU(2)$.) 

The ring of trigonometric polynomials 
in the noncommutative torus is ${\mathbb C}_t[U^{\pm 1},V^{\pm 1}]$,
the ring of Laurent polynomials in the variables $U$ and
$V$ subject to the noncommutation relation $UV=t^2VU$. The
noncommutative torus itself is  a C$^*$-algebra in which 
${\mathbb C}_t[U^{\pm 1},V^{\pm 1}]$ is dense,
viewed as a deformation quantization of the algebra of smooth functions
on the torus \cite{rieffel}, \cite{connes}.  The group
algebra with coefficients in ${\mathbb C} $ of the Heisenberg group
$\heis ({\mathbb Z})$ is isomorphic to the ring of trigonometric 
polynomials in the noncommutative torus, with the isomorphism given by
\begin{eqnarray*}
(p,q,k) \rightarrow t^{k-pq}\, U^pV^q,\mbox{ for }p,q,k\in {\mathbb Z}.
\end{eqnarray*}

If we set $U^N=V^N=1$ and $t=e^{\frac{\pi i}{N}}$, we  obtain the 
noncommutative torus at a root of unity $\widetilde{{\mathbb C}_t}
[U^{\pm 1},V^{\pm 1}]$.  To summarize:

\begin{proposition}
The algebra, $\grpalg$,  of the Weyl quantizations of trigonometric polynomials
on the torus is isomorphic to $\widetilde{{\mathbb C}_t}[U^{\pm 1},V^{\pm 1}]$.
\end{proposition}

As explained in \cite{gelcauribe2}, the Schr\"{o}dinger representation
can be described as the left regular action of the group algebra of
the finite Heisenberg group on a quotient of itself. The construction is
similar to that for the metaplectic representation in the abstract setting
from \S\ref{sec:2.2}.

\subsection{Classical theta functions from a topological perspective}\label{sec:3.3}

In \cite{gelcauribe2} the  theory of classical theta functions
was shown to admit a reformulation  in purely 
 topological language. Let us recall the facts.

Let $M$ be an orientable
 $3$-dimensional manifold. A  framed
link in $M$ is a smooth embedding of a disjoint union of finitely many
annuli. We consider framed oriented links, where the orientation of a
link component is an orientation of one of the boundary components of
the annulus. We draw all diagrams in the blacboard framing, meaning that
the annulus is parallel to the plane of the paper. 

\begin{figure}[h]
\centering
\scalebox{.25}{\includegraphics{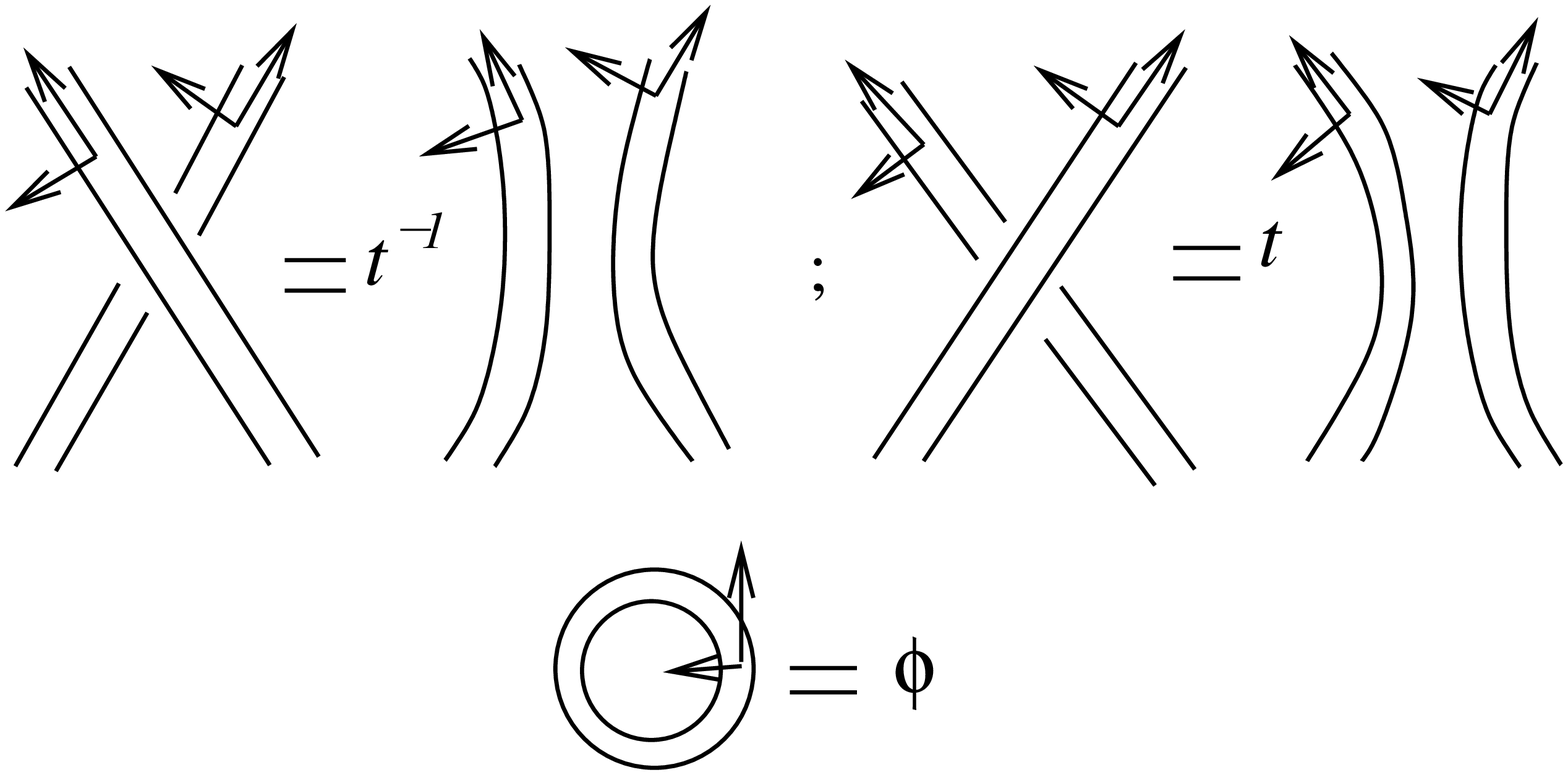}}
\caption{}
\label{skeins}
\end{figure}

Consider 
the free ${\mathbb C}[t,t^{-1}]$-module with basis the set of isotopy classes
of framed oriented links in $M$, including the empty link $\emptyset$. 
Factor it by all equalities  of the form depicted in 
 {Figure~\ref{skeins}}. 
In each of these diagrams,
the two links are  identical except for an embedded ball
in $M$, inside of which they look as shown. This means that in each link
we are allowed to smoothen a crossing provided that we add a coefficient of
$t$ or $t^{-1}$, and any trivial link component can be ignored. These are
called {\em skein relations}. For normalization reasons, we
 add the skein relation that
identifies the trivial knot with $\emptyset$. 
The skein relations are considered for
all possible embeddings of a ball. 
 The result of the factorization
is called  the {\em linking number skein module of $M$},   denoted  by 
${\mathcal L}_t(M)$. 

If $M$ is a $3$-dimensional sphere, then each link $L$
is, as an element of ${\mathcal L}_t(S^3)$, equivalent to the empty link
with the coefficient equal to $t$ raised to the sum of the  linking numbers
of ordered 
pairs of components and the writhes of the components, hence the name. 

These skein modules were first introduced by Przytycki in \cite{przytycki2}.
He pointed out
that they represent one-parameter deformations of the group ring of 
$H_1(M,{\mathbb Z})$ and computed them for all $3$-dimensional 
manifolds. 

For the fixed  positive integer $N$
we define the {\em reduced} linking number skein module of $M$,
denoted by $\widetilde{\mathcal L}_t(M)$, as
the quotient of ${\mathcal L}_t(M)$ obtained by setting $t=e^{\frac{i\pi}{N}}$
and  $\gamma^N=\emptyset$ for every skein $\gamma$
consisting of one link component, where $\gamma^N$  denotes
$N$ parallel copies of $\gamma$. 
 As a rule followed throughout the paper, in a skein module $t$ is a
free variable, while in a reduced skein module it is a root of unity.

If $M={\mathbb T}^2\times [0,1]$,
the topological operation of  gluing one cylinder on top of another
induces a multiplication in $\ctor$ which turns $\ctor$ into an algebra, the
{\em linking number skein algebra} of the cylinder over the torus.
This multiplication descends to $\rctor$. 
We want to explicate its structure. 

For $p$ and $q$  coprime integers,  orient the curve $(p,q)$
by the vector that joins the origin to the point $(p,q)$, and frame it
 so that the annulus is parallel to the torus. Call this the zero framing,
or the {\em blackboard framing}. 
Any other
 framing of the curve $(p,q)$ can be represented by an integer $k$, where
$|k|$ is the number of full twists that are inserted on this curve, with
$k$ positive if the twists are positive, and $k$ negative if the twists
are  negative. Note that in $\ctor$,
$(p,q)$ with framing $k$ is equivalent to $t^k(p,q)$. 

If   $p$ and $q$ are not coprime and  $n$ is their greatest common divisor,
we denote by  $(p,q)$ the framed link consisting of $n$ parallel copies of 
$(p/n,q/n)$, namely $(p,q)=(p/n,q/n)^n$.
Finally, $\emptyset=(0,0)$ is the empty link,  the multiplicative
identity of $\ctor$.

\begin{theorem}\cite{gelcauribe2}
The  algebra $\ctor$ is isomorphic to the 
group algebra ${\mathbb C}[\heis({\mathbb Z})]$, with 
 the isomorphism induced by 
\begin{eqnarray*}
 t^k(p,q)\rightarrow (p,q,k).
\end{eqnarray*}
This map descends to an isomorphism between $\rctor$ 
and the algebra ${\grpalg}$
of Weyl quantizations of trigonometric polynomials.
\end{theorem}
 
 \begin{remark}
The determinant is the sum of the algebraic intersection numbers
of the curves in $(p,q,k)$ with the curves in $(p',q',k')$, so the
multiplication rule of the Heisenberg group is defined using the
algebraic intersection number of curves on the torus. 
\end{remark}

Identifying the group algebra of the Heisenberg group with integer entries with
${\mathbb C}_t[U^{\pm 1}, V^{\pm 1}]$, we obtain the following 

\begin{cor}
The linking number skein algebra of the cylinder over the torus is isomorphic to the 
ring of trigonometric polynomials in the noncommutative torus.
\end{cor}

Let us look at the skein module 
of the solid torus $\cstor$. Let $\alpha$ be
the curve that is the core of the solid torus, with a certain choice of 
orientation and framing.  The
 reduced linking number skein module $\rcstor$ has basis $\alpha^j$,
$j=0,1,\ldots, N-1$.

Let $h_0$
be a homeomorphism of the torus to the boundary of the solid torus
that maps the first generator of the fundamental group
 to a curve isotopic to $\alpha$ (a {\em longitude})
and the second generator
to the curve on the boundary of the solid torus that 
 bounds a disk in the solid torus (a {\em meridian}). 
 The operation of gluing ${\mathbb T}^2\times [0,1]$ to 
the boundary of $S^1\times{\mathbb D}^2$ 
via $h_0$ induces a left action of $\ctor $ onto
$\cstor$. This descends to a left action of  $\rctor$ onto $\rcstor$. 

It is important to observe that $\cstor$ and $\rcstor$ are quotients
of $\ctor$ respectively $\cstor$, with two framed curves equivalent
on the torus if they are isotopic in the solid torus. 

\begin{theorem}\cite{gelcauribe2}\label{thetalink}
There is an isomorphism that intertwines the 
action of the algebra of Weyl quantizations of trigonometric
polynomials  on the space of theta functions
and the representation
of $\rctor$ onto $\rcstor$, and  which maps 
the theta series $\theta_j^\tau(z)$ to $\alpha^j$ for all $j=0,1,\ldots, N-1$.
\end{theorem}

\begin{remark}
The number
\begin{eqnarray*}
-qj=\left|\begin{array}{cc}
p & q\\
j&0
\end{array}\right|
\end{eqnarray*}
is the sum of the linking numbers of the curves in the system $(p,q)$
and those in the system $\alpha^j$. So the Schr\"{o}dinger representation
is defined in terms of the linking number of curves.
\end{remark}

\begin{remark}
The choice of generators of $\pi_1({\mathbb T}^2)$
completely determines the homeomorphism
$h_0$, allowing us to identify the Hilbert space of the quantization
with the vector space with basis $\alpha^0=\emptyset,
\alpha,\ldots , \alpha^{N-1}$.
As we have seen above, these basis elements are the theta series.
\end{remark}


\subsection{The discrete Fourier transform for classical theta functions from a
  topological perspective}\label{sec:3.4}

The symmetries of  classical theta functions are  an instance
of the Fourier-Mukai transform,  the discrete Fourier transform. 
Following \cite{gelcauribe2}, we put them in a topological perspective.

An element 
\begin{eqnarray}\label{sl2z}
h=\left(
\begin{array}{rr}
a& b\\
c& d
\end{array}
\right)\in\mbox{SL}(2,{\mathbb Z})
\end{eqnarray}
of the mapping class group of the torus ${\mathbb T}^2$ 
induces the biholomorpic mapping between
the Jacobian variety with complex structure
defined by $\tau$ and the Jacobian variety with complex structure defined by
 $\tau '=\frac{a\tau +b}{c\tau +d}$.  The mapping is 
$z'=\frac{z}{c\tau +d}$. Identifying the two tori by this
mapping, we deduce that $h$ induces
a linear symplectomorphism  of $\jacob$ (with the same matrix)
 that preserves the complex structure. 

The map $h$ induces an action of the mapping class group
on the Weyl quantizations of exponentials given by
\begin{eqnarray*}
h\cdot \exp(pP+qQ+kE)=\exp[(ap+bq)P+(cp+dq)Q+kE].
\end{eqnarray*}
 This action is easy to describe in the skein theoretical setting,
it just maps every framed link $\gamma$ on the torus
to $h(\gamma)$. 

\begin{theorem}\label{defofphi}
There is a projective representation $\rho$ of the mapping class group
of the torus on the space of theta functions that satisfies the
exact Egorov identity 
\begin{eqnarray*}
h\cdot \exp(pP+qQ+kE)=\rho(h)\exp(pP+qQ+kE)\rho(h)^{-1}.
\end{eqnarray*}  
Moreover, for every $h$, $\rho(h)$ is unique up to multiplication
by a constant.
\end{theorem}

\begin{proof}
We will exhibit two proofs of this well-known 
result, to which we will refer when
discussing non-abelian Chern-Simons theory.\\ 
{\em Proof 1:} The map that associates to $\exp(pP+qQ+kE)$
the operator that acts on theta functions as 
\begin{eqnarray*}
\theta_j^\tau\rightarrow \exp[(ap+bq)P+(cp+dq)Q+kE]\theta_j^\tau
\end{eqnarray*}
is also a unitary irreducible representation of the finite Heisenberg group
which maps $\exp(kE)$ to multiplication by $e^{\frac{i\pi}{N}}$. By the 
Stone-von Neumann theorem, this representation is unitary equivalent to the
Schr\"{o}dinger representation. This proves the existence of $\rho(h)$ 
satisfying the exact Egorov identity.
By Schur's lemma, 
the map $\rho(h)$ is unique up to multiplication by a constant.
Hence, if $h$ and $h'$ are two elements of the mapping class
group, then $\rho(h'\circ h)$ is a constant multiple of 
$\rho(h')\rho(h)$. It follows that $\rho$ defines a projective
representation.  \\
{\em Proof 2:} The map $\exp(pQ+qQ+kE)\rightarrow h\cdot \exp (pP+qQ+kE)$ 
extends to an automorphism of the algebra ${\mathbb C}[\heis ({\mathbb Z})]$.
Because the ideal by which we factor to obtain $\grpalg$
is invariant under the action of the mapping class group, this 
automorphism induces an automorphism 
\begin{eqnarray*}
\Phi:\grpalg\rightarrow \grpalg,
\end{eqnarray*}
which maps each scalar multiple of the identity to itself. Since, 
by Proposition~\ref{allspace}, $\grpalg$ is
the algebra of all linear operators on the $N$-dimensional space
of theta functions,  $\Phi$ is inner \cite{waerden}, meaning that 
there is $\rho(h):\grpalg\rightarrow \grpalg$ such that $\Phi(x)=
\rho(h)x\rho(h)^{-1}$. In particular 
\begin{eqnarray*}
h\cdot \exp(pP+qQ+kE)=\rho(h)\exp(pP+qQ+kE)\rho(h)^{-1}.
\end{eqnarray*}
The Schr\"{o}dinger representation of $\grpalg$ is obviously irreducible,
so again we apply  Schur's lemma and conclude that $\rho(h)$ is unique up to
multiplication by a constant and  $h\rightarrow \rho(h)$ is a projective 
representation.  
\end{proof}

The representation $\rho$ is the well-known Hermite-Jacobi action given
by discrete Fourier transforms. 

As a consequence of Proposition~\ref{allspace}, for any
 element $h$ of  the 
mapping class group, the linear map  $\rho(h)$ is in $\rctor$, hence it can
be represented by a skein ${\mathcal F}(h)$. This skein satisfies
\begin{eqnarray*}
h(\sigma){\mathcal F}(h)={\mathcal F}(h)\sigma
\end{eqnarray*}
for all $\sigma\in \rctor$. Moreover ${\mathcal F}(h)$ is unique
up to multiplication by a constant. We recall from \cite{gelcauribe2} how
 to find a formula for  ${\mathcal F}(h)$. 
 
We start with the simpler case of
\begin{eqnarray*}
h=T=\left(
\begin{array}{cc}
1&1\\
0&1
\end{array}
\right),
\end{eqnarray*}
  the positive twist. 
Then $h((0,j))=(0,j)$ for all $j$, and $h((1,0))=(1,1)$.
 The equality ${\mathcal F}(T)(0,j)=(0,j){\mathcal F}(T)$ for all $j$ implies
that we can write
 ${\mathcal F}(T)=\sum_{j=0}^{N-1}c_j(0,j)$ for some coefficients $c_j$. 
The equality 
\begin{eqnarray*}
(1,1)\sum_{j=0}^{N-1}c_j(0,j)=\sum_{j=0}^{N-1}c_j(0,j)(1,0)
\end{eqnarray*} 
yields
\begin{eqnarray*}
\sum_{j=0}^{N-1}t^jc_j(1,j+1)=\sum_{j=0}^{N-1}t^{-j}c_j(1,j).
\end{eqnarray*}
It follows that $t^jc_j=t^{-j-1}c_{j+1}$ for all  $j=0,1,\ldots,
N-2$. Normalizing so that $c_0>0$ and ${\mathcal F}(T)$ defines a unitary map,
we obtain $c_j=N^{-1/2}t^{1+3+\cdots +(2j-1)}=N^{-1/2}t^{j^2}$. 
We conclude that  
$${\mathcal F}(T)=N^{-1/2}\sum_{j=0}^{N-1}t^{j^2}(0,j).$$

To better understand this formula, 
we recall a few basic facts in low dimensional topology
(see \cite{rolfsen} for details).

Every $3$-dimensional manifold is the boundary of a $4$-dimensional
manifold  obtained by adding $2$-handles ${\mathbb D}^2\times
{\mathbb D}^2$ to a $4$-dimensional ball along the solid tori
${\mathbb D}^2\times S^1$. On  the boundary $S^3$ of the ball, the operation
of adding handles gives rise to surgery on a framed link.
Thus any $3$-dimensional manifold can be obtained as follows.
Start with  framed link $L\subset S^3$. Take a regular
neighborhood of $L$ made out  of disjoint solid tori, each with a framing curve
on the boundary such that the core of the solid torus and this curve
determine the framing of the corresponding
link component. Remove these tori, then
glue them back in so that the meridians are glued to the framing curves. 
The result is the desired $3$-dimensional manifold.

\begin{figure}[h]
\centering
\scalebox{.30}{\includegraphics{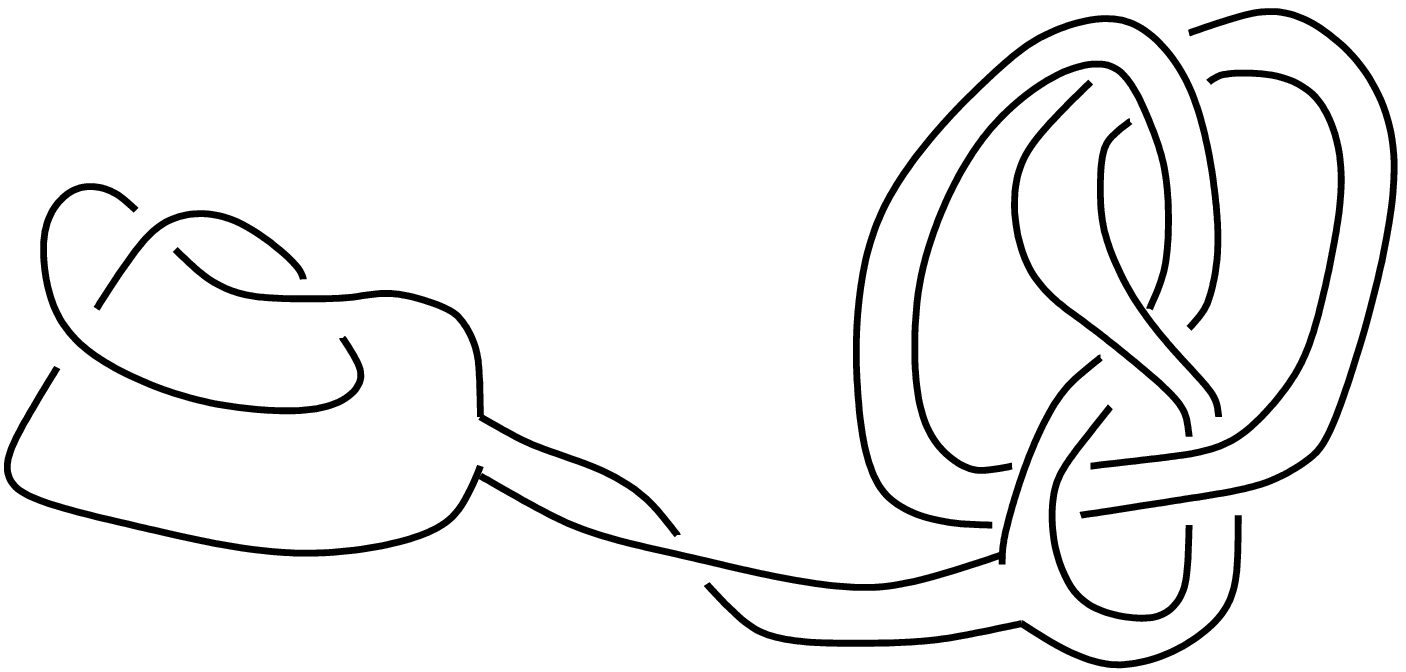}}
\caption{}
\label{slide}
\end{figure}

The operation of sliding one $2$-handle over another corresponds to 
sliding one link component along another using a Kirby band-sum move 
\cite{kirby1}. 
A {\em slide} of $K_1$ along $K$, denoted by $K_1\#K$, is obtained as 
by cutting open the two knots and then joining the ends along the
opposite sides of an embedded rectangle. The result of sliding a trefoil knot
along a figure-eight knot, both with the blackboard framing, 
is shown in Figure~\ref{slide}. For framed knots one should join the
annuli along the opposite faces of an embedded cube (making sure that
the result is an embedded annulus, not an embedded M\"{o}bius band).
The band sum is not unique.

An element of the mapping class group of the torus can
also be described by surgery.
The twist  $T$ is obtained by surgery on the curve
$(0,1)$ with framing $1$. Explicitly,
the mapping cylinder of $T$ is obtained from ${\mathbb T}^2\times [0,1]$ by
removing a solid torus that is a regular neighborhood of 
$(0,1)\times \{\frac{1}{2}\}$ and gluing it back  such 
that its meridian (the homologically
trivial curve on the boundary) is mapped to the framing curve. 
The result  is  homeomorphic to ${\mathbb T}^2\times 
[0,1]$, so that the restriction of the homeomorphism to
${\mathbb T}^2\times \{0\}$ is the identity map and the restriction
to ${\mathbb T}^2\times \{1\}$ is $T$.

We introduce the element
\begin{eqnarray*}
\Omega_{U(1)}=
N^{-1/2}\sum_{j=0}^{N-1}\alpha^j\in \rcstor,
\end{eqnarray*} 
Here $U(1)$ stands for the gauge group $U(1)$ (see \S~\ref{sec:4.1}), which is
related to classical theta functions by Chern-Simons theory.
There is a well known analogue for the group $SU(2)$, to be discussed
in \S~\ref{sec:6.1}.

The skein ${\mathcal F}(T)$ is obtained by coloring the framed surgery
curve of $T$ by $\Omega_{U(1)}$.
 This means that we replace the surgery curve by the skein
 $\Omega_{U(1)}$ such that the curves $\alpha^j$ in the solid torus are 
 parallel to the framing.
In general, for a framed link $L$ we  denote by $\Omega_{U(1)}(L)$
the skein obtained by replacing every  link component  by  
 $\Omega_{U(1)}$ such that $\alpha$ becomes  the framing.

Using the fact that each element of the mapping class group is a product
of twists \cite{lickorish}, we obtain the 
following skein theoretic description of the discrete Fourier transform. 

\begin{theorem}\label{skeinfourier}\cite{gelcauribe2}
Let $h$ be an element of the mapping class group of the torus obtained by 
performing
surgery on a framed link $L_h$ in ${\mathbb T}^2\times [0,1]$. 
The discrete Fourier transform
$\rho(h):\rcstor\rightarrow \rcstor$ is given by 
\begin{eqnarray*}
\rho(h)\beta=\Omega_{U(1)}(L_h)\beta.
 \end{eqnarray*}
\end{theorem}

\begin{remark}
For a framed curve $\gamma$ on the torus, $h(\gamma)$ is the slide of 
$\gamma$ along the components of  $L_h$. 
The  Egorov identity for $\Omega_{U(1)}(L_h)$
means in topological language that we are allowed to perform slides
in the cylinder over the torus along curves  colored
by $\Omega_{U(1)}$.  This points to a surgery
formula for $U(1)$-quantum invariants of 3-manifolds (see \cite{gelcauribe2}). 
\end{remark}


Let us recall the classical description
of the Hermite-Jacobi action. For  $h$ as in (\ref{sl2z}),
\begin{eqnarray*}
\rho(h)\theta_j^\tau(z)=
\exp\left(-\frac{\pi i Ncz^2}{c\tau+d}\right)
\theta_j^{\tau '}\left(\frac{z}{c\tau+d}\right).
\end{eqnarray*}
The exponential factor is introduced to enforce the
periodicity conditions of theta functions for the function
on the right-hand side (see Capter I \S 7 in \cite{mumford}).
For those familiar with the subject, there are no parity restrictions
on $a,b,c,d$ because $N$ being even, 
$SL_\theta(2,{\mathbb Z})=SL(2,{\mathbb Z})$.
  
In particular, for the  generators
\begin{eqnarray*}
S=\left(
\begin{array}{rr}
0& 1\\
-1 &0
\end{array}
\right), \quad T=\left(
\begin{array}{rr}
1& 1\\
0 &1
\end{array}
\right)
\end{eqnarray*}
of $SL(2,{\mathbb Z})$ one has the Jacobi identities
\begin{alignat*}{1}
\rho(S)\theta_j^\tau(z) &=\left(\frac{-i\tau}{N}\right)^{1/2}
\exp\left(\frac{z^2}{2\tau}\right)\theta_j^{-1/\tau}\left(\frac{z}{\tau}
\right) =\left(\frac{i\tau}{N}\right)^{1/2}
\sum_{k=0}^{N-1}e^{-\frac{2\pi i }{N}kj}\theta_k^\tau(z)\\ 
\rho(T)\theta_j^\tau(z)&=\theta_j^{\tau+1}(z)=e^{\frac{2\pi i}{N}j^2}
\theta_j^\tau (z).
\end{alignat*}
We  normalize $\rho(S)$ to a unitary operator dividing  by 
$(-i\tau)^{1/2}$. Note that $$\Omega_{U(1)}=S\emptyset
\in \rcstor.$$

Alternatively, in the real polarization, 
$S$ and $T$ act on $L^2({\mathbb Z}_{N})$ by 
\begin{eqnarray*}
(S f)(j)=
N^{-1/2}\sum_{k\in {\mathbb Z}_{N}}f(k)e^{-\frac{2\pi i}{N}jk} \mbox{ and }
(Tf)(j)= e^{\frac{2\pi i}{N}k^2}f(j),
\end{eqnarray*}
 where the first is the discrete (or finite) Fourier transform and the
second is interpreted as a partial discrete Fourier transform.


Like for the metaplectic representation, the 
Hermite-Jacobi representation  can be made into a true representation
by passing to an extension  of the mapping class group of the torus.
While a ${\mathbb Z}_2$-extension would suffice, we consider
a ${\mathbb Z}$-extension instead, in order to show the similarity with
the non-abelian theta functions. 

Let ${\bf L}$ be a subspace of $H_1({\mathbb T}^2,{\mathbb R})$ spanned
by a simple closed curve. 
Define  the ${\mathbb Z}$-extension of the mapping class group of the torus by
the multiplication rule on $SL(2,{\mathbb Z})\times {\mathbb Z}$,
\begin{eqnarray*}
(h',n')\circ(h,n)=(h'\circ h, n+n'-{\boldsymbol \tau}({\bf L}, h({\bf L}),h'\circ 
h({\bf L})).
\end{eqnarray*}
where ${\boldsymbol \tau}$ is the Maslov index \cite{lionvergne}.
 Standard results 
in the theory of theta functions show that the Hermite-Jacobi action
lifts to a representation of this group.

\section{Non-abelian theta functions from geometric considerations}
\label{sec:4}

\subsection{Non-abelian theta functions from geometric quantization}\label{sec:4.1}
Let $G$ be a compact  simple Lie group, $\lieg$ its Lie algebra,  
and    $\Sigma_g$ be a closed oriented 
surface of genus $g\geq 1$. 
Consider the moduli space of  $\lieg$-connections
on $\Sigma_g$, which
 is the quotient of the affine space of all $\lieg$-connections 
on $\Sigma_g$ (or rather on the trivial principal $G$-bundle $P$ on
$\Sigma_g$) by
the group ${\mathcal G}$ of gauge transformations
$A\rightarrow \phi^{-1}A\phi +\phi^{-1}d\phi$, with
$\phi:\Sigma_g \rightarrow G$  a smooth function.
The space of all connections has a symplectic $2$-form given by 
\begin{eqnarray*}
\omega(A,B)=-\int_{\Sigma_g}\mbox{tr}\, (A\wedge B),
\end{eqnarray*}
where $A$ and $B$ are connection forms in its tangent space. According
to \cite{atiyahbott}, this induces a symplectic form, denoted also
by $\omega$, on the
moduli space, which further defines a  Poisson bracket.
The group of gauge transformations acts on the space of all connections in
a Hamiltonian fashion, with moment map the curvature.
Thus
the moduli space of {\em flat} 
$\lieg$-connections
\begin{eqnarray*}
{\mathcal M}_g=\{A\, |\, A :\, \mbox{flat }\lieg -\mbox{connection}\}/{\mathcal
  G}.
\end{eqnarray*}
arises as the symplectic reduction of the space of all connections by the 
group of gauge transformations.
This space  is the same as the moduli space of semi-stable $G$-bundles
on $\Sigma_g$, and also the character variety of $G$-representations
of the fundamental group of $\Sigma_g$.  It is an affine algebraic
set over the real numbers. 

Each curve $\gamma$ on the surface and each  irreducible representation $V$
 of $G$ define a classical observable on this space
\begin{eqnarray*}
 A\rightarrow \mbox{tr}_{V}\mbox{hol}_\gamma(A),
\end{eqnarray*}
called Wilson line,
by taking the trace of the holonomy of the connection in the
given irreducible representation of $G$. Wilson lines 
are regular functions on the moduli space. When $G=SU(2)$ we let the
Wilson line for the $n$-dimensional irreducible representation be
$W_{\gamma,n}$.  When $n=2$,
we denote $W_{\gamma,2}$
by $W_{\gamma}$. The $W_{\gamma}$'s span the algebra 
of regular functions on ${\mathcal M}_g$. 

\begin{figure}[htbp]
\centering
\scalebox{.30}{\includegraphics{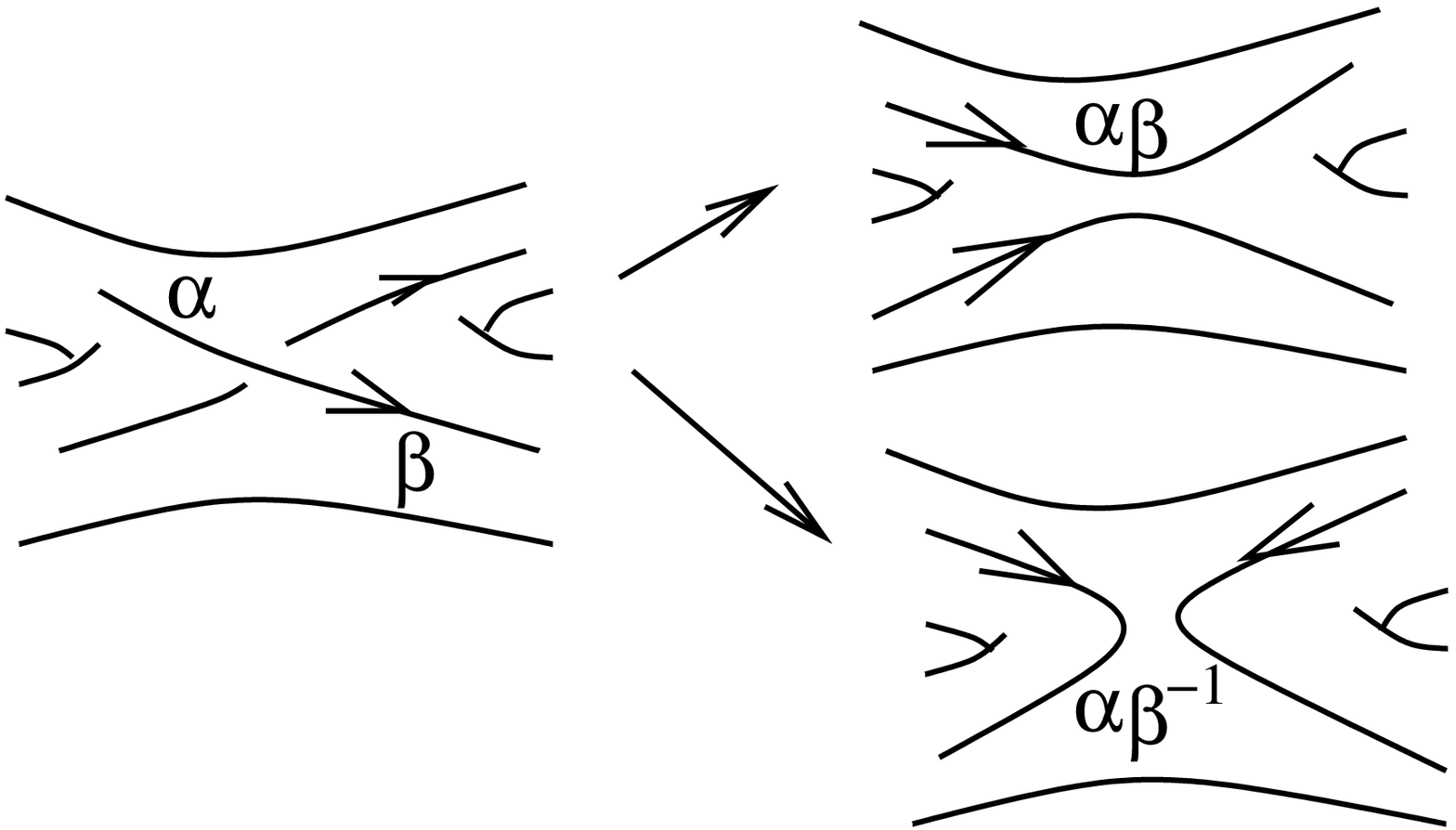}}
\caption{}
\label{alphabeta}
\end{figure}

The form $\omega$ induces   a Poisson bracket, 
which in the case of the gauge group
$SU(2)$ was found by Goldman \cite{goldman} to be 
\begin{eqnarray*}
\{W_{\alpha},W_{\beta}\}= \frac{1}{2}
\sum_{x\in \alpha\cap \beta}
\mbox{sgn}(x)(W_{\alpha\beta^{-1}_x}-W_{\alpha\beta_x})
\end{eqnarray*}
where  $\alpha\beta_x$ and $\alpha\beta^{-1}_x$ are computed as elements of
the fundamental group with base point $x$ (see Figure~\ref{alphabeta}), and
$\mbox{sgn}(x)$ is the signature of the crossing $x$; it is positive
if the frame given by the tangent vectors to $\alpha$ respectively
$\beta$ is positively oriented (with respect to the orientation
of $\Sigma_g$), and negative otherwise.

The moduli space ${\mathcal M}_g$, or rather the smooth part of it,
can be quantized in the direction of  Goldman's Poisson bracket
 as follows.
First, set Planck's constant $\hbar =\frac{1}{N}$, where $N$ is an even
positive integer. 

 The Hilbert space can be obtained using
the method of geometric quantization as the space of sections of
a line bundle over ${\mathcal M}_g$. The general procedure
is to obtain the line bundle as the
tensor product of a line bundle
 with curvature $-2\pi iN\omega$ and
a half-density \cite{snyaticki}. The half-density is
a square root of the canonical line bundle.
 Because the moduli space has a natural complex structure,
it is customary to work in the complex polarization, in which case
the Hilbert space consists of the holomorphic sections
of the line bundle. 

Let us briefly recall how each complex
structure on the surface induces a complex structure on the
moduli space. 
The tangent space to ${\mathcal M}_g$ at a nonsingular point $A$ is
the first cohomology group $H^1_A(\Sigma_g ,\mbox{ad }P)$ of the complex 
of $\lieg$-valued forms
\begin{eqnarray*}
\Omega ^0(\Sigma_g, \mbox{ad }P)\stackrel{d_A}{\rightarrow}
 \Omega ^1(\Sigma_g, \mbox{ad }P)\stackrel{d_A}{\rightarrow}
 \Omega ^2(\Sigma_g, \mbox{ad }P).
 \end{eqnarray*}
Here  $P$ denotes the trivial principal $G$-bundle
over $\Sigma_g$. Each complex structure on $\Sigma_g$ induces a
Hodge $*$-operator on the space of connections on $\Sigma_g$, hence
a $*$-operator on $H^1_A(\Sigma_g, \mbox{ad }P)$. The complex structure
on ${\mathcal M}_g$ is $I:H^1_A(\Sigma_g, \mbox{ad }P)
\rightarrow H^1_A(\Sigma_g, \mbox{ad }P)$, $IB = -*B$. 
For more details we refer the reader to \cite{hitchin}.
This complex structure turns the smooth part of ${\mathcal M}_g$
 into a complex  manifold. It is important to point out
that the complex structure
 is compatible with the symplectic form $\omega$, 
in the sense that $\omega (B, IB )\geq 0$ for all
$B$. 

As said, the Hilbert space consists of the
holomorphic sections of the line bundle of the quantization. 
These are 
the  {\em non-abelian theta functions}. 


The analogue of the group algebra of the finite Heisenberg group 
is the algebra of operators
that are the quantizations of 
Wilson lines. They  arise in the theory of the
Jones polynomial \cite{jones} 
as outlined by Witten \cite{witten}, being defined 
heuristically in the framework of quantum field theory. They are
integral operators with kernel 
\begin{eqnarray*}
<A_1|\op(W_{\gamma, n})|A_2>=\int_{{\mathcal M}_{A_1A_2}}e^{iN{
    L}(A)}
W_{\gamma, n}(A){\mathcal D}A,
\end{eqnarray*}
where $A_1,A_2$ are conjugacy classes of flat connections on
$\Sigma_g$ modulo the gauge group, 
$A$ is a conjugacy class under the action of the gauge group 
on ${\Sigma_g}\times [0,1]$ such that
$A_{{\Sigma_g}\times\{0\}}=A_1$ and 
$A_{{\Sigma_g}\times\{1\}}=A_2$, and
\begin{eqnarray*}
{ L}(A)=\frac{1}{4\pi}\int_{\Sigma_g\times [0,1]}
\mbox{tr}\left(A\wedge dA+\frac{2}{3}A\wedge A\wedge A\right)
\end{eqnarray*}
is the Chern-Simons Lagrangian.
The operator quantizing a Wilson line is
  defined by a Feynman path integral, which 
does not have a rigorous mathematical formulation. It is
thought as an average of the Wilson line computed over
all connections that interpolate between $A_1$ and $A_2$.

The skein theoretic approach to classical theta functions outlined in
\S~\ref{sec:3.3}
 can be motivated by the Chern-Simons-Witten field theory point of view. 
%
%
%
Indeed, the Wilson lines
can be quantized either by using one of the classical methods
for quantizing the torus, or by using
the Feynman path integrals as above. The Feyman path integral approach
allows  localizations of  computations to small balls, in
which  a single crossing shows up. Witten \cite{witten} has explained
that in each such ball  skein relations hold, in this case
the skein relations from Figure 2, which compute the linking number.
As such  the path integral quantization gives rise to the skein theoretic 
model.

On the other hand, Witten's quantization is symmetric under the
action of the mapping class group of the torus, a property shared by
Weyl quantization. And indeed, we have seen in \S~\ref{sec:3.3} that
Weyl quantization and the skein theoretic quantization are the same.
The relevance of Weyl quantization to Chern-Simons theory was first
pointed out in \cite{gelcauribe} for the gauge group $SU(2)$. For
the gauge group $U(1)$, it was noticed in \cite{andersen}.
The abelian Chern-Simons theory from the  skein theoretic point
of view was described in detail in \cite{gelcauribe2}.

\subsection{The Weyl quantization of the moduli space of flat
  $SU(2)$-con\-nec\-tions on the torus}\label{sec:4.2}
The moduli space  ${\mathcal M}_1$ of flat 
$SU(2)$-connections on the torus is the same
as the character variety of $SU(2)$-representations of the 
fundamental group of the  torus. 
It is, therefore, parametrized by the set 
 of pairs of matrices $(A,B)\in SU(2)\times SU(2)$
 satisfying $AB=BA$, modulo
conjugation. Since commuting matrices can be diagonalized simultaneously, and
the two diagonal entries can be permuted simultaneously, 
 the moduli space can be identified with
the quotient of the
torus $\{(e^{2\pi i x}, e^{2\pi i y}), \, x,y\in {\mathbb R}\}$ by
 the ``antipodal'' map $x\rightarrow -y$, $y \rightarrow -y$.
This space is called the {\em pillow case}.

The pillow case is the quotient of
${\mathbb R}^2$ by horizontal and vertical integer translations
and by the symmetry $\sigma $ with respect to the origin.
Except for  $4$ singularities, ${\mathcal M}_1$ is a symplectic
 manifold,
with symplectic form $\omega =2\pi i dx\wedge dy$.
The associated Poisson bracket is 
\begin{eqnarray*}
\{f,g\}=\frac{1}{2\pi i}\left(\frac{\partial f}{\partial x}
\frac{\partial g}{\partial y}-\frac{\partial f}{\partial y}
\frac{\partial g}{\partial x}\right).
\end{eqnarray*}

The Weyl quantization of ${\mathcal M}_1$ in the complex polarization
has been described in
\cite{gelcauribe} for one particular complex structure. 
We do it now in general. 
Again Planck's constant
is the reciprocal of an even integer $\hbar=\frac{1}{N}
=\frac{1}{2r}$.

The tangent space at an arbitrary  point on the pillow case
is spanned by the vectors $\frac{\partial}{\partial x}$
and $\frac{\partial}{\partial y}$. In the formalism of \S 4.1,
these vectors are identified respectively with the cohomology classes
of the $su(2)$-valued $1$-forms 
\begin{eqnarray*}
\left(
\begin{array}{rr}
1&0\\
0&-1
\end{array}
\right)dx \quad \mbox{and}\quad 
\left(
\begin{array}{rr}
1&0\\
0&-1
\end{array}
\right)dy.
\end{eqnarray*}
It follows that a complex structure on the original torus induces
exactly the same complex structure  on the pillow case. So we can
think of the pillow case as the quotient of the complex plane by the
group generated by ${\mathbb Z}+ {\mathbb Z}\tau$ ($\mbox{Im }\tau>0$)
 and the symmetry $\sigma$
with respect to the origin. 
As before, we denote by $(x,y)$
the coordinates in the basis $(1,\tau)$ and by $z=x+\tau y$ the
complex variable. A fundamental domain for the group action in the
$(x,y)$-coordinates is ${\mathcal D}=[0,\frac{1}{2}]\times[0,1]$.

As  seen in \cite{gelcauribe}, 
a holomorphic line bundle ${\mathcal L}_1$  with curvature 
$4\pi i rdx\wedge dy$ on the pillow case
is defined by the cocycle
$\Lambda_1:{\mathbb R}^2\times {\mathbb Z}^2\rightarrow {\mathbb C}^*$, 
\begin{eqnarray*}
&&\Lambda_1((x,y),(m,n))=e^{4\pi i r(\frac{\tau}{2} n^2-2n(x+\tau y))
}=e^{4\pi i r(\frac{\tau}{2} n^2-2nz)}\\
&&\Lambda_1((x,y),\sigma)=1.
\end{eqnarray*}

The square root of the canonical form is no longer the trivial line bundle, 
since  for example  the form $dx$ is not 
defined globally on the pillow case.
The obstruction for $dx$ to be globally defined
 can be incorporated in a
line bundle ${\mathcal L}_2$ defined by the  cocycle
$\Lambda_2:{\mathbb R}^2\times {\mathbb Z}^2\rightarrow {\mathbb C}^*$,
\begin{eqnarray*} 
\Lambda_2((x,y),(m,n))=1, \quad \Lambda_2((x,y),\sigma)=-1.
\end{eqnarray*}
This line bundle can then be taken as the half-density.

The line bundle of the quantization is therefore  ${\mathcal L}_1\otimes 
{\mathcal L}_2$, defined by the cocycle $\Lambda_1\Lambda_2$. The
Hilbert space ${\mathcal H}_r({\mathbb T}^2)$
of non-abelian theta functions on the torus consists of the 
holomorphic sections of this line bundle.
Hence the Hilbert space consists of the odd theta functions (this was
discovered in \cite{ADW}).

 Because Weyl quantization
of the pillow case is equivariant Weyl quantization of the torus,
to specify a basis of ${\mathcal H}_r({\mathbb T}^2)$ we need a
pair of generators of the fundamental group.
 This complex structure and generators of $\pi_1({\mathbb T}^2)$ determine
a point in the Teichm\"{u}ller space of the torus,   specified
by the complex number $\tau$  mentioned before. 
The orthonormal basis of the Hilbert space is 
\begin{eqnarray*}
\zeta_j^\tau(z)=(\theta_j^\tau(z)-\theta_{-j}^\tau(z)), \quad
j=1,2,\ldots, r-1,
\end{eqnarray*}
where $\theta_j^\tau(z)$ are the  theta series from \S 3.1.
The definition of $\zeta_j^\tau(z)$ can  be extended
to all indices by the conditions $\zeta_{j+2r}^\tau(z)=
\zeta_j^\tau (z)$, $\zeta_0^\tau(z)=0$, and $\zeta_{r-j}^\tau(z)=
-\zeta_{r+j}^\tau(z)$.

The space ${\mathcal H}_r({\mathbb T}^2)$ can be embedded isometrically
into $L^2({\mathcal D})$, with the inner product 
\begin{eqnarray*}
\left<f,g\right>=2(-2i r(\tau-\bar{\tau}))^{1/2}
\iint_{\mathcal D}f(x,y)\overline{g(x,y)}
e^{-2\pi ir(\tau-\bar{\tau}) y^2}dxdy
\end{eqnarray*}
The Laplacian is given by the formula (\ref{delta}) (with $N=2r$).

The pillow case is the quotient of the plane by a discrete group, so
again we can apply the Weyl quantization procedure.
If $p$ and $q$ are coprime integers, then the Wilson line of the
curve $(p,q)$ of slope $p/q$ on the torus  for the $2$-dimensional
irreducible representation is 
\begin{eqnarray*}
W_{(p,q)}(x,y)
=\frac{\sin 4\pi (px+qy)}{\sin 2\pi (px+qy)}=2\cos 2\pi (px+qy),
\end{eqnarray*}
 when viewing the pillow case as a quotient of the plane.
This is because the character of the $2$-dimensional irreducible
representation is $\sin 2x/\sin x$.
In general, if  $p$ and $q$ are arbitrary integers, the function
$$f(x,y)=2\cos 2\pi (px+qy)$$ is a linear combination of Wilson
lines. Indeed, if $n=\mbox{gcd}(p,q)$ then 
\begin{eqnarray*}
2\cos 2\pi (px+qy)=\frac{\sin [2\pi (n+1)(\frac{p}{n}x+\frac{q}{n}y)]}{
\sin 2\pi (\frac{p}{n}x+\frac{q}{n}y)}-
\frac{\sin[ 2\pi (n-1)(\frac{p}{n}x+\frac{q}{n}y)]}{\sin 2\pi (\frac{p}{n}x
+\frac{q}{n}y)},
\end{eqnarray*}
so $2\cos 2\pi (px+qy)=W_{\gamma, n+1}-W_{\gamma, n-1}$ where $\gamma $ is the curve
of slope $p/q$ on the torus. This formula also shows that
Wilson lines are linear combinations of cosines, so it suffices to
understand the quantization of the cosines.

Because
\begin{eqnarray*}
2\cos 2\pi (px+qy)=e^{2\pi i (px+qy)}+e^{-2\pi i (px+qy)},
\end{eqnarray*}
 the Weyl quantization of cosines can be 
obtained by taking the Schr\"{o}dinger representation
of the quantum observables that are 
invariant under the map $\exp P\rightarrow \exp (-P)$ 
and $\exp Q\rightarrow \exp (-Q)$, and restrict it
to odd theta functions.
We obtain  the formula
\begin{eqnarray*}
\op (2\cos 2\pi (px+qy))\zeta_j^\tau(z)=e^{-\frac{\pi i}{2r}  pq}
\left(e^{\frac{\pi i}{r} qj}
\zeta_{j-p}^\tau(z)+e^{-\frac{\pi i}{r} qj}\zeta_{j+p}^\tau(z)\right).
\end{eqnarray*}
 
In particular the $\zeta_j^\tau$'s are the eigenvectors of 
$\op (2\cos 2\pi y)$, corresponding to the holonomy
along the curve which cuts the torus into an annulus.
This shows that they are correctly identified as the analogues
of the theta series.  

\section{Non-abelian theta functions from quantum groups}\label{sec:5}

\subsection{A review of the quantum group $U_\hbar(sl(2,{\mathbb C}))$}\label{sec:5.1}

 For the gauge group $SU(2)$,  Reshetikhin and Turaev 
\cite{reshetikhinturaev} constructed  rigorously, by using
quantum groups, a topological quantum field theory
that fulfills Witten's programme.  Within this
theory, for each surface there is a vector space, an
algebra of quantized Wilson lines, and a projective finite-dimensional 
representation of the mapping class group, namely the 
 Reshetikhin-Turaev representation. 
The quantum group quantization 
has the advantage over geometric quantization that it does
not depend on any additional structure, such as a polarization.

We set $\hbar=\frac{1}{N}=\frac{1}{2r}$, and furthermore $r>1$.
 Let $t=e^{\frac{i\pi}{2r}}$ and, for an integer $n$, let $[n]
=\frac{t^{2n}-t^{-2n}}{t^2-t^{-2}}=\sin
  \frac{n\pi}{r}/\sin\frac{\pi }{r}$, called the quantized integer.

The quantum group associated to  $SU(2)$,  
denoted $U_{\hbar}(sl(2,{\mathbb C}))$
 is obtained by passing to the complexification $SL(2,{\mathbb C})$ of $SU(2)$,
 taking the universal enveloping algebra of its Lie algebra,
then deforming this algebra with respect to $\hbar$. 
It is the Hopf algebra over ${\mathbb C}$ with generators $X,Y,K, K^{-1}$ 
subject to the relations
\begin{eqnarray*}
KK^{-1}=K^{-1}K=1, \, KX=t^2XK, \, KY=t^{-2}YK,\, XY-YX=\frac{\textstyle{K^2-K^{-2}}}{
\textstyle{t^2-t^{-2}}}.
\end{eqnarray*}
At the root of unity, namely when $N=2r$, $r$ integer, one has the
additional factorization relations
$X^r=Y^r=0, K^{4r}=1$\footnote{In this case the quantum 
group is denoted by $U_t$ in
  \cite{reshetikhinturaev} and by ${\mathcal A}$ in \cite{kirbymelvin}.}.

As opposed to $SU(2)$, $U_{\hbar}(sl(2,{\mathbb C}))$ has 
only finitely many irreducible representations, among which
we distinguish a certain family
$V^1$, $V^2$, $\ldots$, $V^{r-1}$ (for
 details see \cite{reshetikhinturaev} or \cite{kirbymelvin}). 
For each $k$, the space $V^k$ has basis 
$e_{j}$, $j=-k_0, \ldots, k_0-1,k_0$, where
$k_0=\frac{k-1}{2}$, and the quantum group acts on it by
\begin{eqnarray*}
Xe_j=[k_0+j+1]e_{j+1},\quad 
Ye_j=[k_0-j+1]e_{j-1},\quad 
Ke_j=t^{2j}e_j.
\end{eqnarray*}
The highest weight vector of this representation is $e_{k_0}$; it spans
the kernel of $X$, is a cyclic vector for $Y$, and an eigenvector of $K$. 

The Hopf algebra structure of $U_{\hbar}(sl(2,{\mathbb C}))$ makes
duals and tensor products of representations be representations themselves.
The quantum group acts on the dual $V^{k*}$ of $V^k$ by
\begin{eqnarray*}
Xe^j=-t^2[k_0+j]e^{j-1},\quad
Ye^j=-t^{-2}[k_0-j]e^{j+1},\quad
Ke^j=t^{-2j}e^j,
\end{eqnarray*}
where $(e^j)_j$ is the basis dual to $(e_j)_j$.
There is a (non-natural) isomorphism $D:V^{k*}\rightarrow V^k$,
\begin{eqnarray*}
D(e^j)=\frac{[k_0-j][k_0-j-1]\cdots [1]}{[2k_0][2k_0-1]\cdots
  [k_0-j+1]}(-t^2)^j
e_{-j}.
\end{eqnarray*}
A Clebsch-Gordan theorem holds,
\begin{eqnarray*}
V^m\otimes V^n=\bigoplus_pV^p\oplus B,
\end{eqnarray*}
where $p$ runs among all indices that satisfy $m+n+p$ odd and
 $|m-n|+1\leq p\leq
\min(m+n-1,2r-1-m-n)$ and $B$ is a representation that is 
 ignored because
it has no effect on computations. 

A corollary of the Clebsch-Gordan theorem is the following formula
\begin{eqnarray*}
V^n=\sum_{j=0}^{\lfloor n/2\rfloor }(-1)^j{{n-j}\choose j}(V^2)^{n-2j}=S_{n-1}(V^2), \quad
\mbox{for }n=1,2,\cdots, r-1.
\end{eqnarray*}
Here $S_n(x)$ is  the {\em Chebyshev polynomial of second kind}
 defined  by
\begin{eqnarray*}
S_{n+1}(x)=xS_n(x)-S_{n-1}(x), \quad S_0(x)=1, \quad S_1(x)=x. 
\end{eqnarray*}

We define the representation ring $R(U_\hbar(sl(2,{\mathbb C})))$ as the
ring generated by $V^j$, $j=1,2,\ldots, r-1$ with multiplication
$V^m\otimes V^n=\sum_pV^p$, where the sum is taken over all indices $p$ that
satisfy the conditions from the Clebsch-Gordan theorem.

\begin{proposition}\label{repring}
The representation ring $R(U_\hbar(sl(2,{\mathbb C})))$ is 
isomorphic to ${\mathbb C}[V^2]/S_{r-1}(V^2)$.
If we define $V^n=S_{n-1}(V^2)$ in this ring for all $n\geq 0$, then 
$V^{r+n}=-V^{r-n}$, $V^{r}=0$, and $V^{n+2r}=V^n$ for all $n>0$. 
\end{proposition}

\subsection{The quantum group quantization  of the moduli space of flat
  $SU(2)$-connections on a surface of genus greater than $1$}\label{sec:5.2}
The definition of the
quantization of the moduli space ${\mathcal M}_g$ of flat $SU(2)$-connections
on a genus $g$ surface $\Sigma_g$
 uses {\em ribbon  graphs and framed links} embedded in
$3$-dimensional manifolds. A ribbon graph consists of the embeddings 
in the 3-dimensional manifold of finitely many 
connected components, each of which  is homeomorphic to either an annulus or
 a tubular $\epsilon$-neighborhood
in the plane of a planar trivalent
 graph with small $\epsilon>0$. As such, while the edges of
a classical graph are $1$-dimensional, those of a ribbon graph are
2-dimensional, an edge being homeomorphic to either a rectangle,
or  an annulus. Intuitively, 
 edges are ribbons, hence the name. When embedding
the ribbon graph in a $3$-dimensional manifold, the framings keep 
track of the twistings of edges. A framed link is a particular 
case of a ribbon graph.
The link components and the edges of ribbon graphs are oriented.
All ribbon graphs  
used depicted below are taken with the ``blackboard framing'', meaning that the 
$\epsilon$-neighbor\-hood is in the plane of the paper. 

\begin{figure}[h]
\centering
\scalebox{.40}{\includegraphics{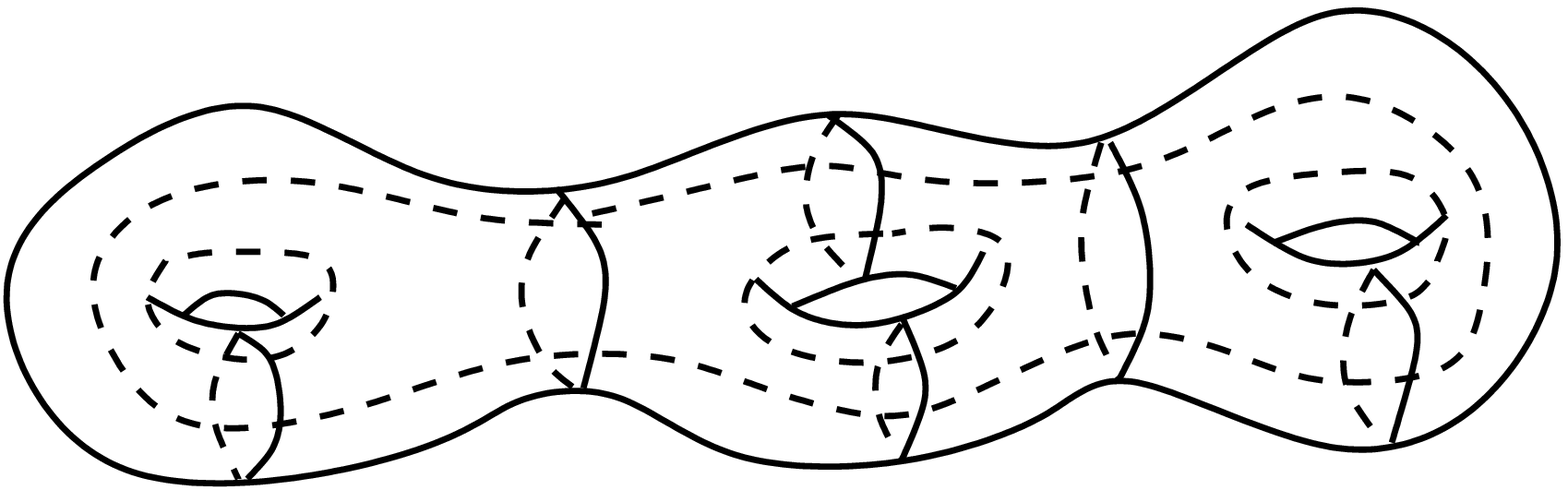}}
\caption{}
\label{rigsurface}
\end{figure}

With these conventions at hand, let us quantize the moduli space 
${\mathcal M}_g$. 
The Hilbert space ${\mathcal H}_r(\Sigma_g)$ is defined by specifying
a basis, the analogue of the theta series. Exactly how in order
 to specify a basis of the space of theta functions one needs 
a pair of generators of $\pi_1({\mathbb T}^2)$, here one needs  additional
structure on $\Sigma_g$, which comes in the form of an
{\em  oriented rigid structure}.

A rigid structure on a surface  
is a collection of simple closed curves that
decompose it into pairs of pants, together
with ``seams'' that keep track of the twistings.
The seams are simple closed curves that, when restricted
to any pair of pants, give 3 nonintersecting arcs that connect pairwise 
the boundary components.  
An oriented rigid structure is one in which the decomposing curves
 are oriented. An example
is shown in {Figure~\ref{rigsurface}}, with  decomposing
curves  drawn with  continuous line, and  seams  with dotted line.

Map $\Sigma_g$ to the boundary
of a handlebody $H_g$ such that the decomposition curves bound disks in
$H_g$. The disks cut $H_g$ into balls. Consider
the   trivalent graph that is the core of $H_g$,
 with a  vertex at the center of each ball and  an
 edge drawn for each  disk. The framing of edges should be 
parallel to the seams (more precisely, to
the region of the surface that lies between the seems). The disks 
are oriented by the decomposition curves on the boundary, and the
orientation of the edges of the graph should agree with that of the disks. 

\begin{figure}[h]
\centering
\scalebox{.30}{\includegraphics{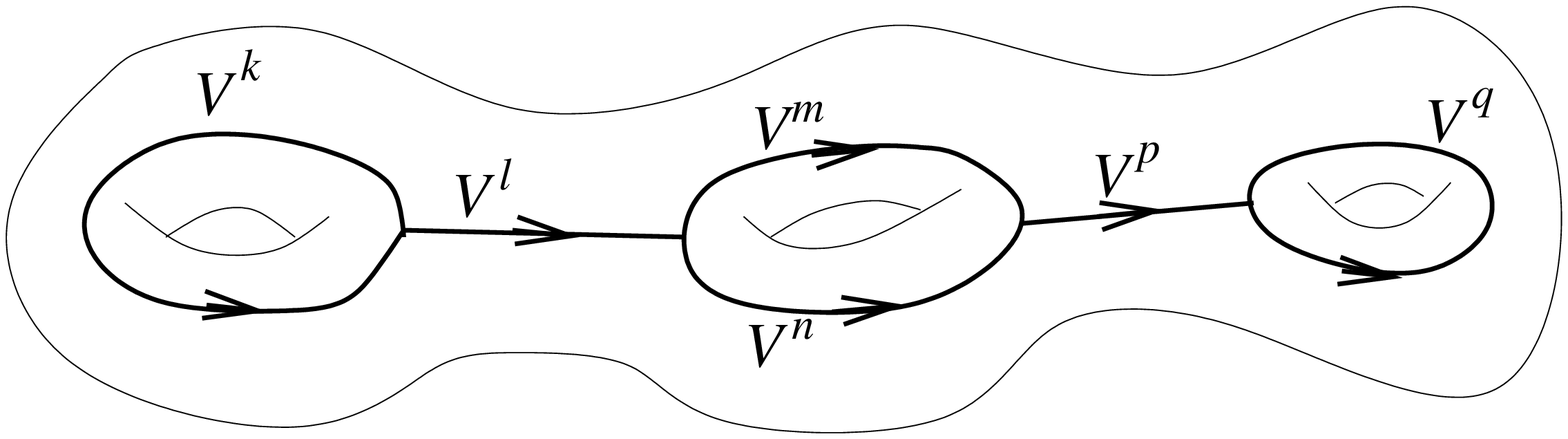}}
\caption{}
\label{resturbas}
\end{figure}

The vectors forming an 
orthonormal basis of ${\mathcal H}_r(\Sigma_g)$ consist of  all 
 possible  colorings of this framed oriented trivalent graph
by $V^j$'s such that at each vertex the three indices satisfy the conditions
from the Clebsch-Gordan theorem (note that the  double inequality is
invariant under  permutations of $m,n,p$). Such a coloring is
called {\em admissible}. In genus $3$
and for the rigid structure from Figure~\ref{rigsurface}, 
a basis element is shown in Figure~\ref{resturbas}.
The inner product $\left<\cdot, \cdot\right>$ is defined so that these 
basis elements are orthonormal. 

This is a nice combinatorial description of the non-abelian
theta functions for the Lie group $SU(2)$, which 
obscures their geometric properties and the  origin of the name.
The possibility to represent non-abelian theta functions as such graphs
follows from the relationship found by Witten between
theta functions and conformal field theory \cite{witten}.

\begin{figure}[h]
\centering
\scalebox{.35}{\includegraphics{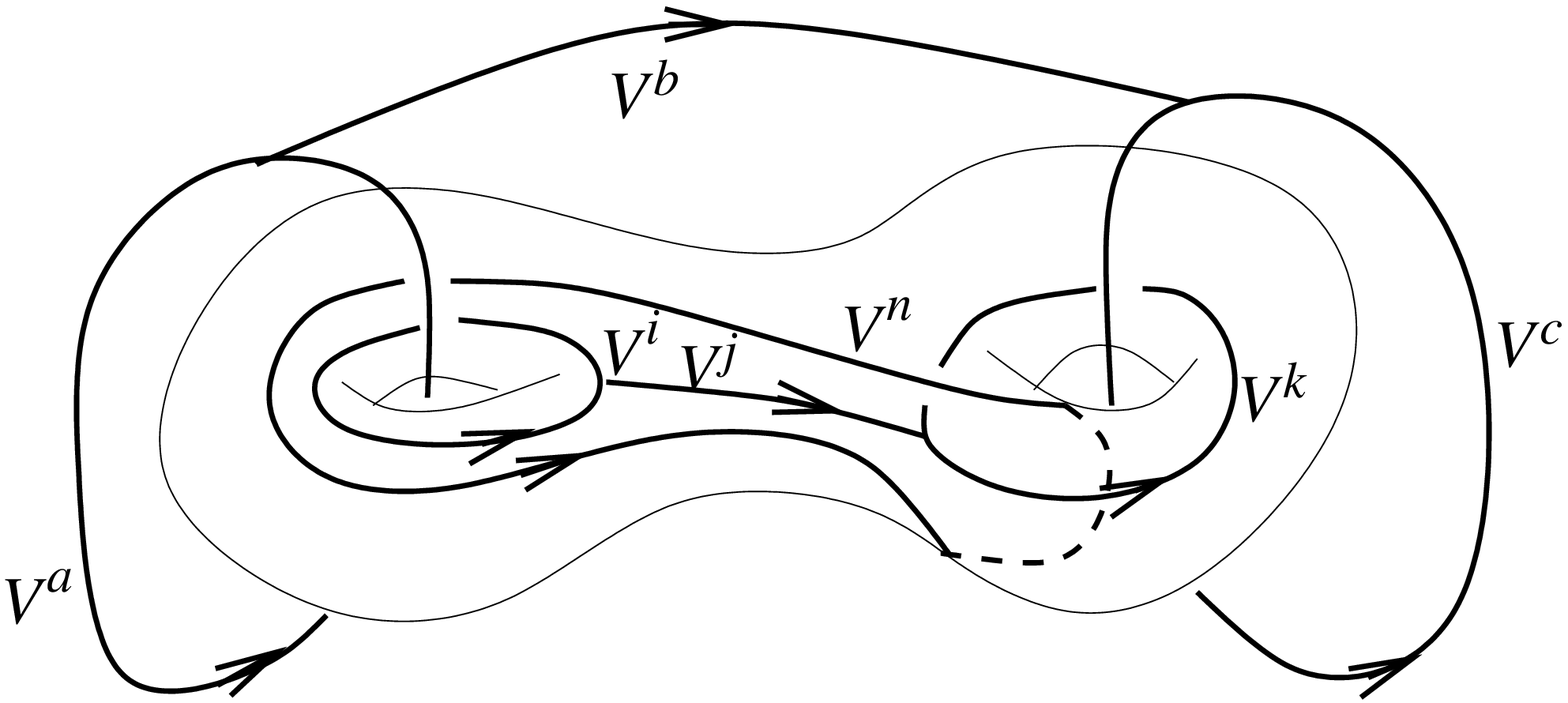}}
\caption{}
\label{resturmat}
\end{figure}

The matrix of the operator $\op(W_{\gamma, n})$ associated to the Wilson line 
\begin{eqnarray*}
W_{\gamma,n}:A\rightarrow
\mbox{tr}_{V^n}\mbox{hol}_\gamma(A)
\end{eqnarray*}
is computed as follows. First, let $1\leq n\leq r-1$. 
Place  the  surface $\Sigma_g$ in standard position in the
$3$-dimensional sphere so that it bounds a genus $g$ handlebody on each side.
Draw the curve $\gamma$ on the surface, give it the framing
parallel to the surface, then color it  by the representation $V^n$ of
$U_\hbar(sl(2,{\mathbb C})$. 
Add two basis elements $e_p$ and $e_q$, viewed as admissible colorings by
irreducible representation of the cores of the interior, respectively
exterior  handlebodies (see Figure~\ref{resturmat}). The interior and exterior
handlebodies should be  copies of the same handlebody with the same oriented
rigid structure on the boundary (thus giving rise to the same non-abelian
theta series), and these copies are glued along $\Sigma_g$ by an orientation
reversing homeomorphism as to obtain $S^3$. 

\begin{figure}[h] 
\centering
\scalebox{.30}{\includegraphics{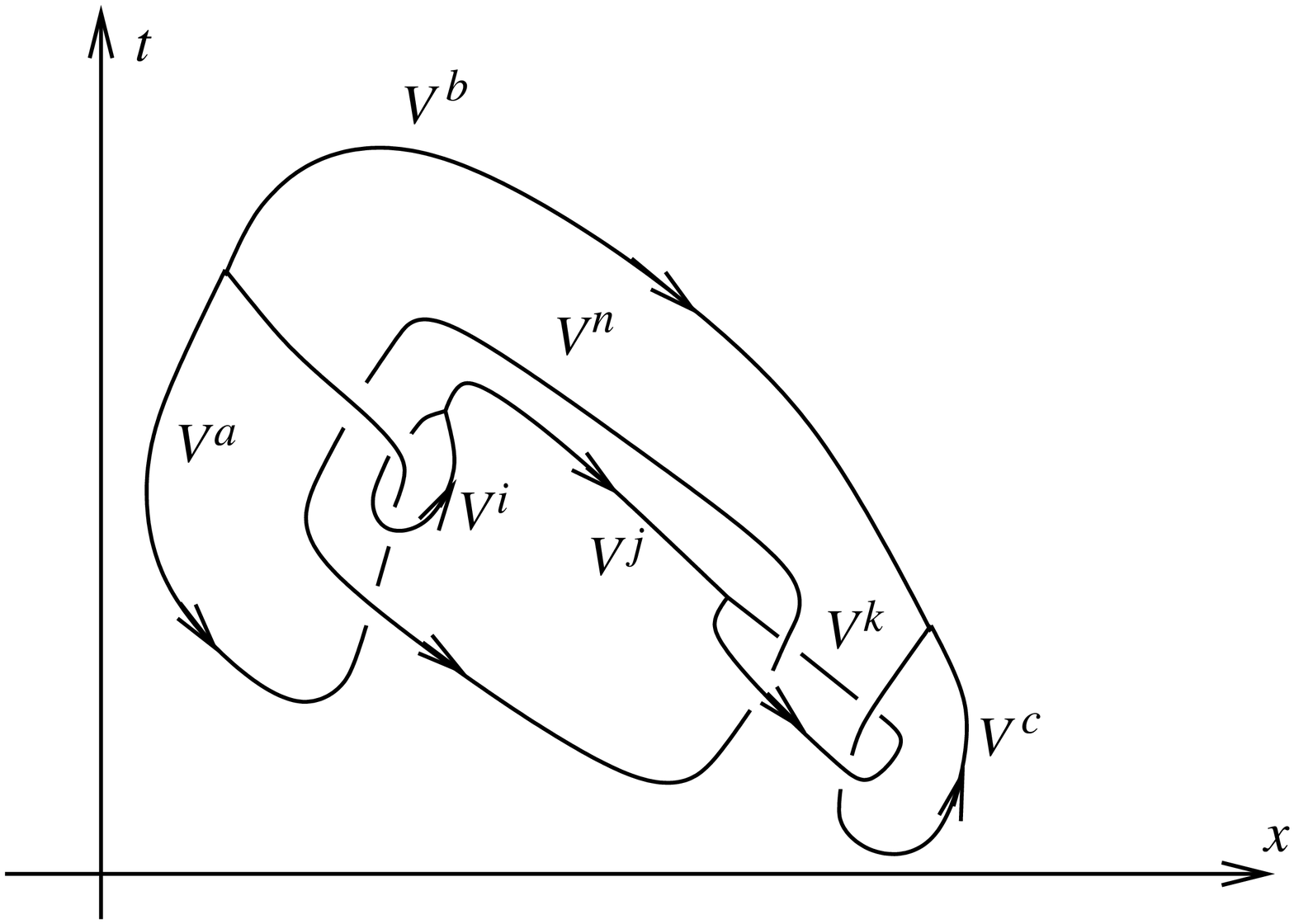}}
\caption{}
\label{particles}
\end{figure}

Erase  the surface to obtain an oriented  tangled  ribbon
graph in $S^3$ whose edges are decorated by irreducible
 representations of $U_{\hbar}(sl(2,{\mathbb C})$ (Figure~\ref{particles}).
Project this  graph onto a plane while 
keeping track of the crossings. 
The Reshetikhin-Turaev theory \cite{reshetikhinturaev}
 gives a way of associating a number to this ribbon graph, which  is 
independent of the
particular projection and  is called the Reshetikhin-Turaev invariant of the
ribbon graph.  

\begin{figure}[h] 
\centering
\scalebox{.35}{\includegraphics{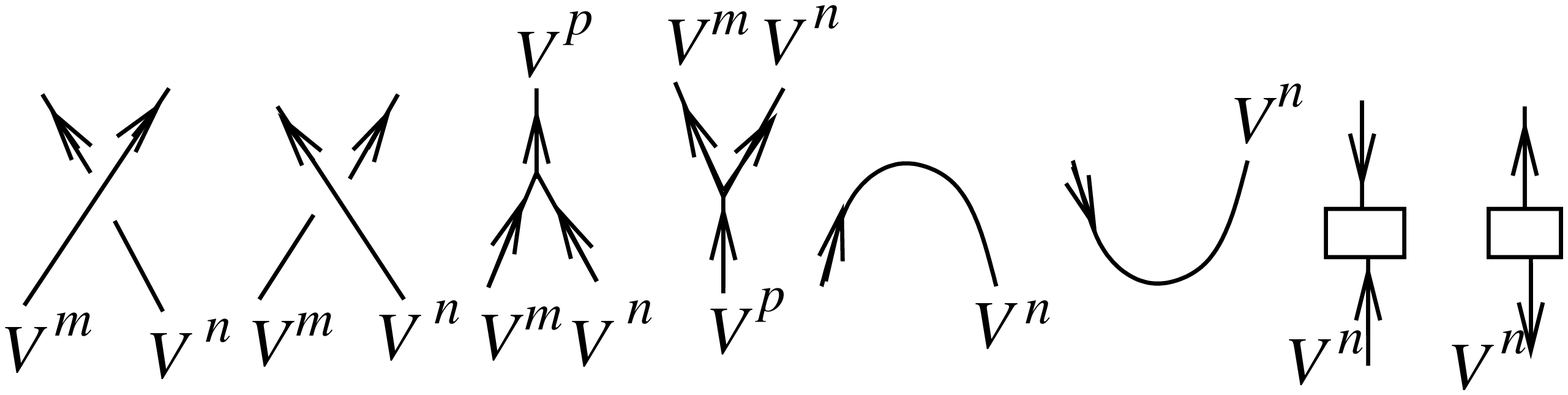}}
\caption{}
\label{phenomena}
\end{figure}

In short, the Reshetikhin-Turaev invariant is computed as follows. 
The ribbon graph should be mapped by an isotopy to one whose projection
can be cut by finitely many
 horizontal lines into slices, each of which
containing one of the phenomena from Figure~\ref{phenomena} and some vertical
strands. To each  horizontal line that slices the graph one associates
the tensor product of the representations that color the crossing strands, 
when pointing downwards, or their duals, when pointing upwards. 
\addtocounter{figs}{-1}To the 
phenomena from {Figure~\ref{phenomena}} one associates, in order, the following
operators: 
\begin{itemize}
\item the flipped universal $R$-matrix $\check{R}:V^m\otimes V^n\rightarrow
V^n\otimes V^m$ (obtained by composing the universal $R$-matrix
with the flip $v\otimes w\rightarrow w\otimes v$), 
\item the inverse $\check{R}^{-1}$ of $\check{R}$,
\item  the projection operator $\beta^{mn}_p:V^m\otimes V^n\rightarrow V^p$,
whose existence and uniqueness is guaranteed by the Clebsch-Gordan theorem,
\item the inclusion  $\beta^p_{mn}:V^p\rightarrow V^m\otimes V^n$, 
\item the contraction $E:V^{n*}\otimes V^{n}\rightarrow {\mathbb C}$,
$E(f\otimes x)=f(x)$
\item its dual  $N:{\mathbb C}\rightarrow V^n\otimes V^{n*}$, 
$N(1)=\sum_je_j\otimes e^j$,
\item the
isomorphism $D:V^{n*}\rightarrow V^{n}$,
\item and its dual $D^*:V^n\rightarrow V^{n**}=V^n$ (see Lemma 3.18 in
\cite{kirbymelvin} for the precise identification of $V^{n**}$ with $V^n$).
\end{itemize}
 One then composes these
operators from the bottom  to the top of the diagram, to obtain a
 linear map from ${\mathbb C}$ to ${\mathbb C}$, which must be of the form
$z\rightarrow \lambda z$. The number $\lambda $ is the Reshetikhin-Turaev
invariant of the ribbon graph. The blank coupons, i.e. the maps
 $D$, might be
required in order to change the orientations of the three edges that
meet at a vertex, 
to make  them   look as depicted in 
{Figure~\ref{phenomena}}. For example
the map $V^{p*}\rightarrow V^m\otimes V^n$ is defined by identifying
$V^{p*}$ with $V^p$ by the isomorphism $D$. 

\begin{figure}[h] 
\centering
\scalebox{.30}{\includegraphics{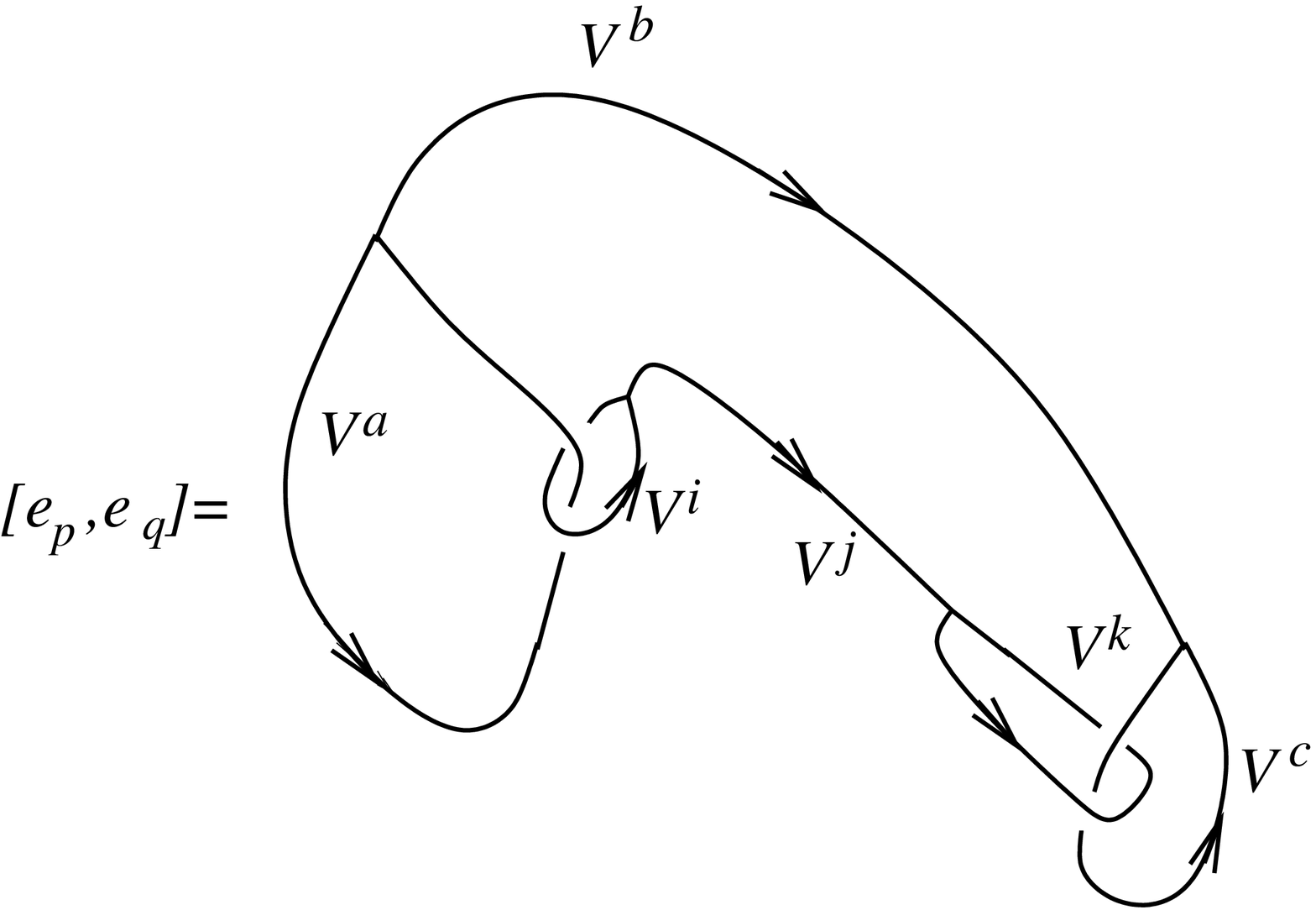}}
\caption{}
\label{pairing}
\end{figure}

Returning to the quantization of Wilson lines, 
the Reshetikhin-Turaev invariant of the graph
 is equal to $[\op(W_{\gamma, n})e_p,e_q]$, where  $[\cdot,\cdot]$ is
the nondegenerate bilinear pairing on ${\mathcal H}_r(\Sigma_g)$ defined in
Figure~\ref{pairing}. One can think of this as being the $p,q$-entry 
of the matrix of the operator, although this is not quite true because
the bilinear pairing is not the inner product.  
But because the pairing is nondegenerate (see \S~\ref{sec:6.3}), 
the above formula
completely determines the operator associated to the Wilson line.

In view of Proposition \ref{repring}, this definition of 
 quantized  Wilson lines is extended to  arbitrary
$n$ by the conventions 
\begin{eqnarray*}
\op(W_{\gamma,r})=0, \quad 
\op(W_{\gamma, r+n}) =-\op(W_{\gamma,r-n}), \quad
\op(W_{\gamma, n+2r})
=-\op(W_{\gamma,n})
\end{eqnarray*}
It can be shown that this quantization  is in the direction of
Goldman's Poisson bracket \cite{alexeevschomerus}. 

\subsection{Non-abelian theta functions from skein modules}\label{sec:5.3}

We re\-phrase the construction from \S~\ref{sec:5.2} in the language
of skein modules. The goal is to express the quantum
group quantization of Wilson lines
 as the left representation of a skein algebra on a quotient
of itself, in the same way that the Schr\"{o}dinger representation
was described in \S~\ref{sec:3.3} as the left 
representation of the reduced linking number skein algebra of 
the cylinder over the torus on a quotient of itself. 

One usually associates to $SU(2)$ Chern-Simons theory
the skein modules of the Kauffman  bracket. The
Reshetikhin-Turaev topological quantum field theory has a Kauffman bracket
analogue constructed in  \cite{BHMV}. However,
the Kauffman bracket skein relations introduce sign discrepancies
in the computation of the desired left action. And since
Theorem~\ref{samequantization} in \S~\ref{sec:5.4} 
brings evidence that  the quantum group quantization 
is the non-abelian analogue of Weyl quantization we 
will define our  modules by 
the skein relations found by Kirby and Melvin in \cite{kirbymelvin} 
for the Reshetikhin-Turaev version of the Jones polynomial. 

We first replace the oriented framed ribbon graphs and links colored
by irreducible representations of $U_{\hbar}(sl(2,{\mathbb C}))$ by 
formal sums of oriented framed links colored by the 2-dimensional 
irreducible representation. Two technical results are needed.

\begin{figure}[h] 
\centering
\scalebox{.30}{\includegraphics{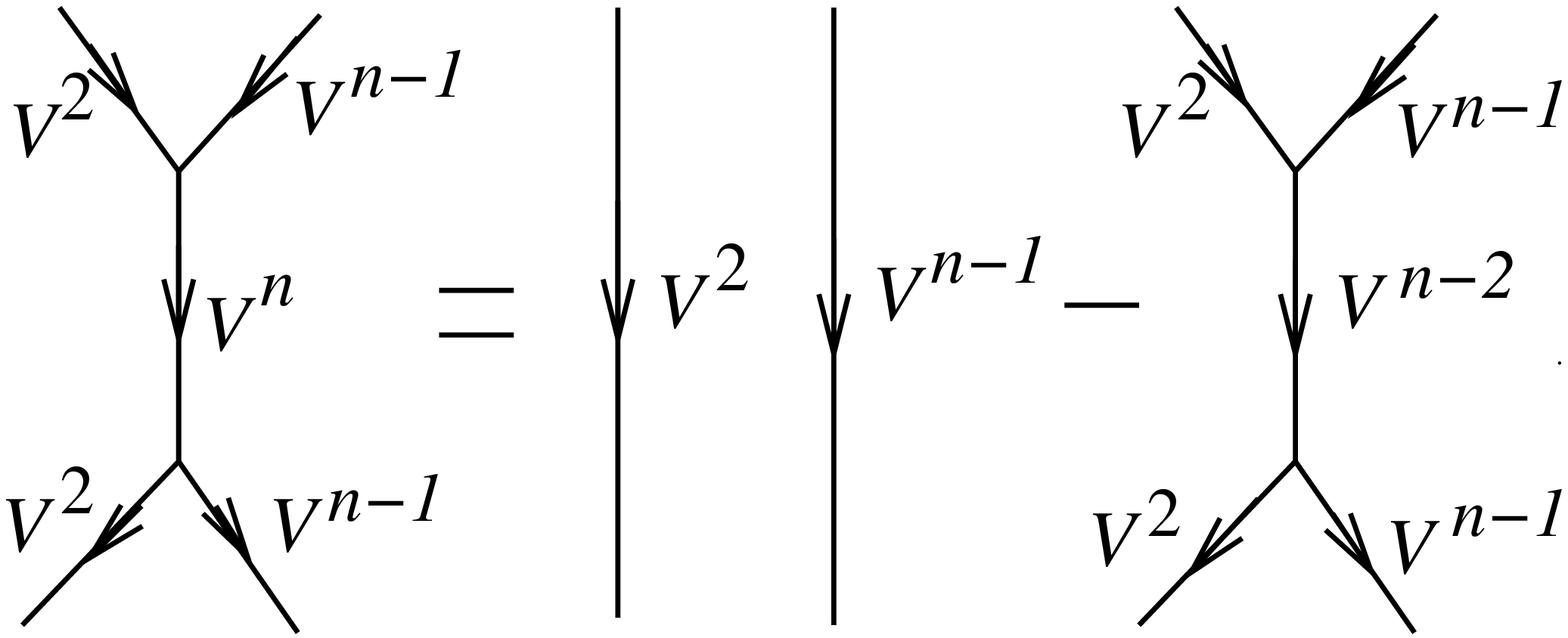}}
\caption{}
\label{clebschgordan}
\end{figure}

\begin{lemma}\label{unu}
For all $n=3,4,\ldots, r-1$ the identity from Figure~\ref{clebschgordan} 
holds.
\end{lemma}

\begin{proof}
This is a corollary of the particular case of the Clebsch-Gordan theorem
$V^n=V^2\otimes V^{n-1}-V^{n-2}$.
\end{proof}

\begin{figure}[h] 
\centering
\scalebox{.35}{\includegraphics{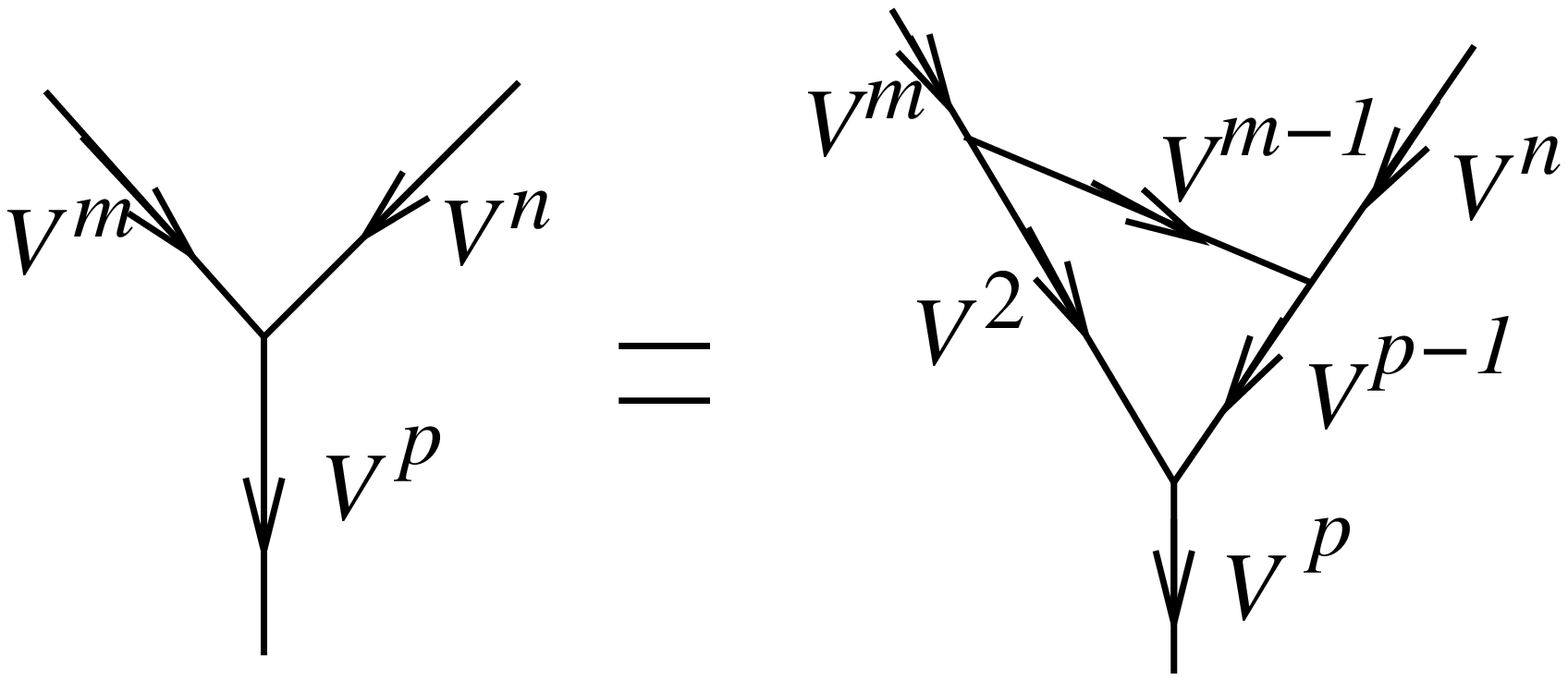}}
\caption{}
\label{vertex}
\end{figure}

\begin{lemma}\label{doi}
For any integers $m,n,p$ satisfying  $m+n+p$ odd and $|m-n|+1\leq p\leq
\min(m+n-1,2r-1-m-n)$, the identity from {Figure~\ref{vertex}} holds.
\end{lemma}

\begin{proof}
We assume familiarity with the proof of the quantum Clebsch-Gordan
theorem in \cite{reshetikhinturaev}.
Set $m_0=\frac{m-1}{2}$, $p_0=\frac{p-1}{2}$.
The morphism described by the 
 diagram on the right is the composition of maps
\begin{eqnarray*}
&&V^p\xrightarrow{\beta^p_{2,p-1}}V^2\otimes V^{p-1}
\xrightarrow{1\otimes \beta^{p-1}_{mn}}V^2\otimes \left(
V^{m-1}\otimes V^n\right)\\ &&=
\left(V^2\otimes V^{m-1}\right)\otimes V^n
\xrightarrow{\beta^{2,m-1}_m\otimes 1}V^m\otimes V^n.
\end{eqnarray*}
Because of Schur's lemma and the quantum Clebsch-Gordan theorem,
this composition is either
the zero or the identity map. To show that 
is not the zero map, we look how the highest weight vector $e_{p_0}$
in  $V^p$ transforms. We have
\begin{eqnarray*}
&&e_{p_0}\rightarrow e_{\frac{1}{2}}\otimes
e_{p_0-\frac{1}{2}}\rightarrow e_{\frac{1}{2}}\otimes
\sum_{i+j=p_0}c_{ij}e_{i}\otimes e_j\\
&&=\sum_{i+j=p_0}c_{ij}
e_{\frac{1}{2}}\otimes e_i\otimes e_j\in V^2\otimes V^{m-1}\otimes V^n.
\end{eqnarray*}

The Clebsch-Gordan coefficients $c_{ij}$ are nonzero,
and in the sum there is a term $c_{m_0-\frac{1}{2},j}e_{\frac{1}{2}}\otimes
e_{m_0-\frac{1}{2}}\otimes e_j$.

On the other hand,  the inclusion  
$\beta^m_{2,m-1}:V^m\rightarrow V^2\otimes V^{m-1}$
 maps the highest weight vector
$e_{m_0}$  in $V^m$ 
to $e_{\frac{1}{2}}\otimes e_{m_0-\frac{1}{2}}$, which is
the product of the vectors
of highest weights in $V^2$ respectively $V^{m-1}$. Hence if
the $\frac{1}{2},m_0-\frac{1}{2}$-component of a vector
 $v$ written in the basis $e_i\otimes e_j$ of 
$ V^2\otimes V^{m-1}$ is nonzero, then $\beta_{2,m-1}^mv$ is nonzero in $V^m$.

In particular,  the above
sum maps to a nonzero vector in $V^m\otimes V^n$. It follows that
the diagram on the right of Figure~\ref{vertex} equals the 
inclusion  $\beta^p_{mn}: V^p
\rightarrow V^m\otimes V^n$, proving the identity.
\end{proof}

\begin{proposition}\label{replacement}
There is an algorithm for replacing each connected
ribbon graph $\Gamma$ colored by irreducible
representations of $U_{\hbar}(sl(2,{\mathbb C}))$ 
by a  sum of oriented framed links
colored by $V^2$ that lie in an $\epsilon$-neighborhood of
the graph, such that if in any ribbon graph $\Gamma '$ that has
$\Gamma$ as a connected component we replace  $\Gamma$ by this sum
of links, we obtain a ribbon graph with the same Reshetikhin-Turaev
invariant as $\Gamma '$. 
\end{proposition}

\begin{figure}[h] 
\centering
\scalebox{.30}{\includegraphics{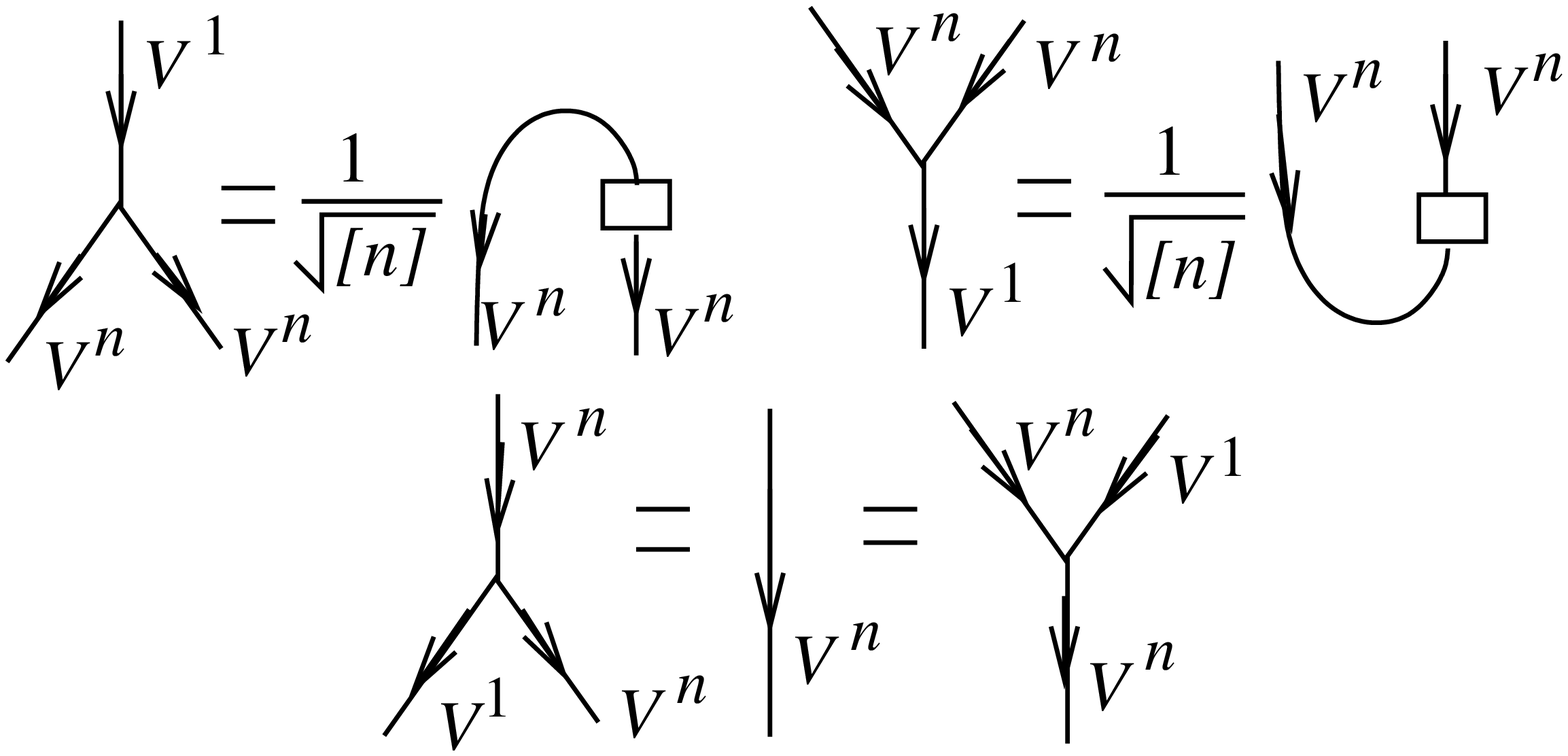}}
\caption{}
\label{vnvone}
\end{figure}

\begin{proof}
For framed knots, the property follows from the cabling formula
given in Theorem 4.15 in \cite{kirbymelvin}; a knot colored by
$V^n$ is replaced by $S_{n-1}(V^2)$.

If the connected ribbon graph has vertices, 
 then by using the isomorphism $D$ to identify 
irreducible representations of $U_{\hbar}(sl(2,{\mathbb C})$ with
their duals, we can obtain the identity from 
Figure~\ref{vertex}  with the arrows reversed.
Also, by taking the adjoint of the map described by the diagram,
we can turn it upside down, meaning that we can write
a similar identity for $\beta^{mn}_p$. 

Based on the two lemmas, the algorithm works as follows. First,
use the identities from {Figure~\ref{vnvone}} to remove all edges colored by 
$V^1$. 
Then apply repeatedly Lemma~\ref{doi} until  
at each vertex of the newly obtained ribbon graph
at least one of the three edges is colored by
$V^2$.  
Next, use Lemma~\ref{unu} to obtain a sum of graphs with the property that,
at each vertex, two
of the three edges are colored by $V^2$ and
one is colored by $V^1$. Finally, use 
the identities from 
Figure~\ref{vnvone} for $n=2$ to transform everything
into a sum of framed links whose edges are colored by $V^2$. 
Each of the  links in the sum
has and even number of blank  coupons (representing the isomorphism
$D$ or its dual)
on each component. Cancel the coupons on each link component in pairs, 
adding a factor of $-1$ each time the two coupons are separated
by an odd number of maxima on the link component. The result is a formal
sum of framed
links with components colored by $V^2$.
\end{proof}

\begin{figure}[h] 
\centering
\scalebox{.35}{\includegraphics{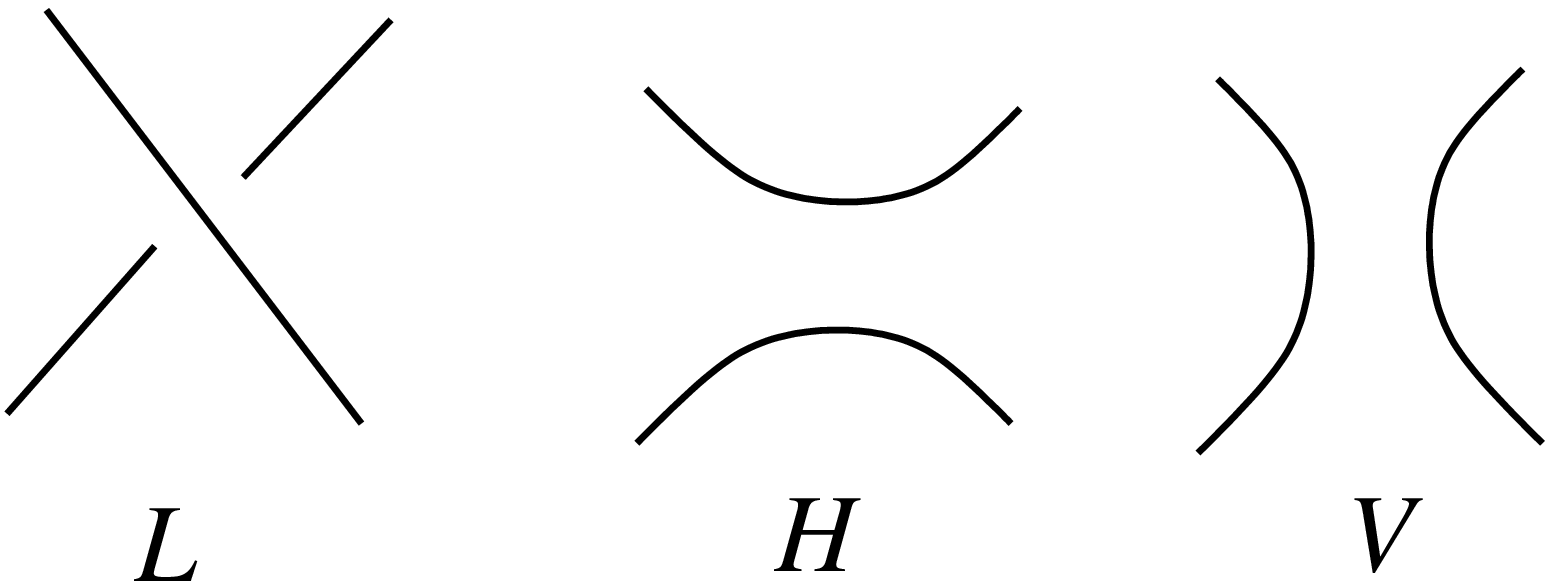}}
\caption{}
\label{crossings}
\end{figure}

Theorem 4.3 in \cite{kirbymelvin} allows us to compute the Reshetikhin-Turaev
invariant of a framed link whose components are colored by $V^2$
using skein relations. First, forget about the orientation of
links. Next, if three
framed links $L,H,V$ in $S^3$ colored by $V^2$ coincide except
in a ball where they look like in Figure~\ref{crossings}, then their 
Reshetikhin-Turaev invariants, denoted by $J_L,J_H$, and $J_V$
satisfy
\begin{eqnarray*}
J_L=tJ_H+t^{-1}J_V \mbox{ or } J_L=\epsilon (tJ_H-t^{-1}J_V)
\end{eqnarray*}   
depending on whether 
 the two crossing strands come from different components or not. 
Here $\epsilon$
is the sign of the crossing, obtained after orienting that link component
(either orientation produces the same sign). Specifically, if the
tangent vectors to the over and under strand form a positive frame
then the sign is positive, otherwise it is negative.
Additionally if a link component bounds a disk inside a ball disjoint
from the rest of the link, then it is replaced by a factor
of $t^2+t^{-2}$.

When $t=1$, namely when $\hbar=0$, it no longer matters whether
one has an undercrossing or  overcrossing,
and both skein relations express  the trace identity of $SU(2)$,
\begin{eqnarray*}
\mbox{tr}(A)\mbox{tr}(B)=\mbox{tr}(AB)+\mbox{tr}(A^{-1}B),
\end{eqnarray*} 
for Wilson lines
\begin{eqnarray*}
W_\alpha W_\beta=W_{\alpha\beta}+W_{\alpha^{-1}\beta}.
\end{eqnarray*}
 For arbitrary $t$, the skein relations are the trace identity for the quantum
group $U_\hbar(sl(2,{\mathbb C}))$, 
\begin{eqnarray*}
t\mbox{tr}(AB)+t^{-1}\mbox{tr}(S(A)B)=\sum_i\mbox{tr}(s_iA)\mbox{tr}(t_iB)
\end{eqnarray*}
where $\sum_i s_i\otimes t_i$ is the universal $R$-matrix of
$U_h(sl(2,{\mathbb C}))$ (see \cite{dougcharliejoanna}). One should
 observe that these skein relations
correspond to the trace identity, while the Kauffman
bracket skein relations correspond to the trace identity for the negative
of the trace.

This prompts us to introduce skein modules defined by these skein relations.
Let for now  $t$  be an abstract variable, rather than the
root of unity  chosen at the beginning of \S~\ref{sec:5.2}.
For an orientable  $3$-dimensional manifold $M$, consider the
 free ${\mathbb C}[t,t^{-1}]$-module
with basis the isotopy classes of framed  links in $M$ including the empty
link. Factor
this module  by the  skein relations 
\begin{eqnarray*}
L=tH+t^{-1}V \mbox{ or } L=\epsilon (tH-t^{-1}V),
\end{eqnarray*}   
depending on whether the two crossing strands come from 
different components or not, 
 where
the links $L, H,V$ are the same except in an embedded ball
in $M$,  inside of which they look as depicted in Figure~\ref{crossings}.
The  same convention for $\epsilon$ is used, with the orientation
of the crossing decided inside the ball.
Additionally, replace any trivial link component that lies inside a ball
disjoint from the rest of the link by
a factor of $t^2+t^{-2}$. We  call the result of the factorization
the {\em Reshetikhin-Turaev skein module} and denote it by $RT_t(M)$. 
One can show that $RT_t(M)$ is isomorphic to  the Kauffman bracket skein
module of $M$.

If $M=\Sigma_g\times [0,1]$ then the homeomorphism
\begin{eqnarray*}
\Sigma_g\times [0,1]\cup_{\Sigma_g}\Sigma_g\times [0,1]\approx
\Sigma_g\times [0,1]
\end{eqnarray*}
induces a multiplication on $R_t(\Sigma_g\times [0,1])$,
turning  it into an algebra, the {\em Reshetikhin-Turaev skein algebra}. 
This algebra is {\em not}
 isomorphic to the Kauffman bracket skein algebra except
in genus one. In higher genus the multiplication rules are
different, as  can be seen by examining the product
of a separating and a nonseparating curve that intersect. 

The operation of gluing $\Sigma_g\times[0,1]$ to the boundary
of a genus $g$ handlebody $H_g$ by a homeomorphism of the
surface induces an $RT_t(\Sigma_g\times[0,1])$-module
structure on $RT_t(H_g)$. Moreover, by gluing
$H_g$ with the empty skein inside to $\Sigma_g\times [0,1]$ we
see that $RT_t(H_g)$ is the quotient of $RT_t(\Sigma_g\times [0,1])$
obtained by identifying the skeins in $\Sigma_g\times [0,1])$
that are isotopic in $H_g$. 

\begin{figure}[h] 
\centering
\scalebox{.30}{\includegraphics{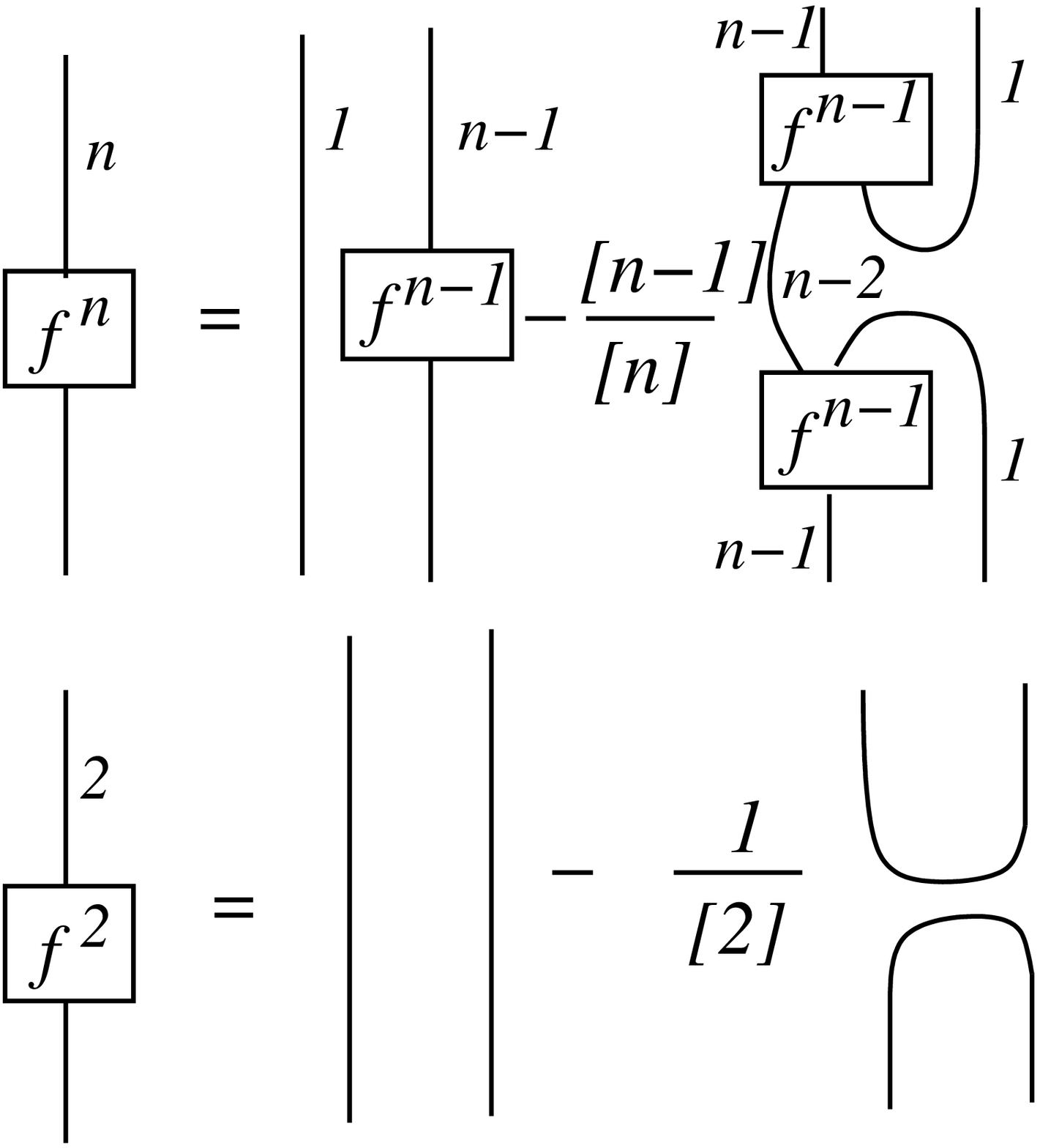}}
\caption{}
\label{jw}
\end{figure}

In view of Lemma~\ref{unu} and the identities from 
Figure~\ref{vnvone}, the irreducible representations
$V^n$ can be represented by  skeins. Explicitly,
$V^n=S_{n-1}(V^2)=f^{n-1}$, where $f^n$ are  defined recursively
in Figure~\ref{jw}. These are the well-known Jones-Wenzl idempotents
\cite{jones1}, \cite{wenzl}. 

The condition $S_{r-1}(V^2)=0$ translates to the condition $f^{r-1}=0$.  
This prompts us to define the {\em reduced Reshetikhin-Turaev
skein module} $\widetilde{RT}_t(M)$ by factoring $RT_t(M)$ by  
$t=e^{\frac{i\pi}{2r}}$ and by the skein relation $f^{r-1}=0$, taken in
every possible embedded ball. 
This reduction is compatible with the multiplicative structure of
$RT_t(\Sigma_g\times [0,1])$ and with its action on $RT_t(H_g)$.

\begin{proposition}\label{rtskein}
The quantum group quantization
of the moduli space of flat $SU(2)$-connections on a surface
$\Sigma_g$ can be represented as  the left multiplication of 
$\widetilde{RT}_t(\Sigma_g\times [0,1])$ on $\widetilde{RT}_t(H_g)$. 
\end{proposition}

\begin{proof}
The proof is based on Proposition \ref{replacement} and 
Proposition \ref{repring}. Because each $f^n$ involves $n$
parallel strands, $RT_t(H_g)$ is a free $C[t,t^{-1}]$-module
with  basis the skeins obtained by 
\begin{itemize}
\item  replacing each edge of the core of 
$H_g$ by a Jones-Wenzl idempotent in such a way that, if $f^m, f^n,f^p$
meet at a vertex, then $m+n+p$ is even,
$m+n\leq p$, $m+p\leq n$, $n+p\leq m$, and
\item replacing the  vertices by the unique collection of strands 
that lie in a disk neighborhood of the vertex and
join the ends of the three Jones-Wenzl idempotents meeting there
 in such a way that there are no crossings.  
\end{itemize}
Because of the Clebsch-Gordan theorem and
Proposition \ref{repring}, in the reduced skein module 
$\widetilde{RT}_t(H_g)$, only edges colored by $f^n$ with $n\leq r-2$ need
to be considered, and also if $f^m,f^n,f^p$ meet at a vertex, then
$m+1,n+1,p+1$ and their cyclic permutations
should satisfy the double inequality from the Clebsch-Gordan theorem.
Each element of this form comes from a basis element in the
quantum group quantization. A more detailed explanation of this can
be found, at least 
for the Kauffman bracket skein modules, in \cite{lickorish3}.

The computation from Figure~\ref{jwchebyshev2}, 
performed in the dotted annulus,
shows that for a simple closed curve $\gamma$ on the torus,
$\op(W_{\gamma,n})$ can be identified with the skein $S_{n-1}(\gamma)
\in \widetilde{RT}_t(\Sigma_g\times [0,1])$. We conclude that
the action of quantum observables on the Hilbert space is modeled
by the action of $\widetilde{RT}_t(\Sigma_g\times [0,1])$
on $\widetilde{RT}_t(H_g)$. 

\begin{figure}[h] 
\centering
\scalebox{.40}{\includegraphics{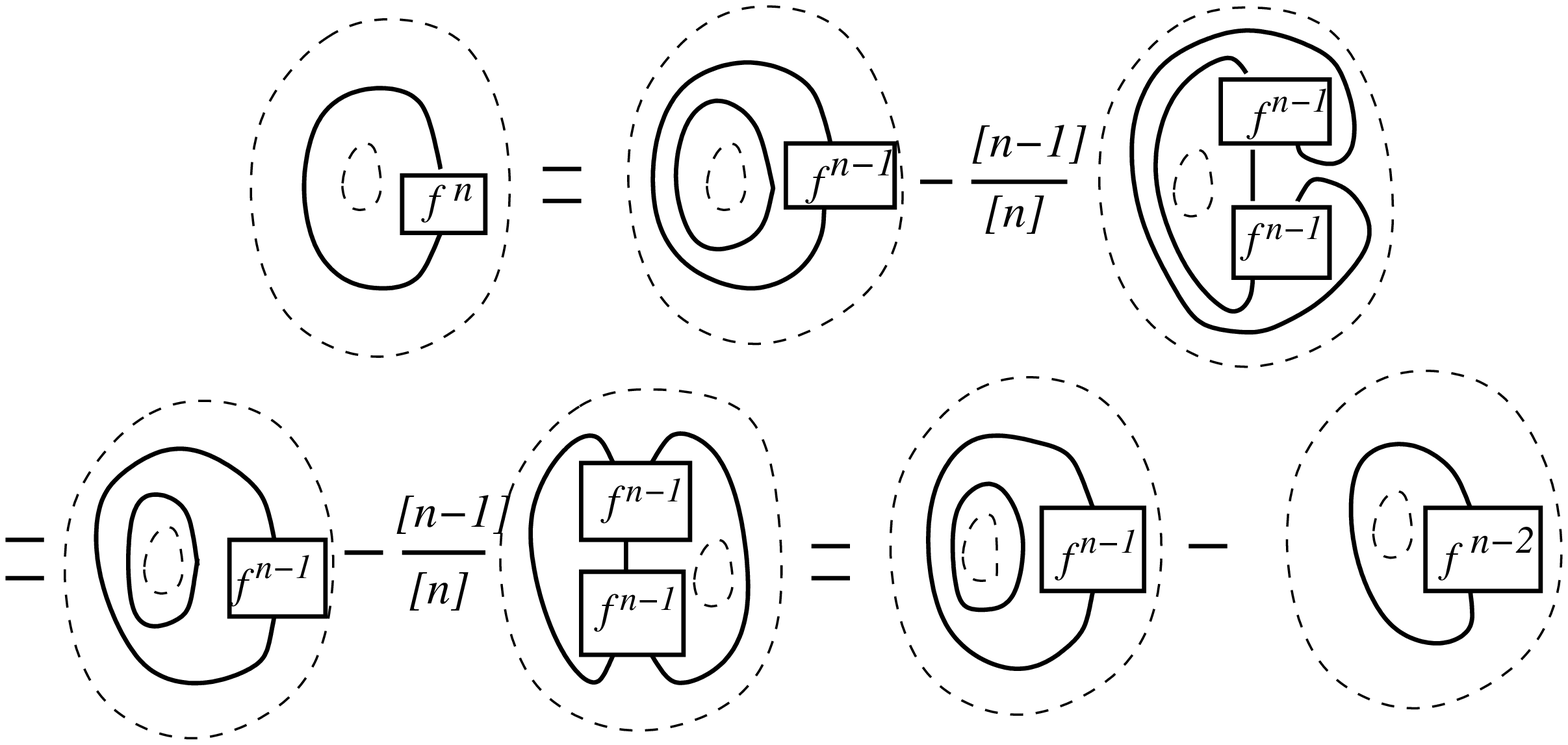}}
\caption{}
\label{jwchebyshev2}
\end{figure}

 To identify the two quantization models,
we also have to prove that the skeins associated to admissible
colorings of the core of the handlebody form a basis, 
namely that they are linearly independent in $\widetilde{RT}_t(H_g)$.

The smooth part of ${\mathcal M}_g$ has real dimension $6g-6$. 
This smooth part is a  completely integrable manifold in the
Liouville sense. Indeed, the Wilson lines $W_{\alpha_i}$, where $\alpha_i$,
$i=1,2,\ldots, 3g-3$, are the curves in Figure~\ref{complint}, form a 
maximal set of Poisson commuting functions (meaning that $\{W_{\alpha_i},
W_{\alpha_j}\}=0$). The quantum group quantization of the moduli space
of flat $SU(2)$-connections is thus a quantum integrable system, with the
operators $\op(W_{\alpha_1})$,
$\op(W_{\alpha_2})$, ..., $\op(W_{\alpha_{3g-3}})$ satisfying the
integrability condition.

\begin{figure}[h]
\centering
\scalebox{.35}{\includegraphics{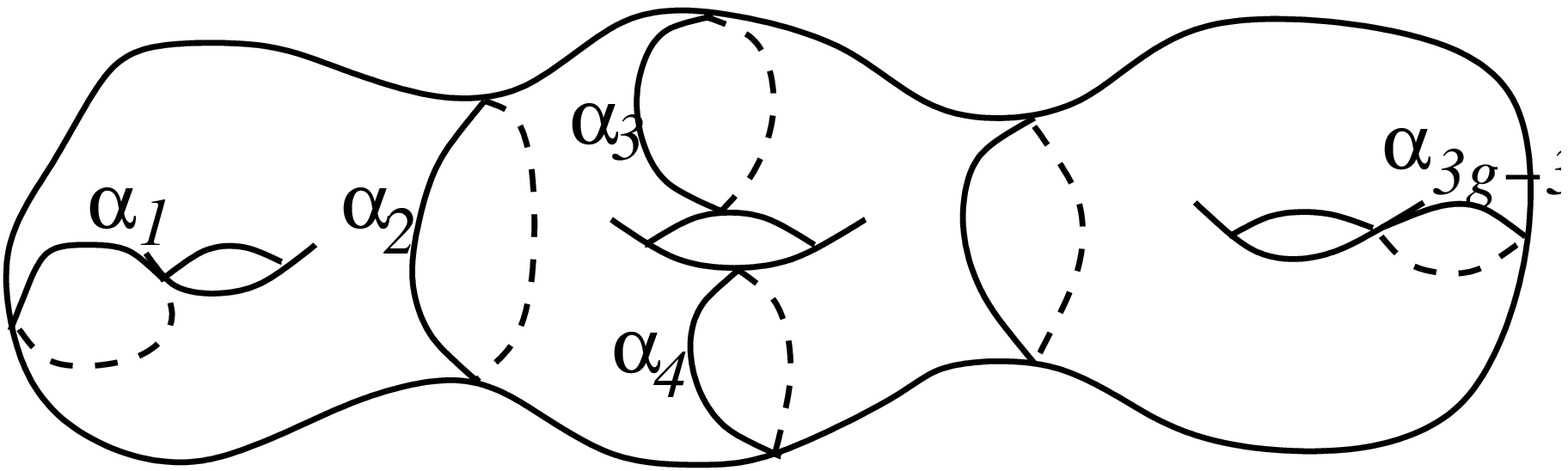}}
\caption{}
\label{complint}
\end{figure}

The identity from Figure~\ref{eigen}, which holds for any choice of orientation
of the strands, implies that
the spectral decomposition of the commuting
$(3g-3)$-tuple of self-adjoint operators
 $$(\op(W_{\alpha_1}), \op(W_{\alpha_2}), ..., \op(W_{\alpha_{3g-3}}))$$ has only 
$1$-dimensional eigenspaces consisting precisely of the colorings
of the edges following the specified rule. Indeed,  the basis elements 
 are as  
described in \S~\ref{sec:5.2} for the case where the curves that
cut the surface into pairs of pants are $\alpha_1,\alpha_2,\cdots,
\alpha_{3g-3}$,
and the identity from {Figure~\ref{eigen}} shows that
the eigenvalues of an $e_j$ with respect to the $3g-3$
quantized Wilson lines completely determine the colors of its edges.
\begin{figure}[h]
\centering
\scalebox{.35}{\includegraphics{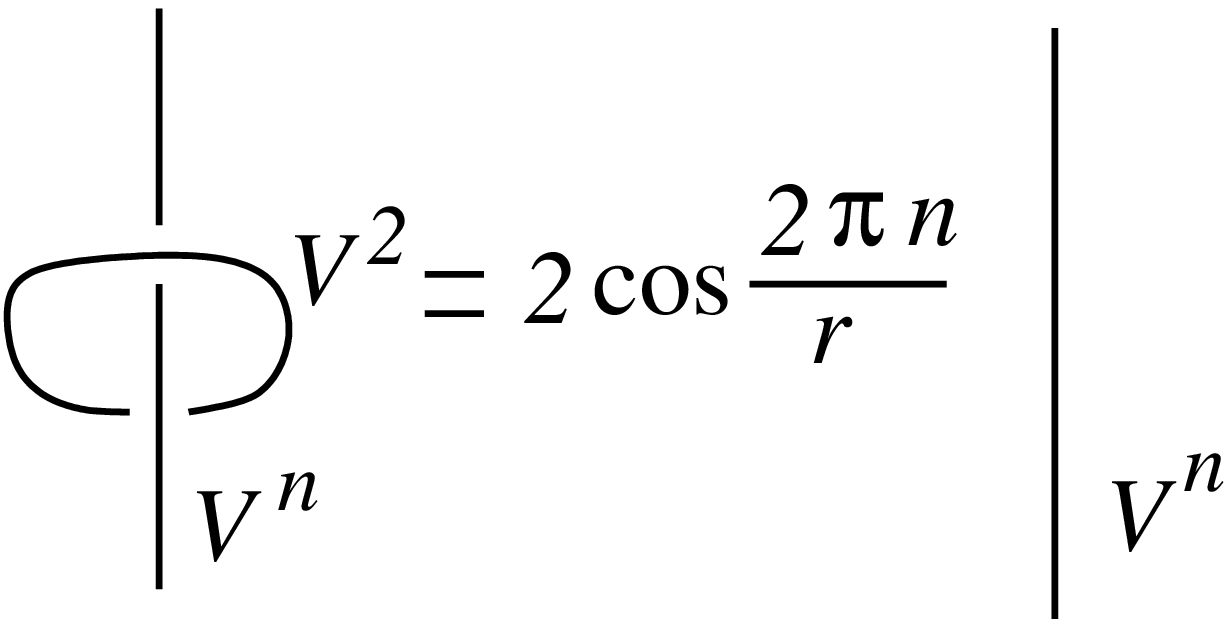}}
\caption{}
\label{eigen}
\end{figure}
This concludes the proof.
\end{proof}

\begin{remark}
Proposition \ref{rtskein} should be compared to 
Theorem \ref{thetalink}. Again the
algebra of quantized observables is a skein algebra, the space
of non-abelian theta functions is a quotient of this algebra, and the
 factorization relation is of topological nature; it is defined by gluing
the cylinder over the surface to a handlebody via a homeomorphism. 
The skein modules ${ RT}_t(\Sigma_g\times [0,1])$ and
$\widetilde{RT}_t(\Sigma_g\times [0,1])$ are 
the analogues, for the gauge group $SU(2)$, of the algebras 
${\mathbb C}[\heis ({\mathbb Z})]$ and $\grpalg$. 
\end{remark}

Since we have not yet proved  that the pairing $[\, ,\, ]$ defined
in \S~\ref{sec:5.2} is nondegenerate, we will take for the moment this 
representation of $\widetilde{RT}_t(\Sigma_g\times [0,1])$ to
actually be the quantum group quantization of the moduli space 
${\mathcal M}_g$. We will prove the  nondegeneracy 
in \S~\ref{sec:6.3}.

The quantum group quantization is more natural than it seems.
Quantum groups were introduced by Drinfeld to solve vertex models,
as means of finding operators that satisfy the Yang-Baxter equation.
They lead to the deformation quantization model for the quantization
of ${\mathcal M}_g$ in \cite{alexeevschomerus}. This gives rise to
the skein algebra of the surface, and by analogy with \S~\ref{sec:3.3} we are
led to consider the skein module of the handlebody. The basis 
consisting of admissible colorings of the core of the handlebody
appears when looking at the spectral decomposition of the system
of commuting operators from the proof of Proposition~\ref{rtskein}.

\subsection{The quantum group quantization of the moduli space of flat
  $SU(2)$-connections on the torus}\label{sec:5.4}

The quantum group quantization of ${\mathcal M}_1$
is a particular case of the
construction in \S~\ref{sec:5.2} and has been described in \cite{gelcauribe}.
 A basis for the Hilbert space 
is specified by an oriented rigid structure on the torus. The curves $a$ and
$b$ in Figure~\ref{rigtorus} define such a structure with $a$ the seam
and $b$ the curve that cuts the torus into an annulus. 
Mapping the torus to the boundary of the solid torus such that
$b$ becomes null homologous and $a$ the   generator of the fundamental group,
 we obtain an
orthonormal basis consisting of the  vectors
$V^1(\alpha), V^2(\alpha),\ldots, V^{r-1}(\alpha)$,
which are the colorings of the core $\alpha$ of the solid torus by 
the irreducible representations $V^1,V^2, \ldots, V^{r-1}$  
of $U_\hbar(sl(2,{\mathbb C}))$.
These are the quantum group analogues of the 
$\zeta_j^\tau$'s. Here, the orientation of the rigid structure,
and hence of the core of the solid torus are irrelevant, reversing
the orientation gives the same results in computations (orientation
of link components is irrelevant \cite{witten}).

The operator associated to the function $f(x,y)=2\cos 2\pi (px+qy)$ is 
computed like for higher genus surfaces. The  
bilinear form on the Hilbert space comes from the Hopf
link and is $[V^j(\alpha),V^{k}(\alpha)]=[jk]$, $j,k=1,2,\ldots, r-1$.  
The value
of $[\mbox{Op}(2\cos 2\pi (px+qy))V^{j}(\alpha),V^{k}(\alpha)]$ is
equal to the Reshetikhin-Turaev invariant of the three-component 
colored framed link consisting the curve of slope $p/q$ 
on the torus embedded in standard position
in  $S^3$, colored by $V^{n+1}-V^{n-1}$
where $n$ is the greatest common divisor of $p$ and $q$, the core
of the solid torus that lies on one side of the torus colored by
$V^j$, and the core of the solid torus that lies on the other side 
colored by $V^k$. Coloring the curve by $V^{n+1}-V^{n-1}$ is the same
as coloring it by 
\begin{eqnarray*}
T_n(V^2)=\sum_{j=0}^{\lfloor \frac{n}{2}\rfloor}(-1)^j\frac{n}{n-j}{{n-j}
\choose {j}}(V^2)^{n-2j},
\end{eqnarray*} 
where $T_n(x)$ is the Chebyshev polynomial of the first kind defined
recursively by $T_0(x)=2$, $T_1(x)=x$, $T_{n+1}(x)=xT_{n}(x)-T_{n-1}(x)$,
for $n\geq 1$. 
Again, the quantum group quantization can be modeled by the
action of the reduced Reshetikhin-Turaev skein algebra of
the torus on the reduced Reshetikhin-Turaev skein module of the solid
torus.

It has been shown in 
\cite{gelcauribe} that the quantum group quantization of the 
pillow case is
unitary equivalent to Weyl quantization. However, that proof
makes use of the Reshetikhin-Turaev representation of the mapping
class group of the torus, and  does not
serve our purpose of showing {\em how} the Reshetikhin-Turaev representation
arises from quantum mechanical considerations. For that reason
we  give a different proof of this result using
the structure of the Reshetikhin-Turaev skein algebra of the torus. 

For a pair of integers $p,q$, let $n$ be their common divisor and
define $(p,q)_T=T_n((p/n,q/n))\in RT_t({\mathbb T}^2\times [0,1])$.
The proof of the following result is identical to that of
Theorem 4.1 in \cite{frohmangelca}, which covers the case of the Kauffman
bracket.
 
\begin{theorem}\label{producttosum}
For any integers $p,q,p',q'$ the following product-to-sum formula
holds
\begin{eqnarray*}
(p,q)_T(p',q')_T=t^{\left|\begin{smallmatrix}p&q\\p'&q'\end{smallmatrix}\right|}(p+p',q+q')_T+
t^{-\left|\begin{smallmatrix}p&q\\p'&q'\end{smallmatrix}\right|}(p-p',q-q')_T.
\end{eqnarray*}
\end{theorem}

As we can see,  the Reshetikhin-Turaev  and the Kauffman bracket
skein algebras of the torus are isomorphic.

\begin{theorem}\label{samequantization}\cite{gelcauribe}
The Weyl quantization and the  quantum group quantization of the 
moduli space of flat $SU(2)$-connections on the torus are unitary 
equivalent.
\end{theorem}

\begin{proof}
We rephrase the quantum group quantization in terms of 
skein modules. The Hilbert space  is 
$\widetilde{RT}_t(S^1\times{\mathbb D}^2)$. Indeed, this skein
module is spanned by the vectors $S_{j-1}(\alpha)$, $j=1,2,
\ldots, r-1$, and these vectors are linearly independent because
they are eigenvectors with different eigenvalues of the operator
defined by $(0,1)$. 

Considering the projection $\pi:RT_t({\mathbb T}^2\times [0,1])
\rightarrow RT_t(S^1\times {\mathbb D}^2)$ defined by attaching
the cylinder over the torus to the solid torus by the homeomorphism
$h_0$ from \S~\ref{sec:3.3}, and using Theorem~\ref{producttosum} we deduce the
recursive formula
\begin{eqnarray*}
\pi((p+1,q)_T)=t^{-q}\alpha \pi((p,q)_T)-t^{-2q}\pi((p-1,q)_T).
\end{eqnarray*}
Also $\pi((0,q)_T)=t^{2q}+t^{-2q}$, and $\pi((1,q)_T)=t^{-2q}\alpha$.
Solving the recurrence we obtain
\begin{eqnarray*}
\pi((p,q)_T)=t^{-pq}(t^{2q}S_{p}(\alpha)-t^{-2q}S_{p-2}(\alpha)).
\end{eqnarray*}
Using again  Theorem~\ref{producttosum} we have 
\begin{eqnarray*}
 (p,q)_TT_j(\alpha)=\pi[(p,q)_T(j,0)_T]
=\pi[t^{-jq}(p+j,q)_T+t^{jq}(p-j,q)_T]\\
=t^{-pq}[t^{-(2j-2)q}S_{p+j}(\alpha)+t^{(2j+2)q}S_{p-j}(\alpha)-
t^{-(2j-2)q}S_{p+j-2}(\alpha)\\-t^{(2j-2)q}S_{p-j-2}(\alpha).
\end{eqnarray*}
Since $T_n(x)=S_n(x)-S_{n-2}(x)$ for all $n$, we have
\begin{eqnarray*}
(p,q)_TS_{j-1}(\alpha)=t^{-pq}(t^{-2qj}S_{p+j-1}(\alpha)+
t^{2qj}S_{p-j+1}(\alpha), \mbox{ for }j>0.
\end{eqnarray*}
Reducing to the relative skein modules and using
the fact that $S_{j-1}(\alpha)=V^j(\alpha)$, we obtain
\begin{eqnarray}\label{actiontorus}
\begin{array}{r}
\op(2\cos 2\pi (px+qy))V^j(\alpha)=e^{-\frac{\pi i}{2r}  pq}\left(e^{\frac{\pi i}{r} qj}
V^{j-p}(\alpha)\right.\\
\left.+e^{-\frac{\pi i}{r} qj}V^{j+p}(\alpha)\right),
\end{array}
\end{eqnarray}
with the conventions $V^r(\alpha)=0$, $V^{j+2r}(\alpha)=V^j(\alpha)$,
$V^{r+j}(\alpha)=-V^{r-j}(\alpha)$ if the indices
 leave the range $1,2,\ldots, r-1$. This is the formula for
the Weyl quantization of the pillow case given in \S~\ref{sec:4.2}, which
proves the theorem.
\end{proof}

\subsection{A Stone--von Neumann theorem on the moduli space of flat
  $SU(2)$-connections
on the torus}\label{sec:5.5}

 Weyl quantization
yields a representation of
 $\widetilde{RT}_t({\mathbb T}\times [0,1])$ such that
$t$ acts as multiplication by $e^{\frac{i\pi}{2r}}$ and
every  simple closed curve on the torus  acts as
 a self-adjoint operator.
The algebra $\widetilde{RT}_t({\mathbb T}\times [0,1])$ is a non-abelian
analogue of the group algebra of the finite Heisenberg group. 
A Stone-von Neumann theorem holds also in this case.

\begin{theorem}\label{SvNpillow}
The representation of the reduced Reshetikhin-Turaev skein algebra of the
torus defined by the Weyl quantization of the moduli
space of flat $SU(2)$-connections on the torus is the 
 {\em unique} irreducible representation of  this algebra that maps
simple closed curves to self-adjoint operators and $t$ to multiplication
by $e^{\frac{\pi i}{2r}}$. Moreover, quantized Wilson lines span the algebra
of  all linear operators on the Hilbert space of the quantization. 
\end{theorem}

\begin{proof} 
We prove irreducibility by
showing  that any vector is cyclic.
Because the eigenspaces of each quantized Wilson line are $1$-dimensional,
in particular those of $\op(2\cos 2\pi y)$, 
it suffices to check this property
 for the eigenvectors of this operator, namely
for $\zeta_j^\tau$, $j=1,2,\ldots, r-1$.
And because 
\begin{eqnarray*}
&&\op (2\cos2\pi x)\zeta_j^\tau=\zeta_{j-1}^\tau+\zeta_{j+1}^\tau\\
&&\op(2\cos 2\pi (x+y))\zeta_j^\tau=t^{-1}(t^2\zeta_{j-1}^\tau+t^{-2}\zeta_{j+1}^\tau),  
\end{eqnarray*}
by taking linear combinations we see that from $\zeta_j^\tau$ we can generate
both $\zeta_{j+1}^\tau$ and $\zeta_{j-1}^\tau$. Repeating, we can generate
the entire basis. This shows that $\zeta_j^\tau$ is cyclic for each $j=1,2,
\ldots, r-1$, hence  the representation is irreducible.

To prove uniqueness, 
consider an irreducible representation of $\widetilde{RT}_t({\mathbb
  T}^2\times [0,1])$ with the required  properties. 
The condition  $S_{r-1}(\gamma)=0$ for any
simple closed curve $\gamma$ on the torus
implies, by the spectral mapping theorem,  that the
eigenvalues of the operator associated to
$\gamma$ are among the numbers $2\cos \frac{k\pi}{r}$,
$k=1,2,\ldots, r-1$.  

We write for  generators $X=(1,0)$, $Y=(0,1)$, and $Z=(1,1)$
of   $\widetilde{RT}_t({\mathbb T}^2\times [0,1])$ the relations
\begin{eqnarray*}
&&tXY-t^{-1}YX=(t^2-t^{-2})Z\\
&&tYZ-t^{-1}ZY=(t^2-t^{-2})X\\
&&tZX-t^{-1}XZ=(t^2-t^{-2})Y\\
&&t^2X^2+t^{-2}Y^2+t^2Z^2-tXYZ-2t^2-2t^{-2}=0,
\end{eqnarray*}
by analogy with
 the presentation of the Kauffman bracket skein algebra of the torus
found by Bullock and Przytycki in \cite{bullockprzytycki}.


In fact $\widetilde{RT}_t({\mathbb T}^2\times [0,1])$
 is generated by just $X$ and $Y$, since we can substitute $Z$ from
the first equation.  
This gives the equivalent presentation
\begin{equation}\label{relations}
\begin{array}{ll}
&(t^2+t^{-2})YXY-(XY^2+Y^2X)=(t^4+t^{-4}-2)X\\
&(t^2+t^{-2})XYX-(YX^2+X^2Y)=(t^4+t^{-4}-2)Y\\
&(t^6+t^{-2}-2t^2)X^2+(t^{-6}+t^2-2t^{-2})Y^2+XYXY+YXYX\\
&\quad - t^2YX^2Y-t^{-2}XY^2X=2(t^6+t^{-6}-t^2-t^{-2}).
\end{array}
\end{equation}
Setting  $t=e^{\frac{i\pi}{2r}}$ the first equation in (\ref{relations}) becomes
\begin{eqnarray*}
2\cos \frac{\pi}{r}YXY-(XY^2+Y^2X)=4\sin ^2\frac{\pi}{r}Y.
\end{eqnarray*}

Let $v_k$ be an eigenvector of $Y$ with  eigenvalue  
$2\cos \frac{k\pi}{r}$ for some $k\in \{1,2,\ldots, r-1\}$. 
We wish to generate a basis of 
the representation by acting repeatedly on $v_k$ by $X$.  
For this set $Xv_k=w$. The above relation yields
\begin{eqnarray*}
2\cos \frac{\pi}{r}\cdot 2\cos \frac{k\pi}{r}Y w-4\cos ^2\frac{k\pi}{r}w-Y^2w=
4\sin ^2\frac{\pi}{r}w.
\end{eqnarray*}
Rewrite this as 
\begin{eqnarray*}
\left[Y^2-4\cos \frac{k\pi}{r}\cos \frac{\pi}{r}Y-4\left(\sin ^2\frac{\pi}{r}+\cos ^2
\frac{k\pi}{r}\right)\right]w=0.
\end{eqnarray*}
It follows that either $w=0$ or $w$ is in the kernel of the operator 
\begin{equation}\label{operator}
Y^2-4\cos \frac{k\pi}{r}\cos \frac{\pi}{r}Y-4\left(\sin ^2\frac{\pi}{r}+\cos ^2
\frac{k\pi}{r}\right)Id.
\end{equation}
The second equation in
(\ref{relations}) shows that if $Xv_k=w=0$ then $Yv_k=0$.
This  is impossible because of the
third relation in (\ref{relations}). Hence $w\neq 0$, so  $w$ lies
in the kernel of the operator from (\ref{operator}). Note that if 
 $\lambda$ is an eigenvalue of $Y$ which satisfies
\begin{eqnarray*}
\lambda ^2-4\cos \frac{k\pi}{r}\cos \frac{\pi}{r}\lambda -4\left(\sin ^2\frac{\pi}{r}+4\cos ^2\frac{k\pi}{r}\right)=0,
\end{eqnarray*}
then necessarily $\lambda=2\cos \frac{(k\pm 1)\pi}{r}$. It follows that 
\begin{eqnarray*}
Xv_k=v_{k+1}+v_{k-1},
\end{eqnarray*}
where $Yv_{k\pm 1}=2\cos \frac{(k\pm 1)\pi}{4}v_{k\pm 1}$, 
and $v_{k+1}$ and $v_{k-1}$ are not simultaneously equal to zero. 
We wish to enforce $v_k$, $v_{k+1}$, and $v_{k-1}$ to be elements of a basis.
For that we need to check that $v_{k+1}$ and $v_{k-1}$  are nonzero, and we also
need to understand the action of $X$ on them. 

Set $Xv_{k+1}=\alpha v_k+v_{k+2}$ and $Xv_{k-1}=\beta v_k+v_{k-2}$, where
$Yv_{k\pm 2}=2\cos \frac{(k\pm 2)\pi}{r}v_{k\pm 2}$. It might be possible that
the scalars $\alpha $ and $\beta$ are zero. The vectors $v_{k+2}$, $v_{k-2}$ 
might as well be zero; if they are not zero,  then they are eigenvectors 
of $Y$, and their respective eigenvalues are as specified (which can be seen
by repeating the above argument). 

Applying both sides of the second equation in (\ref{relations}) to
$v_k$ and comparing the $v_k$ coordinate of the results we obtain
\begin{eqnarray*}
\cos \frac{\pi}{r}\cos \frac{(k+1)\pi}{r}\alpha +
\cos \frac{\pi}{r} \cos \frac{(k-1)\pi}{r}\beta -\cos \frac{k\pi}{r}(\alpha +\beta)
\\=  \cos \frac{2\pi}{r} \cos \frac{k\pi}{r}-\cos \frac{k\pi}{r}.
\end{eqnarray*}
This is equivalent to
\begin{eqnarray*} 
\left(\cos \frac{(k+2)\pi}{r}+\cos \frac{k\pi}{r}\right)(\alpha -1)+
\left(\cos \frac{(k-2)\pi}{r}+\cos \frac{k\pi}{r}\right)(\beta -1)=0
\end{eqnarray*}
that is 
\begin{eqnarray*}
\sin \frac{(k+1)\pi}{r}(\alpha -1)+\sin \frac{(k-1)\pi}{r}(\beta -1)=0.
\end{eqnarray*}
For further use, we write this as 
\begin{eqnarray}\label{furtheruse}
(t^{4k+4}-1)(\alpha -1)+(t^{4k}-t^4)(\beta -1)=0.
\end{eqnarray}

Applying the two sides of the  last equation in (\ref{relations}) to
$v_k$ and comparing the $v_k$ coordinate of the results we obtain
\begin{eqnarray*}
&&(t^6+t^{-2}-2t^2)(\alpha+\beta)+(t^{-6}+t^2-2t^{-2})4\cos ^2\frac{k\pi }{r}\\
&&+8\cos \frac{k\pi }{r}\cos \frac{(k+1)\pi}{r}\alpha+
8\cos \frac{k\pi}{r}\cos \frac{(k-1)\pi}{r}\beta
-4t^2\cos^2\frac{k\pi}{r}(\alpha+\beta)\\
&&-4t^{-2}\cos^2\frac{(k+1)\pi}{r}\alpha-
4t^{-2}\cos^2\frac{(k-1)\pi}{r}\beta=2(t^6+t^{-6}-t^2-t^{-2}).
\end{eqnarray*}
This can be rewritten as
\begin{eqnarray*}
&& (t^{6}+t^{-2}-2t^{2})(\alpha +\beta)+4\cos \frac{(2k+1)\pi}{r}\alpha 
+4\cos \frac{\pi}{r}\alpha  +4\cos \frac{(2k-1)\pi}{r}\beta\\ 
&&+4\cos \frac{\pi}{r}\beta -2t^{2}\cos \frac{2k\pi}{r}\alpha -2t^{2}\cos
\frac{2k\pi}{r}\beta -2t^{2}\alpha -2t^{2}\beta\\
&&-2t^{-2}\cos \frac{(2k+2)\pi}{r}\alpha -2t^{-2}\alpha
  -2t^{-2}\cos \frac{(2k-2)\pi}{r}\beta
  -2t^{-2}\beta \\
&& =2(t^6+t^{-6}-t^2-t^{-2}) -2(t^{-6}+t^{2}-2t^{-2})
  -2(t^{-6}+t^{2}-2t^{-2})\cos \frac{2k\pi}{r}.
\end{eqnarray*}
Using the fact that $t=\cos \frac{k\pi}{2r}+i\sin \frac{k\pi}{2r}$
 we can transform this into 
\begin{eqnarray*}
&&(t^{-4k-6}+t^{-4k+2}+2t^{2}-2t^{-4k-2}-t^{6}-t^{-2})(\alpha -1)\\
&&+(t^{4k-6}+t^{4k+2}+2t^{2}-2t^{4k-2}-t^{6}-t^{-2})(\beta -1)=0.
\end{eqnarray*}
Dividing through by $t^{-6}+t^{2}-2t^{-2}$ we obtain 
\begin{eqnarray*}
(t^{-4k}-t^4)(\alpha -1)+(t^{4k}-t^4)(\beta -1)=0.
\end{eqnarray*}
Combining this with (\ref{furtheruse}), we obtain the
system  
\begin{alignat*}{1}
(t^{4k+4}-1)u +(t^{4k}-t^4)v&=0\\
(t^{-4k}-t^4)u+(t^{4k}-t^4)v&=0
\end{alignat*}
in the unknowns $u=\alpha -1$ and $v=\beta -1$. 
Recall that $t=e^{\frac{i\pi}{2r}}$. 
 
The coefficient of $v$ equals zero if and only if $k=1$, in which case
we are forced to have $\beta =0$, because $0$ is not an eigenvalue of $Y$. 
The coefficient of $u$ in one of the equations is equal to zero if and
only if $k=r-1$, in which case we are forced to have $\alpha =0$, 
because $-1$ is not an eigenvalue of $Y$. 
 
In any other situation, by subtracting the equations we obtain 
\begin{eqnarray*}
(t^4-t^{-4k})(t^{4k}+1)u=0.
\end{eqnarray*}
This can happen only if $t^{4k}=-1$, namely if $2k=r$. 
 
So, if $k\neq \frac{r}{2}$, then $Xv_k=v_{k+1}+v_{k-1}$ with  $v_{k+1}$
and $v_{k-1}$ eigenvectors of $Y$ with eigenvalues $2\cos \frac{(k+1)\pi}{r}$
respectively $2\cos \frac{(k-1)\pi}{r}$, and $Xv_{k\pm 1}=v_k+v_{k\pm 2}$,
where $v_{k\pm 2}$ lie in the eigenspaces of $Y$ of the eigenvalues 
$2\cos \frac{(k\pm 2)\pi}{r}$.
 
What if $k=\frac{r}{2}$? One of $v_{k+1}$ and $v_{k-1}$ is
not zero, say $v_{k+1}$. Applying the above considerations to $v_{k+1}$ we
have $Xv_{k+1}=\alpha v_k+v_{k+2}$ and $X\alpha v_k=v_{k+1}+v'_{k-1}$, for some
$v'_{k-1}$ in the eigenspace of $Y$ of 
the eigenvalue $2\cos \frac{(k-1)\pi}{r}$. Then on the one hand 
 $Xv_k=v_{k+1}+v_{k-1}$ and on the other  $\alpha Xv_k=v_{k+1}+v'_{k-1}$. 
This shows that $\alpha =1$, and because $(\alpha -1)+(\beta -1)=0$, 
it follows that $\beta =1$
as well. A similar conclusion is reached if $v_{k-1}\neq 0$.
 
 Repeating the argument we conclude that the irreducible representation, 
which must be the span
 of $X^mY^nv_k$ for $m,n\geq 0$,  has the basis $v_1,v_2,\ldots, v_{r-1}$,
 and $X$ and $Y$ act on these vectors by 
 \begin{eqnarray*}
 Xv_j=v_{j+1}+v_{j-1}, \quad Yv_j=2\cos \frac{j\pi}{r},
 \end{eqnarray*}
 with the convention $v_0=v_r=0$. 
And we recognize the representation defined by the Weyl quantization of
the moduli space of flat $SU(2)$-connections on the torus.

The fact that the algebra of all quantized Wilson lines
is the algebra of all linear operators on the Hilbert space
of the quantization is a corollary of Theorem 6.1 in \cite{gelca2}.
\end{proof}

\section{The Reshetikhin-Turaev representation 
as a Fourier transform for non-abelian theta functions}\label{sec:6}

 \subsection{The Reshetikhin-Turaev representation of the mapping class group
  of the torus}\label{sec:6.1}

In this section we deduce the existence of the Reshe\-ti\-khin-Turaev
 projective representation of the
mapping class group of the torus from quantum mechanical considerations,
and show that it can be computed  explicitly {\em from these considerations}.
This should be compared with the computations in \S~\ref{sec:3.4}.

There is an action of the mapping class group  of the 
torus on the ring 
of functions on the pillow case, given by
\begin{eqnarray*}
h\cdot f(A)=f(h^{-1}_*A),
\end{eqnarray*}
where  $h_*^{-1}A$ denotes the
pull-back of the connection $A$ by $h$. In particular
the Wilson line of a curve $\gamma$ is mapped to the Wilson
line of the curve $h(\gamma)$.
The action of the mapping class group on functions
on the pillow case
induces an action on the quantum observables by
\begin{eqnarray*}
h\cdot \op(f(A))= \op(f(h^{-1}_*A)),
\end{eqnarray*}
which for Wilson lines is 
\begin{eqnarray*}
h\cdot \op(W_{\gamma})=\op(W_{h(\gamma)}).
\end{eqnarray*}

\begin{theorem}\label{egorov}
There exists a projective representation of the 
mapping class group of the torus that satisfies the exact Egorov identity 
\begin{eqnarray*}
\op(W_{h(\gamma)})=\rho(h)\op(W_{\gamma})\rho(h)^{-1}
\end{eqnarray*}
with the quantum group quantization of Wilson lines.
Moreover, $\rho(h)$ is unique up to multiplication
by a constant. 
\end{theorem}

\begin{proof}
We follow the first proof to Theorem~\ref{defofphi}.
The bijective map  $L\rightarrow h(L)$ on the set of isotopy classes
of framed links in the cylinder over the torus induces an automorphism of
the free ${\mathbb C}[t,t^{-1}]$-module with basis these isotopy
classes of links. Because this map leaves invariant the
ideal defined by the skein relations (for crossings and for the $r-1$st
Jones-Wenzl idempotent),
it defines an automorphism $\Phi:\widetilde{RT}_t({\mathbb T}^2
\times [0,1])\rightarrow \widetilde{RT}_t({\mathbb T}^2\times [0,1])$.
The representation of $RT_t({\mathbb T}^2\times [0,1])$ given
by $V^j(\alpha)\rightarrow \op(W_{h(\gamma)})V^j(\alpha)$ is an irreducible
representation of $\widetilde{RT}_t({\mathbb T}^2\times[0,1])$
which still maps $t$ to multiplication by $e^{\frac{i\pi}{2r}}$ and simple
closed curves to self-adjoint operators. In view of 
 Theorem~\ref{SvNpillow} this representation is equivalent to the
standard representation. This proves the existence of the map $\rho(h)$
that satisfies the exact Egorov identity with quantizations of Wilson
lines. Schur's
lemma implies that $\rho(h)$ is unique up to multiplication by a constant
and that $\rho$ is a projective representation of the mapping class group.
 Let us mention that
a computational proof
of uniqueness was given in \cite{gelca2}.
\end{proof}

Also from Theorem~\ref{SvNpillow} we deduce that for every  $h$ in the mapping
class group, the map $\rho(h)$ can be represented as multiplication
by a skein ${\mathcal F}(h)\in\widetilde{RT}_t({\mathbb T}^2\times [0,1])$. 
We want to find ${\mathcal F}(h)$ explicitly.
We consider first the case of the positive twist $T$ along
the curve $(0,1)$. 
Since the twist leaves the curve $(0,1)$  invariant,
$${\mathcal F}(T)(0,1)V^k(\alpha)=(0,1){\mathcal F}(T)V^k(\alpha), \quad
\mbox{ for all }k.$$ And because the eigenspaces of
 $\op(W_{(0,1)})$ are $1$-dimensional,
the linear operator defined by ${\mathcal F}(T)$ on the Hilbert space is 
a polynomial in $\op (W_{(0,1)})$. The polynomials
  $S_j(x)$, $0\leq j\leq r-1$ form a basis for ${\mathbb C}[x]/S_{r-1}(x)$, so
  \begin{eqnarray*}
{\mathcal F}(T)=\sum_{j=1}^{r-1}c_jS_{j-1}((0,1)), \quad c_j\in {\mathbb C}.
 \end{eqnarray*}
  
On the other hand,
\begin{eqnarray}\label{fandzeroone}
(1,1){\mathcal F}(T)V^k(\alpha)={\mathcal F}(T)(1,0)V^k(\alpha).
\end{eqnarray}
Using  (\ref{actiontorus}) and the
fact that $S_{j-1}((0,1))V^k(\alpha)=\frac{[jk]}{[k]}V^k(\alpha)$ for
all $j$ and $k$, we  rewrite (\ref{fandzeroone}) as
\begin{eqnarray*}
&&\sum_jc_j\frac{[jk]}{k}t^{-1}(t^{-2k}V^{k+1}(\alpha)+t^{2k}V^{k-1}(\alpha)\\
&&\quad =\sum_jc_j\left(\frac{[j(k+1)]}{[k+1]}V^{k+1}(\alpha)+\frac{[j(k-1)]}{[k-1]}
V^{k-1}(\alpha)\right).
\end{eqnarray*}
Setting the coefficients of $V^{k+1}$ on both sides equal yields
\begin{eqnarray*}
\sum_{j=1}^{r-1}c_j[j(k+1)]=\frac{[k+1]}{[k]}t^{-2k-1}\sum_{j=1}^{r-1}c_j[jk].
\end{eqnarray*}
Denoting $\sum_jc_{j=1}^{r-1}[j]=t^{-1}u$, we obtain the system of equations
in  $c_j$, $j=1,2,\ldots, r-1$,
\begin{eqnarray*}
\sum_{j=1}^{r-1}[kj]c_j=[k]t^{-k^2}u, \quad k=1,2,\ldots, r-1.
\end{eqnarray*}
Recall that $[n]=\sin \frac{n\pi }{r}/\sin \frac{\pi}{r}$,
so the coefficient matrix is a multiple of the matrix of the
discrete sine transform. The square of the discrete sine transform 
is the identity map, so there is a constant $C$ such that 
\begin{eqnarray*}
c_j=C\sum_{k}[jk][k]t^{-k^2}.
\end{eqnarray*}
Standard results in the theory of Gauss sums \cite{lang} show that
 $\sum_k[jk][k]t^{-k^2}=C'[j]t^{j^2}$ where $C'$ is a constant independent
of $j$.
We conclude that ${\mathcal F}(T)$ is a multiple of
$\sum_{j=1}^{r-1}[j]t^{j^2}S_{j-1}((0,1))$. We normalize  ${\mathcal F}(T)$ to
make it unitary
by multiplying  by
 $\eta=\sqrt{\frac{r}{2}}\sin\frac{\pi}{r}=(\sum_{j=1}^{r-1}[j]^2)^{-1/2}$,
and for reasons that will become apparent in a moment we  
also multiply it by $t^{-1}$. 
In view of this formula, and by analogy with 
 \S~\ref{sec:3.4}, we define
\begin{eqnarray*}
\Omega_{SU(2)}=\eta\sum_{j=1}^{r-1}[j]V^j(\alpha)=\eta\sum_{j=1}^{r-1}S_{r-1}(
\alpha)\in \widetilde{RT}_t(S^1\times {\mathbb D}^2).
\end{eqnarray*}

For a framed link $L$, let $\Omega_{SU(2)}(L)$ be the skein obtained
by replacing each component of $L$ by
$\Omega_{SU(2)} $ such that 
the curve $(1,0)$ on the boundary of the solid torus is mapped to the framing.
Then ${\mathcal F}(T)$ is the coloring by $\Omega_{SU(2)}$
of the surgery curve of $T$. This is true for any twist, and
because on the one hand  any element of the mapping class group
is a product of twists, and on the other $\rho(h)$ is unique,
we obtain a description of the map $\rho(h)$ in topological terms, as
an the analogue of Theorem~\ref{skeinfourier}.

\begin{theorem}\label{resetturaev}
Let $h$ be an element of the mapping class group of the torus 
obtained by performing
surgery on a framed link $L_h$ in ${\mathbb T}^2\times [0,1]$. 
The map $\rho(h):\rcstor\rightarrow \rcstor$ is given by 
\begin{eqnarray*}
\rho(h)\beta=\Omega_{SU(2)}(L_h)\beta.
 \end{eqnarray*}
\end{theorem}

\begin{remark}
Everyone familiar with the Witten-Reshetikhin-Turaev theory  has
recognized the element $\Omega_{SU(2)}$, which is a fundamental
building block of that theory. The exact  Egorov identity
in Theorem~\ref{egorov} implies that $\gamma$ can be slid
along  $L_h$ colored by $\Omega_{SU(2)}$. Once this is observed,
it is natural to try slides over knots in general, and to deduce the
Reshetikhin-Turaev formula for $3$-manifold invariants 
\cite{reshetikhinturaev}.    
\end{remark}

The projective representation of the mapping class group from
Proposition~\ref{resetturaev} is the Reshetikhin-Turaev representation.
There is an alternative  representation, with a more geometric flavor.
Since odd and even theta functions
are invariant under the  Hermite-Jacobi action given by 
discrete Fourier transforms, for  an element 
$h$ of the mapping class group of the torus given by (\ref{sl2z})
 we have the action  on the $\zeta_j^\tau$'s:
\begin{eqnarray*}
\rho_{HJ}(h)\zeta_j^\tau(z)=
\exp\left(-\frac{\pi i Ncz^2}{c\tau+d}\right)
\zeta_j^{\tau '}\left(\frac{z}{c\tau+d}\right)
\end{eqnarray*}
where $\tau'=\frac{a\tau+b}{c\tau+d}$. 

We  compare this representations to the Reshetikhin-Turaev
representation. 
Modulo a multiplication by a positive normalization constant, 
\begin{eqnarray*}
&&\rho_{HJ}(S)\zeta^\tau_j(z)=\rho(S)(\theta_j^\tau(z)-\theta_{-j}^\tau(z))=
\sum_{k=0}^{2r-1}\left(e^{-\frac{\pi i}{r}jk}\theta_k^\tau(z)-
e^{\frac{\pi i}{r}jk}\theta_{-k}^\tau(z)\right)\\ 
&&\quad=\sum_{k=1}^{r-1}\left(e^{-\frac{\pi i}{r}jk}\theta_k^\tau(z)-
e^{\frac{\pi i}{r}jk}\theta_{-k}^\tau(z)\right) +\sum_{k=1}^{r-1}
\left(e^{-\frac{\pi i}{r}j(2r-k)}\theta_{2r-k}^\tau(z)\right.\\ &&\quad \left.-
e^{\frac{\pi i}{r}j(2r-k)}\theta_{k}^\tau(z)\right)
=2\sum_{k=1}^{r-1}\left(e ^{-\frac{\pi i }{r}jk}-e^{\frac{\pi
      i}{r}jk}\right)
(\theta^\tau_k(z)-\theta^\tau_{-k}(z))\\
&&\quad = -4i\sum_{k=1}^{r-1}\sin \frac{\pi jk}{r}\zeta_k^\tau(z),\\ 
&&\rho_{HJ}(T)\zeta_j^\tau(z)=\rho(T)(\theta_j^\tau-\theta_{-j}^\tau(z))=
e^{\frac{\pi i}{r}j^2}(\theta_j^\tau(z)-\theta_{-j}^\tau(z))
 =e^{\frac{\pi i}{r}j^2}\zeta_j^\tau(z).
\end{eqnarray*}

The matrix of  $\rho(S)$ 
defined via quantum groups has the
 $j,k$-entry equal to the Reshetikhin-Turaev invariant of the
Hopf link with components colored by $V^j$ respectively $V^k$, which
is $[jk]$. We normalize both $\rho_{HJ}(S)$ and $\rho(S)$   by  multiplying them
by $\eta$. The map
$\rho(T)$ introduces a positive twist 
on each basis element, and as such it is the diagonal matrix with
diagonal entries $e^{\frac{\pi i}{2r}(j^2-1)}$.
We have
\begin{eqnarray*}
\rho(T)=e^{\frac{i\pi}{2r}}\rho_{HJ}(T).
\end{eqnarray*}
This explains why although for
Weyl quantization one can pass to a ${\mathbb Z}_2$-extension of
$SL(2,{\mathbb Z})$ to remove projectiveness and obtain
a true representation, for the Reshetikhin-Turaev
theory one  has to take a full  ${\mathbb Z}$-extension.

The $S$-matrices satisfy
\begin{eqnarray*}
\rho(S)=i\rho_{HJ}(S).
\end{eqnarray*}
We have $\rho(S)^2=Id$ but
$\rho_{HJ}(S)^2=-Id$. This is reflected in the fact that the map
$z\rightarrow -z$ does not change the basis element $V^j(\alpha)$,
while $\zeta_j(-z)=-\zeta_j(z)$. So for the  Weyl quantization
the curve $b$ in Figure~\ref{rigtorus} does have to be oriented, while for
the quantum group quantization it does not. 
Note also the equality
$\Omega_{SU(2)}=\rho(S)\phi.$

The relationship with classical theta functions allows us to
 adapt a formula of Ka\v{c} and Peterson \cite{kacpeterson} 
to obtain an explicit formula for the Reshetikhin-Turaev representation
of the mapping class group of the torus. 
\begin{theorem}
Let 
\begin{eqnarray*}
h=\left(
\begin{array}{rr}
a& b\\
c&d
\end{array}
\right), 
\end{eqnarray*}
be an element of the mapping class group of the 
torus. Then there is a number $c(2r,h)\in
{\mathbb C}$ such that
\begin{eqnarray*}
\rho(h)\zeta_j^\tau(z)=c(2r,h)
\sum_{k}e^{\frac{\pi i}{2r}(cdk^2+abj^2)}[bckj]\zeta_{aj+ck}^\tau(z)
\end{eqnarray*}
where the sum is taken over a family of $j\in {\mathbb Z}$ that
give all representatives of the classes $cj$ modulo $2r{\mathbb Z}$ 
and the square brackets denote a quantized integer. 
 \end{theorem}

\begin{proof}
Because $2r$ is an even integer, 
the group $SL_\theta(2,{\mathbb Z})$ is the whole
$SL(2,{\mathbb Z})$. By Proposition 3.17 in \cite{kacpeterson},
there is a constant $\nu(2r,h)$ such that 
\begin{eqnarray*}
\theta^{\tau '}_j\left(\frac{z}{c\tau +d}\right)=\nu(2r,h)
\exp\left(\frac{2\pi i r
cz^2}{c\tau +d}\right)\sum_ke^{\frac{i\pi }{2r}(
cd k^2+2bckj+abj^2)}\theta^\tau_{aj+ck}(z)
\end{eqnarray*}
with the same summation convention as in the statement of the theorem,
and with $\tau '=\frac{a\tau +b}{c\tau +d}$. The map 
\begin{eqnarray*}
\theta_j^\tau(z)\rightarrow \sum_k
e^{\frac{i\pi }{2r}(cd k^2+2bckj+abj^2)}\theta^\tau_{aj+ck}(z)
\end{eqnarray*}
is, up to multiplication by a constant, 
the unique map that satisfies the exact Egorov identity with
the representation of the Heisenberg group. It follows that
\begin{eqnarray*}
\zeta_j^\tau(z)\rightarrow 
\sum_{k}e^{\frac{\pi i}{2r}(cdk^2+abj^2)}[bcjk]\zeta_{aj+ck}^\tau(z)
\end{eqnarray*}
satisfies the exact Egorov identity with the Weyl quantization of
the pillow case. There is a unique map 
with this property, up to multiplication
by a constant. Hence the conclusion.
\end{proof}

\subsection{The structure of the reduced Reshetikhin-Turaev skein algebra of
  the cylinder over a surface}\label{sec:6.2}

As mentioned above, the element $\Omega_{SU(2)}$ is the fundamental
building block in the construction of  the Witten-Reshetikhin-Turaev
 quantum invariants of $3$-manifolds. Here are some of its well known
properties that will be used in the sequel.

\begin{figure}[h]
\centering
\scalebox{.35}{\includegraphics{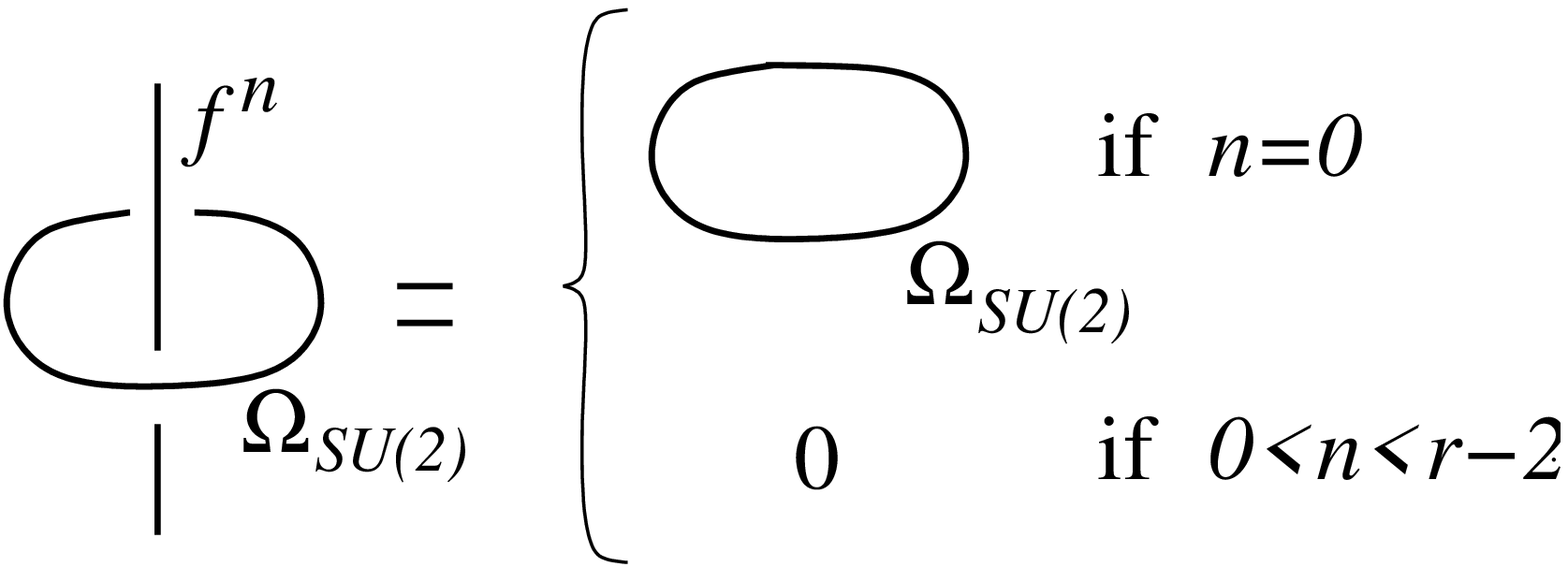}}
\caption{}
\label{identity}
\end{figure}

\begin{proposition}\label{identities}
If $M$ is an orientable compact $3$-manifold, then in $\widetilde{RT}_t(M)$
the following hold:
\begin{itemize}
\item[a)] if $O$ is the framed unknot in $M$, then 
$\Omega_{SU(2)}(O)=\eta^{-1}\emptyset$, 
\item[b)] if $K$ and $K'$ are framed knots in $M$,
then in $\widetilde{RT}_t(M)$, 
\begin{eqnarray*}
K\cup \Omega_{SU(2)}(K)=(K\#K')\cup \Omega_{SU(2)}(K'),
\end{eqnarray*}
(recall that $K\#K'$ denotes the slide of $K$ along $K'$, see 
\S~\ref{sec:3.4}),\\
\item[c)] for all skeins, the skein relation from {Figure~\ref{identity}} 
holds.   
\end{itemize}
\end{proposition} 

\begin{lemma}\label{totspatiul}
The   quantum group quantizations of all 
Wilson lines on a surface generate the algebra of all linear 
operators on the Hilbert space of the quantization.
\end{lemma}

\begin{proof}
The case of the torus was addressed in Theorem~\ref{SvNpillow}. For a higher
genus surface $\Sigma_g$, 
the conclusion  follows if we show that every nonzero vector is  cyclic
for the  algebra generated by quantized Wilson lines.

Recall the curves $\alpha_i$ from Figure 20. 
Given a knot in the handlebody $H_g$, we  can talk about the linking
number of this knot with one of the curves $\alpha_i$; just
embed the handlebody in $S^3$ in standard position. We agree to
take this with a positive sign. The linking number of a link $L$ in $H_g$
with the curve $\alpha_i$ is the sum of the linking numbers of the components.
Associate to $L$ the number $d(L)$ obtained by summing these for
all $i=1,2,\ldots, 3g-3$. Finally, for a skein $\sigma=\sum c_jL_j$, 
where $L_j$ are links and $c_j\in {\mathbb C}$, let $d(\sigma)=
\max_jd(L_j)$. We claim that for each skein $\sigma$ that is
not a multiple of the empty link, there is
a skein $\sigma'$ such that $d(\sigma')<d(\sigma)$ and
$\sigma'$ is in the cyclic representation generated by $\sigma$.

To this end write $\sigma$ in a basis of eigenvectors
of the $3g-3$-tuple $$(\op(W_{\alpha_1}),\op(W_{\alpha_2}),
 \ldots, \op(W_{\alpha_{3g-3}}))$$ as $\sigma=\sum c_je_j$. Because
the spectral decomposition of this  $3g-3$-tuple of operators
 has only $1$-dimensional eigenspaces, each
$e_j$ with  nonzero coefficient is in the cyclic representation
generated by $\sigma$. For each such $e_j$, $d(e_j)\leq d(\sigma)$.
If one of these inequalities is sharp, then the claim is proved. If
not, we show that if $e_k$ is not the empty link (i.e. the trivalent
graph with all edges colored by $V^1$), then in the cyclic representation
generated by $e_k$ there is a skein $\sigma'$ with $d(\sigma')<d(e_k)$. 

After deleting all edges of $e_k$ colored by the trivial representation $V^1$,
the not necessarily connected graph obtained 
has an edge whose endpoints
coincide, which  is colored by some nontrivial representation $V^n$. Let $\beta$
be a framed simple closed  curve on $\sigma _g$ that is isotopic to
this edge and choose an $\alpha_i$ that intersects $\beta$ 
as shown in Figure~\ref{loop}. 

\begin{figure}[h]
\centering
\scalebox{.35}{\includegraphics{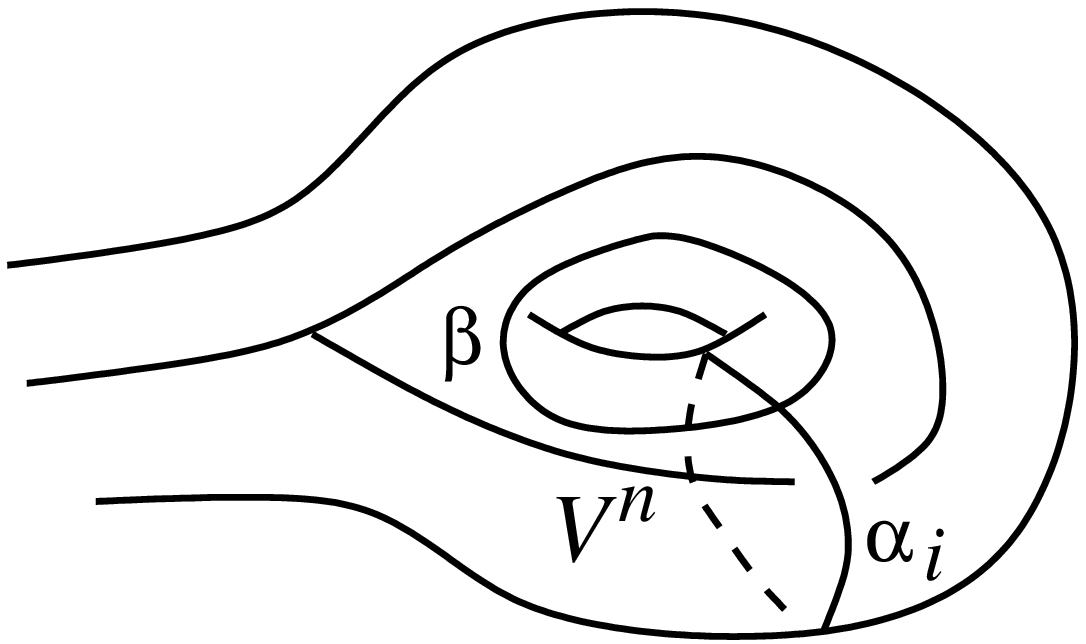}}
\caption{}
\label{loop}
\end{figure}
The recursive formula in Lemma~\ref{unu} implies that
$\op(W_{\beta})e_k$ is the sum of two skeins, $\sigma'$ that has the edge linking
$\alpha_i$ colored by $V^{n-1}$ and $\sigma''$ 
that has the edge linking $\alpha_i$
colored by $V^{n+1}$. It is a standard fact that  $\sigma'$  is an
eigenvector of $\op(W_{\alpha_i})$ with eigenvalue $[2n-2]$, while, if it is
nonzero, then 
$\sigma''$ is an eigenvector of $\op(W_{\alpha_i})$ with eigenvalue $[2n+2]$.  
We can therefore conclude that $\sigma'$ is in the cyclic representation
generated by $e_k$, and therefore in the cyclic representation generated
by $\sigma$. 

Repeating, we  eventually descend to the empty link.
It remains to show that the empty link is cyclic. But this
is obviously true, since each basis element can be represented as a framed
link, hence is the
image of a collection of nonintersecting framed simple closed curves on the 
boundary. This completes the proof.
\end{proof}

\begin{theorem}\label{faithful}
Given a genus $g$  surface $\Sigma_g$, $g\geq 1$, the representation of 
$\widetilde{RT}_t(\Sigma_g\times [0,1])$ on $\widetilde{RT}_t(H_g)$
 is faithful. Moreover, $\widetilde{RT}_t(\Sigma_g\times [0,1])$ is the
algebra of all linear operators on $\widetilde{RT}_t(H_g)$. 
\end{theorem}

\begin{proof}
In view of Lemma~\ref{totspatiul}, 
it suffices to show that the dimension of 
the vector space $\widetilde{RT}_t(\Sigma_g\times [0,1])$ equals the
square of the dimension of $\widetilde{RT}_t(H_g)$. This fact, 
at least for the Kauffman bracket, goes back
to unpublished work of J. Roberts.  In the case of the Kauffman bracket,
part of the proof can be found in \cite{sikora} and the case where 
$r$ is an odd prime can be found in \cite{charliejoanna}. 

For two compact, orientable $3$-dimensional
manifolds $M$ and $N$, we denote by $M\#N$ their connected sum, 
obtained by removing a $3$-dimensional ball from both $M$ and $N$
and then gluing the resulting manifolds along the newly obtained
 boundary spheres.
In $M\#N$, the manifolds $M$ and $N$ are separated by a sphere
$S_{sep}$.
In particular, by turning the handlebody $H_g$ inside out,
we see that $H_g\#H_g$ is an $S^3$ with two handlebodies    removed.

\begin{lemma}
Given the $3$-dimensional manifolds $M$ and $N$, the  map 
\begin{eqnarray*}
\widetilde{RT}_t(M)\otimes \widetilde{RT}_t(N)\rightarrow
\widetilde{RT}_t(M\#N)
\end{eqnarray*}
defined by $(L,L')\rightarrow L\cup L'$,  where $L$
and $L'$ are framed links in $M$ respectively $N$,  
is an  isomorphism of vector spaces.
\end{lemma}

\begin{proof}
The proof of this lemma was inspired by \cite{przytyckisum}.
Any skein in $\widetilde{RT}_t(M\#N)$ can be written as
$\sum_{j=1}^{k}c_j\sigma_j$, where $c_j$'s are complex coefficients and
each $\sigma_j$ is a skein that intersects $S_{sep}$ along the $j$th 
Jones-Wenzl idempotent. Taking a trivial knot colored by $\Omega_{SU(2)}$
and sliding it over $S_{sep}$ we obtain, by using
Proposition~\ref{identities} c),
 the equality
\begin{eqnarray*}
\eta^{-1}\sum_{j=0}^{r-2}c_j\sigma_j=\eta^{-1}c_0.
\end{eqnarray*}  
Hence any skein is equal to a skein that is disjoint from $S_{sep}$.
This proves that the map from the statement is an epimorphism.  
It is also a monomorphism because the skein module of a regular 
neighborhood of $S_{sep}$ is trivial. Hence it is an isomorphism.
\end{proof}

From here we continue as in \cite{sikora}.
The manifold $H_g\times[0,1]$ can be obtained from $H_g\#H_g$ by
surgery on a $g$-component framed link $L_g$ as shown in 
Figure~\ref{surgeryhandle}.
Let $N_1\subset H_g\#H_g$ be a regular neighborhood of $L_g$, which consists
of $g$ solid tori. Let $N_2\subset H_g\times [0,1]$ be the union of
the $g$ surgery tori and $L_g'$ the framed link in $H_g\times [0,1]$
consisting of the cores of these tori. Then $H_g\#H_g$ can be 
obtained from $H_g\times [0,1]$ by performing surgery on $L_g'$.

Every skein in $H_g\#H_g$ respectively $H_g\times [0,1]$ can
be isotoped as to miss $N_1$ respectively $N_2$.
 The homeomorphism  $\phi:(H_g\#H_g)\backslash N_1
\rightarrow (H_g\times [0,1])\backslash N_2$ induces  
an  isomorphism of skein modules
 $$\phi:\widetilde{RT}_t((H_g\#H_g)\backslash N_1)
\rightarrow\widetilde{RT}_t((H_g\times[0,1])\backslash N_2).$$ 
Unfortunately $\phi$ does not  induce an isomorphism between
$\widetilde{RT}_t(H_g\#H_g)$ and $\widetilde{RT}_t(H_g\times [0,1])$, it
does not even give a well defined map 
because a skein in $H_g\#H_g$ can be pushed in many ways off $N_1$,
while the images of these push-offs through $\phi$ are not isotopic. 

\begin{figure}[h] 
\centering
\scalebox{.30}{\includegraphics{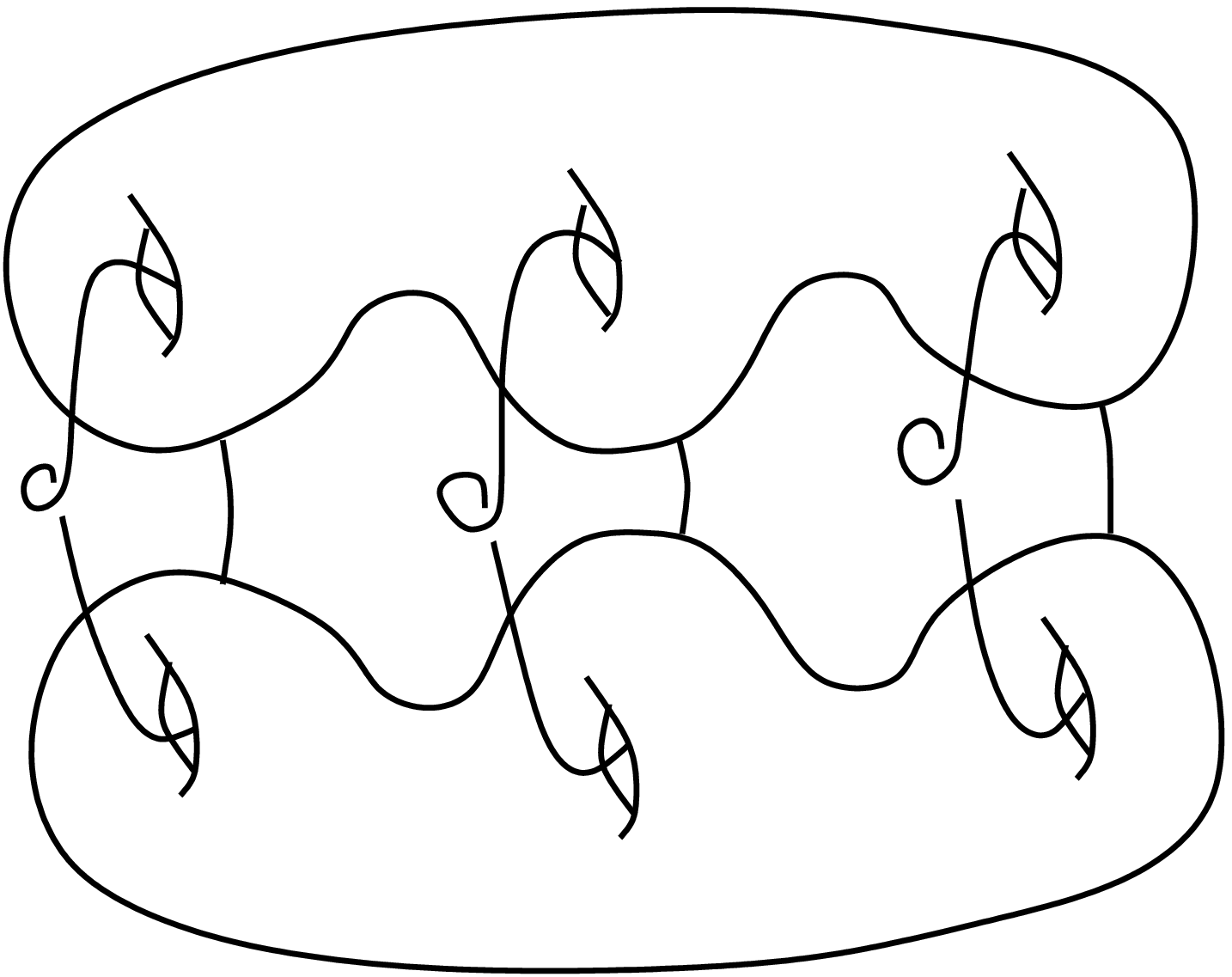}}
\caption{}
\label{surgeryhandle}
\end{figure}

Sikora's idea was to define 
$F_1: \widetilde{RT}_t(H_g\times[0,1])\rightarrow
\widetilde{RT}_t(H_g\#H_g)$ by $F_1(\sigma)=\phi(\sigma)\cup \Omega_{SU(2)}(L_g')$.
By Proposition~\ref{identities} b), we are allowed to slide
$\phi(\sigma)$ along $L_g'$, which implies that this map is well defined.    
Its inverse is $F_2(\sigma)=\phi^{-1}(\sigma)
\cup\Omega_{SU(2)}(L_g)$.  Indeed, to see that these maps are 
the inverse of each
other, push $L_g'$ off $N_2$ in the direction of the framing of $L_g'$. 
Then each of the components of $\phi^{-1}(L_g')$ is the meridian 
of the surgery torus, and it surrounds  once the corresponding 
component of $L_g$. By Proposition~\ref{identities} c), 
$\Omega_{SU(2)}(L_g)\cup \Omega_{SU(2)}(\phi^{-1}(L_g'))=\emptyset 
\in \widetilde{RT}_t(H_g\#H_g)$. Hence $F_2\circ F_1=Id$. 
Similarly we check that $F_1\circ F_2=Id$, and the theorem is proved.
\end{proof}

\begin{remark}
This result is the non-abelian analogue of Proposition~\ref{allspace}.
\end{remark}

\subsection{The quantization of Wilson lines determines the  Reshetikhin-Turaev representation}
\label{sec:6.3}

Like for the torus, there is
an  action of the mapping class group  of a 
surface $\Sigma_g$ on the ring 
of regular functions on the moduli space ${\mathcal M}_g$ of
flat $SU(2)$-connections on $\Sigma_g$ given  by  
$h\cdot f(A)=f(h^{-1}_*A)$.
The action of the mapping class group on regular functions
on ${\mathcal M}_g$
induces an action on the quantum observables, which for Wilson lines
is given by
\begin{eqnarray*}
h\cdot \op(W_{\gamma})=\op(W_{h(\gamma)}).
\end{eqnarray*}

\begin{theorem}\label{egorov2}
There is a  projective representation $\rho$ of the 
mapping class group of a closed
surface that satisfies the exact Egorov identity 
\begin{eqnarray*}
\op(W_{h(\gamma)})=\rho(h)\op(W_{\gamma})\rho(h)^{-1}
\end{eqnarray*}
with the quantum group quantization of Wilson lines.
Moreover, for every $h$, $\rho(h)$ is unique up
to multiplication by a constant. 
\end{theorem}

\begin{proof}
We mimic the second proof to Theorem~\ref{defofphi}.
The bijective map  $L\rightarrow h(L)$ on the set of isotopy classes
of framed links in the cylinder over the torus induces an automorphism of
the free ${\mathbb C}[t,t^{-1}]$-module with basis these isotopy
classes of links. Because the
ideal defined by the skein relations is invariant under this map,
the map defines an automorphism $$\Phi:\widetilde{RT}_t(\Sigma_g
\times [0,1])\rightarrow \widetilde{RT}_t(\Sigma_g\times [0,1]).$$
By Proposition~\ref{faithful} the algebra $\widetilde{RT}_t(
\Sigma_g\times[0,1])$ is the algebra of all linear operators 
on $\widetilde{RT}_t(H_g)$, so the automorphism $\Phi$
is inner \cite{waerden}. This proves the existence of $\rho(h)$.
The fact that $\rho$ is a representation and the uniqueness are
consequences of Schur's lemma.
\end{proof}

Each element of the mapping class group  preserves
the Atiyah-Bott symplectic form, so it induces a
symplectomorphism of the ${\mathcal M}_g$. 
The theorem proves  that the symplectomorphisms of ${\mathcal M}_g$ that
arise from elements of the mapping class group can be quantized.
Their quantization plays the role of the Fourier transform for
non-abelian theta functions.
Of course $\rho$ is (up to multiplication by constants) the 
Reshetikhin-Turaev representation. 
This  result shows that {\em all the information about 
the Reshetikhin-Turaev representation of the mapping class group
is contained in the quantum group quantization of Wilson lines}. 

Once we know that the element $\Omega_{SU(2)}$ allows handle slides,
discovering the formula for $\rho(h)$ is easy. However if we
were able to prove a Stone-von Neumann theorem in higher genus,
or to prove Theorem~\ref{faithful} without using links colored
by $\Omega_{SU(2)}$, then the formula
for $\rho(h)$ could be deduced like in the case of the torus.
The  idea is to write $\rho(h)$ as a composition of twists
along nonseparating curves using the Lickorish twist theorem,
then examine each twist separately. 

If $\gamma $ is such a curve,
with corresponding twist $T_{\gamma}$, and if $\rho(T_{\gamma})$
is represented by a skein ${\mathcal F}(T_{\gamma})$, then
${\mathcal F}(T_{\gamma})$  commutes with all skeins that do not
intersect $\gamma$. One can show that on
each eigenspace of $\op (W_{\gamma})$ these skeins span the algebra
of all linear operators. Hence the skeins commuting with
${\mathcal F}(T_{\gamma})$ span the algebra of all operators that 
commute with $\op(W_{\gamma})$. Consequently, ${\mathcal F}(T_{\gamma})$
is a polynomial in $\gamma$. Next we can restrict ourselves
to a solid torus containing $\gamma$ and follow the
steps from the computation in \S~\ref{sec:6.1} to deduce the formula
for ${\mathcal F}(T_{\gamma})$.

As explained in \cite{turaev} and \cite{walker}, the 
projective representation of the mapping class group can be
made into a true representation by passing to a ${\mathbb Z}$-extension
of the mapping class group. Like for classical theta
functions, the extension can be defined  in terms of either 
the Maslov index. 

For this, fix a rigid structure on the surface
 $\Sigma_g$ and consider the subspace
${\bf L}$ of $H_1(\Sigma_g,{\mathbb R})$ spanned by the curves that dissect
the surface into pairs of pants. Then ${\bf L}$ is a Lagrangian subspace
of $H_1(\Sigma_g,{\mathbb R})$ with respect to the intersection form.
The composition of extended homeomorphisms is defined by
\begin{eqnarray*}
(h',n')\circ (h,n)=(h'\circ h, n+n'-{\boldsymbol \tau}({\bf L},h({\bf L}),h'\circ
 h({\bf L})),
\end{eqnarray*}  
where ${\boldsymbol \tau}$ is the Maslov index with respect 
to the intersection pairing.


For completeness, 
we conclude our discussion with the proof of the following result mentioned
in \S~\ref{sec:5.2} and whose importance was addressed at the end of 
\S~\ref{sec:5.3}.

\begin{proposition}
The bilinear pairing
used in the definition of the quantum group quantization from \S~\ref{sec:5.2}
is nondegenerate.
\end{proposition}

\begin{proof}
 We first give a description of the inner product
on the Hilbert space 
${\mathcal H}_r(\Sigma_g)=\widetilde{RT}_t(H_g)$ by diagrams,
 following an idea in \cite{charliejoanna}.

The handlebody $H_g$ has a natural orientation reversing symmetry $s$
that leaves its core invariant. Glue two copies of $H_g$ along
their boundaries by the restriction of $s$ to the boundary
to obtain a connected sum of $g$ copies of $S^1\times S^2$, 
 denoted $\#_gS^1\times S^2$. This induces a pairing  
\begin{eqnarray*}
\left<\ ,\ \right>_0
:\widetilde{RT}_t(H_g)\times \widetilde{RT}_t(H_g)\rightarrow 
\widetilde{RT}_t(\#_gS^1\times S^2).
\end{eqnarray*}
The manifold $\#_gS^2\times S^2$ is obtained from $S^3$ by
performing surgery on the trivial link with $g$ components.
Identifying $\widetilde{RT}_t(\#_gS^1\times S^2)$ with
$\widetilde{RT}_t(S^3)$ via Sikora's isomorphism
as in the proof of Theorem~\ref{faithful}, we deduce that
the pairing $\left<\ ,\ \right>_0$ takes values in ${\mathbb C}$. 

The pairing
of two basis elements is given by the Reshetikhin-Turaev
invariant of a graph like the one in Figure~\ref{nondeg}. We argue on
this figure, but one should keep in mind that there are
many different graphs that can be the cores of the same handlebody.
By Proposition~\ref{identities} c), in order for this
Reshetikhin-Turaev invariant to be nonzero, in each pair of  edges linked by a
circle colored by $\Omega_{SU(2)}$ the colors  must be equal.
This is because in order for the tensor product
$V^{j_i}\otimes V^{k_i}$ to contain a copy of $V^1$, the dimensions 
of the two irreducible representations must be equal. 
Note also that because we work in $S^3$,
the pairs of edges like the $V^{j_3}$ and 
$V^{k_3}$ are also linked by a circle colored by $\Omega_{SU(2)}$,
namely the circle that links $V^{j_4}$ and $V^{k_4}$. In general, the edges
corresponding to decomposition circles that do not disconnect the surface
fall in this category.

\addtocounter{figs}{-1}
\begin{figure}[ht] 
\centering
\scalebox{.40}{\includegraphics{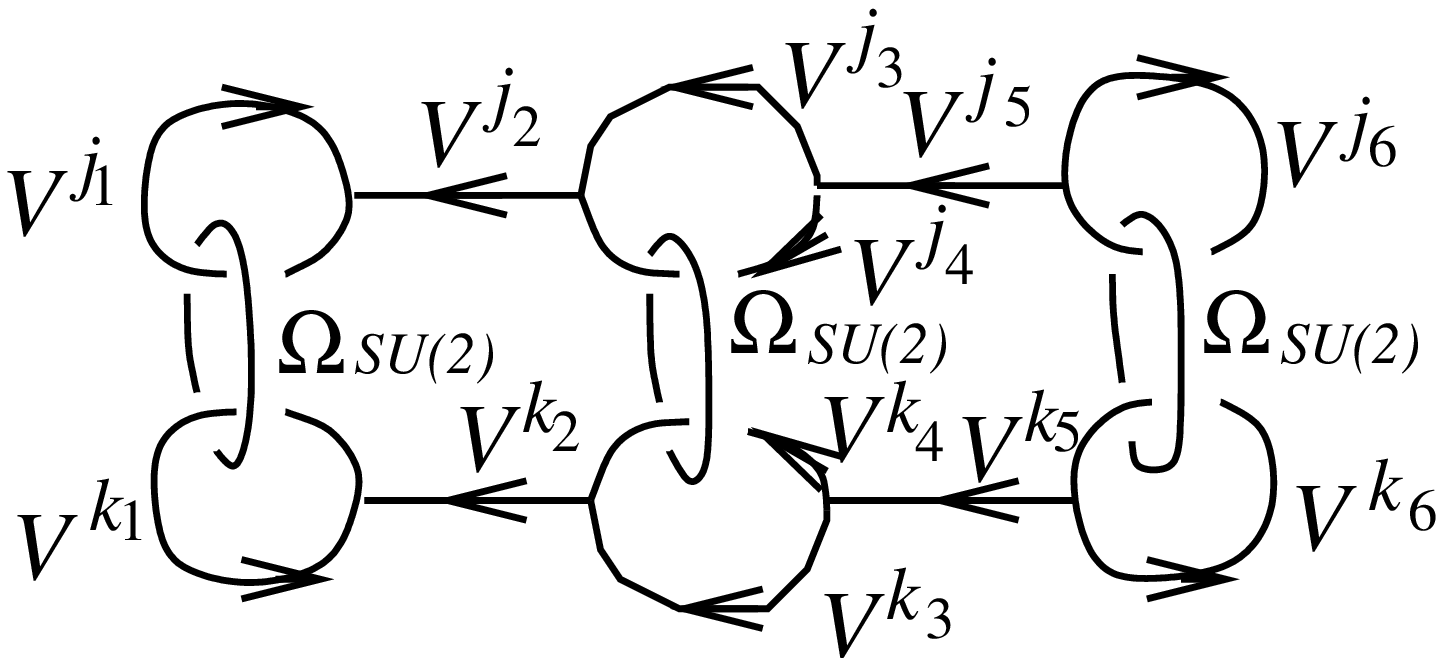}}
\caption{}
\label{nondeg}
\end{figure}

Let us  examine next the pairs of edges that are not linked by surgery
circles, such as those colored by $V^{j_2}$, $V^{k_2}$ in the figure.
In general, the edges that come from decomposition circles that
disconnect the surface fall in this category.
Rotating the graph by $90^\circ$ and evaluating the Reshetikhin-Turaev
invariant by the rules we obtain a homomorphism ${\mathbb C}=V^1
\rightarrow V^{j_2}\otimes V^{k_2}$. This homomorphism is nonzero if
and only if $j_2=k_2$. We conclude that the pairing of two
distinct basis elements is zero. On the other hand, computing the pairing
of a basis element with itself we can trace a $V^1$ from the bottom
to the top, and the value of  the pairing is $\Omega_{SU(2)}(O)=\eta^{-g}$.
Hence $\left<\ ,\ \right>_0=\eta^{-g}\left<\ ,\ \right>$, 
where $\left<\ ,\ \right>$ 
is the inner product.

The bilinear pairing  $[\ ,\ ]$ from \S~\ref{sec:5.2}
 is defined by gluing two copies of $H_g$
along an orientation reversing homeomorphism as to obtain $S^3$. 
The homeomorphism
is of the form $s \circ h$, so $[e_i,e_j]=\left<e_i,\rho(h)e_j\right>$.
Because $\rho(h)$ is an automorphism of the Hilbert space
of the quantization, the pairing is nondegenerate.  
\end{proof}

\end{document}